\documentclass[aps,prd,twocolumn,superscriptaddress,nofootinbib,10pt]{revtex4-2}

\usepackage[T1]{fontenc}
\usepackage[utf8]{inputenc}
\usepackage{amsmath,amssymb,mathtools}
\usepackage{tensor}
\usepackage{physics}
\usepackage{lipsum}
\usepackage{bigints}
\usepackage{verbatim}
\usepackage{braket}
\usepackage{cancel}
\usepackage{booktabs}
\usepackage{color}
\usepackage{graphicx}
\usepackage{subcaption}
\usepackage{hyperref}
\usepackage{nameref}
\usepackage{cleveref}

\makeatletter
\def\l@subsubsection#1#2{}%
\makeatother

\usepackage{graphicx,subcaption}

\newcommand{\roughly}[1]{\mathrel{\raise.3ex\hbox{$#1$\kern-0.85em
\lower1ex\hbox{$\sim$}}}}
\newcommand{\lsim}{\roughly<}

\newcommand{\mfa}{{\mathfrak a}}

\newcommand{\bfg}{{\mathbf{g}}}

\newcommand{\cH}{{\cal H}}

\newcommand{\cO}{{\cal O}}

\newcommand{\ssB}{{\scriptscriptstyle B}}

\newcommand{\ssN}{{\scriptscriptstyle N}}

\newcommand{\ssQ}{{\scriptscriptstyle Q}}

\newcommand{\GN}{G_\ssN}
\newcommand{\MP}{M_p}

\newcommand{\MEW}{M_\EW}
\newcommand{\exd}{{\rm d}}

\newcommand{\EW}{{\scriptscriptstyle EW}}

\newcommand{\ax}{{\mfa}}
\newcommand{\sax}{{\rm ax}}

\newcommand{\nn}{\nonumber}
\newcommand{\be}{\begin{equation}}
\newcommand{\bea}{\begin{eqnarray}}
\newcommand{\ee}{\end{equation}}
\newcommand{\eea}{\end{eqnarray}}

\newcommand{\gdm}{{\zeta}}
\newcommand{\bitemize}[1]{\medskip\noindent \textbf{#1} \hspace{-2 mm}}

\begin{document}

\author{Adam Smith}
\affiliation{School of Mathematical and Physical Sciences, University of Sheffield, Hounsfield Road, Sheffield S3 7RH, United Kingdom}

\author{Maria Mylova}
\affiliation{Kavli Institute for the Physics and Mathematics of the Universe (IPMU)%, The University of Tokyo, Kashiwa, Chiba 277-8583, Japan
}
\affiliation{Perimeter Institute for Theoretical Physics%, 31 Caroline Street North, Waterloo ON, Canada.
}

\author{Carsten van de Bruck}
\affiliation{School of Mathematical and Physical Sciences, University of Sheffield, Hounsfield Road, Sheffield S3 7RH, United Kingdom} 

\author{C.P.~Burgess}
\affiliation{Perimeter Institute for Theoretical Physics%, 31 Caroline Street North, Waterloo ON, Canada.
}
\affiliation{Department of Physics \& Astronomy, McMaster University%, 1280 Main Street West, Hamilton ON, Canada.
}
\affiliation{School of Theoretical Physics, Dublin Institute for Advanced Studies%, 10 Burlington Rd., Dublin,  Co. Dublin, Ireland
}

\author{Eleonora Di Valentino}
\affiliation{School of Mathematical and Physical Sciences, University of Sheffield, Hounsfield Road, Sheffield S3 7RH, United Kingdom}

\date{\today}

%\title{Universal Mass Rescaling from a Light Dilaton: A Self-Consistent Recombination-Era Solution to the Hubble Tension}
\title{
\texorpdfstring
{The Serendipitous Axiodilaton:\\
A Self-Consistent Recombination-Era Solution to the Hubble Tension}
{The Serendipitous Axiodilaton: A Self-Consistent Recombination-Era Solution to the Hubble Tension}
}

\date{\today} % optional

\begin{abstract}
Axio--dilaton cosmology provides a minimal benchmark model for both Dark Matter (DM) and Dark Energy (DE) that is well motivated by fundamental physics. The axion and dilaton arise as pseudo-Goldstone modes of symmetries that predict particle masses to depend on the dilaton, and therefore to evolve cosmologically, leading to correlated modifications of recombination physics, the sound horizon, and late-time expansion and growth histories. We confront this model with Planck 2018 temperature, polarisation, and lensing data, SPT-3G high-$\ell$ measurements, DESI DR2 BAO, and Pantheon$+$ supernovae, assuming that the axion makes up all of the dark matter and that the dilaton plays the role of a dark energy field. We find that it fits the data somewhat better than $\Lambda$CDM cosmology, with the $\chi^2$ lowered by $\simeq 7$ for three additional parameters, and significantly raises the inferred Hubble constant to $H_0 \simeq 69.2~\mathrm{km\,s^{-1}\,Mpc^{-1}}$, reducing the Hubble tension to $\lesssim 3\sigma$ and thereby allowing a joint fit of CMB and SH0ES data. The model fits this enlarged data set as well as the $w_0w_a$ model with an electron mass modified by hand at recombination, though it does so with calculable dynamics. Axio--dilaton self-interactions robustly fake a phantom equation of state in DESI measurements.
There is a sting: cosmology prefers dilaton--matter couplings $|\mathbf{g}|\sim 10^{-2}$--$10^{-1}$, which are large enough to have been detected in solar-system tests of General Relativity. These results show how axio--dilatons can provide a viable cosmology preferred by current data at surprisingly large couplings, within a framework that links dark energy, dark matter, and time-dependent particle masses in a coherent way. They suggest both new observable signals and new theoretical directions, aimed at resolving the apparent inconsistency with non-cosmological observations.
\end{abstract}

\maketitle

%\setcounter{tocdepth}{1}
%\tableofcontents
%\noindent\rule{\linewidth}{0.4pt}

\section{Introduction}

The succinct description of precision cosmological data by the six-parameter $\Lambda$CDM framework is now part of the furniture within the household of modern science, though a range of persistent tensions has begun to make it seem increasingly threadbare. The most prominent such tension is the discrepancy in the inferred Hubble constant $H_0$ between early- and late-time probes, though the same high-precision datasets that sharpen this tension also point towards dynamics in the dark energy sector. $\Lambda$CDM also leaves fundamental questions unanswered, like the microscopic origin of dark energy, the nature of dark matter, and the origin of the striking hierarchy of scales in fundamental physics.

These developments have stimulated many searches for ways to extend standard cosmology, but rather than pointing to a single, universally accepted picture, current data instead allow a broad landscape of possible extensions (for reviews see~\cite{Verde:2019ivm,DiValentino:2021izs,Abdalla:2022yfr,DiValentino:2022fjm,Kamionkowski:2022pkx,CosmoVerse:2025txj}).

In this paper we compare cosmological observations with a class of theories that are minimal (in a sense made clear below), theoretically well-motivated, and {\it not} built with the resolution of cosmic tensions in mind. We find these describe the data remarkably well, resolving the current tensions while introducing new puzzles of their own.

\subsection{Anomalies}

Although cosmologists have been disagreeing on the value of the Hubble constant since before most of us were born, the current version of the Hubble tension rests on a mature comparison between independent analyses of multiple high-precision datasets.

Within $\Lambda$CDM, \textit{Planck} 2018 temperature and polarisation data favour $H_0 \simeq 67\,\mathrm{km\,s^{-1}\,Mpc^{-1}}$~\cite{Planck:2018vyg}, in close agreement with the combined SPT+ACT CMB constraints, which give $H_0 \simeq 66.6\,\mathrm{km\,s^{-1}\,Mpc^{-1}}$ with comparable uncertainties~\cite{SPT-3G:2025bzu}. These measurements infer $H_0$ from the physics of the early Universe, essentially by fitting the CMB acoustic pattern and propagating it forward assuming $\Lambda$CDM.

Late-time “distance-ladder’’ determinations instead measure $H_0$ directly from the local expansion rate, using calibrated standard candles. The SH0ES programme, calibrated with Cepheids and now anchored by multiple geometric distance indicators, reports $H_0 \simeq 73.2\,\mathrm{km\,s^{-1}\,Mpc^{-1}}$ with an uncertainty below $2\%$~\cite{Riess:2021jrx,Breuval:2024lsv}. This is confirmed by the recent distance-network measurements from the H0DN collaboration~\cite{H0DN:2025lyy}, reaching $1\%$ precision.

Although systematic errors are difficult to assess, the discrepancy has survived repeated re-examination and currently exceeds $7\sigma$ when the latest H0DN determination is compared with the Planck+SPT+ACT CMB estimate. It remains robust across different ladder calibrators, Hubble-flow samples, and CMB analyses~\cite{Verde:2019ivm,DiValentino:2020zio,Schoneberg:2021qvd,Abdalla:2022yfr,DiValentino:2024yew,CosmoVerse:2025txj}. This robustness has motivated a broad set of proposals that modify either the pre-recombination expansion (early-time solutions) or the late-time dark-energy sector, or both, in order to reconcile early and late determinations of $H_0$~\cite{Murgia:2016ccp,Pourtsidou:2016ico,Nunes:2016dlj,Kumar:2016zpg,Kumar:2017dnp,DiValentino:2017iww,Yang:2018uae,DiValentino:2019ffd,Yang:2020uga,Lucca:2020zjb,DiValentino:2020leo,Kumar:2021eev,Nunes:2021zzi,Gariazzo:2021qtg,Bernui:2023byc,Mishra:2023ueo,vanderWesthuizen:2023hcl,Zhai:2023yny,Liu:2023kce,Hoerning:2023hks,Pan:2023mie,Castello:2023zjr,Forconi:2023hsj,Yao:2023jau,Garcia-Arroyo:2024tqq,Benisty:2024lmj,Silva:2024ift,Giare:2024ytc,Bagherian:2024obh,Sabogal:2025mkp,DiValentino:2016hlg,DiValentino:2017rcr,Dutta:2018vmq,vonMarttens:2019ixw,DiValentino:2020naf,Yang:2021flj,DiValentino:2021rjj,Heisenberg:2022lob,Giare:2023xoc,Adil:2023exv,Lapi:2023plb,Krolewski:2024jwj,Bousis:2024rnb,Tang:2024gtq,Manoharan:2024thb,Poulin:2018cxd,Smith:2019ihp,Niedermann:2019olb,Krishnan:2020obg,Schoneberg:2021qvd,Ye:2021iwa,Poulin:2021bjr,Niedermann:2021vgd,deSouza:2023sqp,Poulin:2023lkg,Cruz:2023lmn,Niedermann:2023ssr,Vagnozzi:2023nrq,Efstathiou:2023fbn,Cervantes-Cota:2023wet,Garny:2024ums,Giare:2024akf,Giare:2024syw,Poulin:2024ken,Pedrotti:2024kpn,Kochappan:2024jyf,Greene:2023cro,Greene:2024qis,Baryakhtar:2024rky,Lynch:2024hzh,Lee:2022gzh,Seto:2024cgo,Poulin:2023lkg,DiValentino:2019exe,Alestas:2021luu,Ruchika:2023ugh,Frion:2023xwq,Ruchika:2024ymt,Visinelli:2019qqu,Ye:2020btb,Calderon:2020hoc,Sen:2021wld,DiGennaro:2022ykp,Ong:2022wrs,Akarsu:2019hmw,Akarsu:2021fol,Akarsu:2022typ,Akarsu:2023mfb,Anchordoqui:2023woo,Adil:2023exv,Akarsu:2024qsi,Halder:2024uao,Anchordoqui:2024gfa,Gomez-Valent:2023uof,Akarsu:2024eoo,Yadav:2024duq,DiValentino:2020vnx,Silva:2025hxw,Paraskevas:2024ytz,Scherer:2025esj,Gomez-Valent:2024tdb,Akarsu:2025gwi,Soriano:2025gxd,Bouhmadi-Lopez:2025ggl,Pan:2019hac,Yang:2020zuk,Yang:2021eud,Jiang:2024xnu,Gomez-Valent:2024ejh,Specogna:2025guo,Ozulker:2025ehg,Lee:2025pzo,Hart:2017ndk,Hart:2019dxi,Knox:2019rjx,Sekiguchi:2020teg,Schoneberg:2024ynd,Toda:2024ncp}.

In parallel, the first and second data releases of DESI have revealed a statistically independent discrepancy in the universe's late-time expansion history. DESI maps the baryon acoustic oscillation (BAO) feature in the clustering of galaxies and quasars, providing a standard ruler for distances at redshifts $z \sim 0.3$–$2.3$. When these BAO measurements are combined with CMB constraints, the inferred evolution of the dark energy sector departs from a pure cosmological constant at the $\sim 3\sigma$ level, with the significance rising to $\sim 3$–$4\sigma$ once Type Ia supernovae are included~\cite{DESI:2024mwx,DESI:2025zgx,DES:2025sig}.

The emerging picture is that the same data sets that underpin $\Lambda$CDM are beginning to favour non-trivial dynamics in both the early and late Universe. A large body of recent work has therefore focused on phenomenological parameterisations that quantify deviations from $\Lambda$CDM cosmology. Examples include representations of the dark energy equation of state -- such as the $w_0w_a$ (or CPL) parameterisation -- effective rescalings of the lensing amplitude, and models with ad hoc variations in fundamental constants like the electron mass $m_e$ or fine–structure constant $\alpha$~\cite{DESI:2024mwx,DESI:2025zgx,Cortes:2024lgw,Shlivko:2024llw,Luongo:2024fww,Yin:2024hba,Gialamas:2024lyw,Dinda:2024kjf,Najafi:2024qzm,Wang:2024dka,Ye:2024ywg,Tada:2024znt,Carloni:2024zpl,Chan-GyungPark:2024mlx,DESI:2024kob,Bhattacharya:2024hep,Ramadan:2024kmn,Notari:2024rti,Orchard:2024bve,Hernandez-Almada:2024ost,Pourojaghi:2024tmw,Giare:2024gpk,Reboucas:2024smm,Giare:2024ocw,Chan-GyungPark:2024brx,Menci:2024hop,Li:2024qus,Li:2024hrv,Notari:2024zmi,Gao:2024ily,Fikri:2024klc,Jiang:2024xnu,Zheng:2024qzi,Gomez-Valent:2024ejh,RoyChoudhury:2024wri,Li:2025cxn,Lewis:2024cqj,Wolf:2025jlc,Shajib:2025tpd,Giare:2025pzu,Chaussidon:2025npr,Kessler:2025kju,Pang:2025lvh,Roy:2024kni,RoyChoudhury:2025dhe,Paliathanasis:2025cuc,Scherer:2025esj,Giare:2024oil,Liu:2025mub,Teixeira:2025czm,Santos:2025wiv,Specogna:2025guo,Sabogal:2025jbo,Cheng:2025lod,Herold:2025hkb,Cheng:2025hug,Ozulker:2025ehg,Lee:2025pzo,Ormondroyd:2025iaf,Silva:2025twg,Ishak:2025cay,Fazzari:2025lzd,Smith:2025icl}.

Such approaches can be valuable for charting which kinds of departures from $\Lambda$CDM improve the fit to specific datasets and for connecting anomalies in $H_0$ and late-time distances to broad classes of early- or late-time physics. However, parameterisation in itself is not explanatory and is mute about the underlying physics that might be responsible for any particular observed effects. Without a real theory it is impossible to identify the full set of observational and theoretical constraints that a physically consistent alternative to $\Lambda$CDM must obey.

\subsection{Theories vs models}

Varying–electron–mass scenarios provide a particularly clear illustration of this gap. In many implementations, $m_e(z)$ is taken to be a function of redshift, typically chosen with step–like features at $z \gtrsim 50$ to ensure the value at recombination differs from the value measured much nearer by. This modification is then inserted directly into numerical Boltzmann codes such as \textsc{CAMB}~\cite{Lewis:1999bs} or \textsc{CLASS}~\cite{Blas:2011rf} to see whether agreement with the data can be improved.

These studies have been useful for mapping the sensitivity of CMB and BAO observables to the value of $m_e$ at different redshifts, but they leave unspecified the underlying dynamics for why the masses should change in the first place. No explicit dynamical mechanism is provided to generate the variation, whose connection to the gravitational sector is undefined. It is left ambiguous whether the same physics could simultaneously satisfy constraints from local tests, big bang nucleosynthesis (BBN), and measurements of dimensionless observables like the proton-to-electron mass ratio, $\mu := m_p/m_e$.

An alternative approach instead starts with dynamics and studies how cosmology changes within specific field theories, such as scalar–tensor or coupled dark energy models. Typically a light scalar field $\phi$ is coupled to matter and/or dark matter, and its classical field equations are solved to compute how the background expansion and the growth of structure are modified. In practice, constraints are obtained by implementing the modified Einstein–Boltzmann equations in numerical solvers. Most of this literature, however, confines the scalar to the dark sector or focuses on its purely gravitational effects, with no direct couplings to electrons and baryons that could give rise to variations in visible-sector parameters like $m_e$ (and so also to cosmologically relevant quantities like the Thomson cross section or atomic binding energies). Bounds derived for these types of scalars might well be misleading if naively applied to a dynamical theory that predicts evolution for things like $m_e$.

Brans--Dicke scalars~\cite{Jordan:1955, Brans:1961sx, Dicke:1961gz} (for a review see~\cite{Brans:2014}) provide simple examples of dynamical fields that do couple to matter and so provide a natural framework for having field-dependent masses and so also for describing mass evolution during cosmology. They are also well motivated since they often arise within UV completions to gravity, such as within extra-dimensional models (often paired with a second scalar -- an `axion' -- when supersymmetry plays an extra-dimensional role). But Brans--Dicke scalars are often not envisioned to be light enough to evolve significantly during the recent Universe, and if they are light their evolution is governed by a scalar potential whose design to ensure any desired cosmological evolution of masses carries a whiff of being a `just-so' story.

Even worse, approaches using scalars usually ignore what might be the most important clue of all: whatever describes cosmology must arise as the low-energy limit of the fundamental theory describing all of the rest of physics. This is a surprisingly restrictive condition because the scalar potential is both the part of the theory most directly responsible for its phenomenological success and the part that is most notoriously sensitive to quantum influences from high-energy degrees of freedom. Scalar potentials are dangerous in cosmology because they can easily undermine the validity of the low-energy expansion~\cite{Gildener:1976ih, Burgess:2009ea, Adshead:2017srh}, but for gravitational theories it is this same low-energy expansion that ultimately justifies use of the semiclassical methods that everyone actually uses in practice~\cite{Weinberg:1978kz, Burgess:2003jk}.

\subsection{Yoga fitness}

What to do? In this paper we focus on the simplest scalar sector that is motivated by the above discussion. We ask it to capture the minimal physics of the complete dark sector ({\it i.e.} both Dark Energy and Dark Matter) and so require at least two scalars. As argued in~\cite{Smith:2024ayu}, two scalars are also generically attractive when seeking deviations from General Relativity (GR) because two is the minimum number of scalars for which nontrivial two-derivative sigma-model interactions can exist, which the general power-counting arguments of~\cite{Burgess:2009ea, Adshead:2017srh} show are the ones that most naturally compete with the two-derivative interactions of GR at low energies (for a review of the issue see~\cite{Burgess:2025vxs}).

We further choose our two scalars to be an axio-dilaton pair (for which a shift symmetry ensures the two-derivative interactions are independent of one of the fields -- the `axion'). We do so because these arise so generally as model-independent moduli in higher-dimensional models (and so are well motivated), but also because an axio-dilaton combination plays a central role in recent attempts~\cite{Burgess:2021obw} to understand the small size of the Dark Energy density in a way consistent with high-energy quantum effects (doing so through a `natural relaxation' mechanism -- hence the name: Yoga models).

The dilaton part of the axio-dilaton pair plays the role of Dark Energy and arises as the Goldstone boson for an accidental scale invariance very common in extra-dimensional constructions~\cite{Burgess:2020qsc}. This makes it a Brans--Dicke scalar that automatically couples to matter in a way that gives particles field-dependent masses. Yoga models predict its scalar potential has the Albrecht--Skordis form~\cite{Albrecht:1999rm} first considered long ago (including implications for evolving particle masses~\cite{Albrecht:2001xt}), in a manner not motivated at all by the current Hubble tensions. The axion of the axio-dilaton pair is the Dark Matter, though its absence of direct couplings to matter means it cannot be the QCD axion~\cite{Weinberg:1977ma, Wilczek:1977pj} that is hypothesized to play a role in understanding the Strong CP problem.

The cosmology of this axio-dilaton system was recently explored~\cite{Smith:2024ibv} and found to be broadly viable. More interestingly, the theory's underlying scale invariance tends to cause the dilaton to evolve in `tracker' solutions during cosmology that follow the evolution of the dominant energy density at any given epoch. This tracker solution very generically must change at radiation-matter equality as the dominant energy density changes, causing a transient excursion that is still in progress at recombination (providing a robust reason for why masses at recombination might differ from their present-day and BBN values).

This framework naturally realises cosmologies with varying electron mass without ad hoc prescriptions, using dynamics motivated by plausible low-energy approximate symmetries common to extra–dimensional constructions together with a relaxation mechanism aimed at understanding the technical naturalness of dark energy. It provides a unified setting in which dark energy, dark matter, and varying masses arise from a single scalar sector.\footnote{The biggest problem with these models is the predicted coupling strength of the dilaton to ordinary matter, which is large enough to be ruled out in {\it e.g.}~solar-system tests of GR (more about which below). Viability of the model seems to require some sort of screening mechanism, perhaps along the lines studied in~\cite{Brax:2023qyp}.}

Following~\cite{Smith:2024ibv} we implement the axio–dilaton model in \texttt{CLASS}, evolving the background and scalar perturbations alongside the standard cosmological components, and feed the resulting redshift–dependent electron mass $m_e(z)$ self–consistently using the HyRec framework~\cite{Ali_Ha_moud_2011}. The resulting mass evolution modifies the hydrogen binding energy, Thomson cross section, and related atomic rates in parallel with the usual scalar–tensor effects. 

We then confront the results with the full suite of modern cosmological data: \textit{Planck} 2018 temperature, polarization, and lensing~\cite{Planck:2018vyg}, ACT DR6 lensing~\cite{ACT:2023kun} and SPT-3G high–$\ell$ measurements~\cite{SPT-3G:2023flv}, DESI DR2~\cite{DESI2025_DR2ResultsII} or SDSS BAO~\cite{Alam:2016hwk}, and Pantheon+ supernovae~\cite{Brout:2022vxf}. For context we compare the axio–dilaton constraints directly to those obtained in $\Lambda$CDM, $\Lambda$CDM\,+\,varying-$m_e$, and the $w_0w_a$ (or CPL)~\cite{Chevallier:2000qy, Linder:2002et, Caldwell:2005tm} parameterizations of Dark Energy (both with and without varying $m_e$), to allow a consistent comparison with phenomenological approaches commonly used in the literature.

We find that axio-dilaton models reduce the Hubble tension to below $3\sigma$, low enough to allow a joint fit to combined CMB and late-time Cepheid data for the calibration of the supernovae. The fit to these data is as good as for standard varying–electron–mass models, though with the enormous advantage of having calculable dynamics that correlate effects at early and late times (which, for instance, explains why masses need not differ at BBN even if they do at recombination). The improvement in the fit is significantly more than would have been expected just by the addition of the few new parameters that arise.

%\begin{table*}
%\centering
%\begin{tabular}{lccccccc}
%\hline\hline
%Model & $H_0$ & $\mathbf{g}$ & $\zeta$   & $\Delta\chi^2$  \\
%\hline
%Yoga-VI & $69.17^{+0.66}_{-0.76}$ (69.13) & $-0.001 \pm 0.095$ (-0.052) & $0.003 \pm 0.050$ (-0.003) & -7.2 \\
%EXP & $69.17 \pm 0.67$ (69.99) & $0.002 \pm 0.093$ (0.104) & $-0.001 \pm 0.044$ (-0.022) & -7.3 \\
%$\Lambda\text{CDM} + m_e$ & $69.74 \pm 0.66$ (69.39) & -- & -- & -6.7 \\
%$\text{CPL} + m_e$ & $68.32 \pm 0.83$ (67.95) & -- & -- & -12.1 \\
%\hline
%\end{tabular}
%\end{table*}

\begin{table*}
\centering
\makebox[\textwidth][c]{%
\begin{tabular}{lcccc}
\hline\hline
Model &  $H_0$ & $\mathbf{g}$ & $\zeta$ & $\Delta \chi^2$ \\
\midrule
Yoga VI  & $70.92 \pm 0.61$ (70.86) & $-0.01^{+0.19}_{-0.17}$ (0.099) & $0.006 \pm 0.059$ (-0.031) &  -19.2 \\
EXP & $70.80 \pm 0.58$ (71.13) & $0.00 \pm 0.16$ (-0.199) & $-0.001 \pm 0.061$ (0.082) & -19.5 \\
$\Lambda$CDM+$m_e$ & $70.97 \pm 0.57$ (71.34) & -- & -- &  -19.2 \\
$w_0w_a$+$m_e$ & $70.51 \pm 0.73$ (70.37) & -- & -- &  -19.4 \\
\bottomrule
\end{tabular}
}
\caption{Posterior means ($\pm 1\sigma$) -- with best-fit values given in parentheses -- for the main parameters of the axio–dilaton models: the inferred Hubble constant $H_0$, the baryon–dilaton coupling $\mathbf{g}$, and the dark-sector axion–dilaton coupling $\zeta$, as inferred by fitting to the CMB-A\,DESI\,PP\,SH0ES dataset. The last column gives the improvement in best-fit $\chi^2$ relative to $\Lambda$CDM (which is perverse because, strictly speaking, the Hubble tension makes $\Lambda$CDM too poor a description of the data to combine the entire dataset). This table is a subset of the results shown in Table~\ref{tab:ppshoes} in later sections. What this table does not show is the bimodal nature of the likelihood function that prefers nonzero dilaton--matter couplings but does not care much about their sign -- see {\it e.g.}~\Cref{fig:Exp Shoes} -- a feature that makes the marginalized allowed regions misleadingly concentrate around zero coupling.}
\label{tab:Summary Constraints}
\end{table*}

Axio-dilatons are only slightly worse than $w_0w_a$ models with a changed electron mass in terms of goodness-of-fit to the full dataset combinations considered here, though this comparison is fraught because it is difficult to know how many parameters are actually hidden inside an ad hoc change in $m_e$ within the $w_0w_a$+$m_e$ framework. Axio-dilaton success turns out also to be perfectly consistent with apparently phantom equations of state being favoured in $w_0w_a$ analyses of observations (such as DESI) despite satisfying all the positive-energy conditions that forbid $w < -1$ (see~\cite{Smith:2024ibv} for details).

Effective dilaton-baryon couplings of order $|\mathbf{g}|\sim\mathcal{O}(10^{-2}$–$10^{-1})$ are not only cosmologically allowed but modestly {\it preferred} over $\Lambda$CDM, and this is compatible with early- and late-time cosmological constraints once the full dynamics are taken into account. This demonstrates that including both gravitational and microphysical effects substantially shifts the qualitative picture of what is allowed for light, conformally coupled scalars. But it also means that cosmology favours couplings that are too large to have been missed in solar-system tests of GR, indicating that the couplings seen in cosmology should differ from those experienced in our immediate neighbourhood. It is an as-yet unsolved problem as to how these couplings might differ like this, though model-building efforts are underway.

The preference for nonzero couplings is well hidden by the summary given in Table~\ref{tab:Summary Constraints}, which gives a concise summary of the constraints and best-fit values (taken from Table~\ref{tab:ppshoes} given below). The allowed ranges shown in this table are obtained by marginalising over the other variables, but what the table does not show is the bimodal nature of the likelihood function (see~\Cref{fig:Exp Shoes}). The bimodality arises because the fits prefer nonzero couplings but do not much care about their sign, a situation that misleadingly leads marginalized intervals to be centered about zero.

The remainder of the paper is structured as follows.
In~\Cref{sec:axiodilaton} we introduce the axio–dilaton framework, define the model, and summarise the relevant field equations. \Cref{sec:Minimal_cosmology} develops the corresponding cosmology, covering the background evolution, linear perturbations, and the quasi–static regime in which the axion behaves as CDM with a time–dependent mass. We switch gears in \Cref{sec:data_analysis} to comparing with observations. \Cref{ssec:analysis} outlines the datasets and the methodology used in our joint likelihood analysis. Our main results are presented in \Cref{sec:results}, where we compare the Yoga-style and exponential axio–dilaton models, assess the impact of including SPT-3G data, and study the effect of the SH0ES prior. \Cref{sec:discussion} discusses the interpretation of these results, including the relation to phenomenological $m_e(z)$ models and to CPL extensions. We conclude in \Cref{sec:conclusions} with a brief summary. In the appendices, we provide some details of calculations; we review the observational probes sensitive to universal mass rescalings, including BBN, recombination physics, BAO and large–scale structure, 21 cm signals, astrophysical environments, spectroscopy, and local tests; and we summarize the constraints on all sampled cosmological parameters.

\section{Theoretical framework}\label{sec:axiodilaton}

We here define the model and specify the equations governing its cosmological evolution, largely following the detailed discussion given in~\cite{Smith:2024ibv}. Aficionados more interested in comparisons with the data should feel free to jump directly to \Cref{sec:data_analysis}.

\subsection{Axio-dilatons}

As motivated above, the model supplements the usual core theory (the Standard Model + General Relativity) with two new scalar fields: the dimensionless scalar dilaton denoted $\chi$ and the dimensionless axion denoted $\mfa$. These gravitate ({\it i.e.}~minimally couple to GR) and couple to one another through two-derivative sigma-model interactions, and the dilaton also couples to ordinary matter through a Brans--Dicke type coupling.

\subsubsection{Definition of the model}

The Einstein–frame action we study is given by\footnote{Unless explicitly stated otherwise we use fundamental units $c=\hbar=1$ throughout and define the reduced Planck mass in terms of Newton's constant by $\MP^{-2} = 8\pi \GN$.}
\begin{align}\label{eq:Action}
    S = \!\int \exd^4x \sqrt{-g}
    \Biggl\{
        \frac{\MP^2}{2}
        \Bigl[
            R
            &- \partial_\mu\chi\,\partial^\mu\chi
            - W^2(\chi)\,\partial_\mu\mfa\,\partial^\mu\mfa
        \Bigr]
        \nn\\&
        - V(\chi,\mfa)
    \Biggr\}
    + S_m(\Psi,\tilde g)\,,
\end{align}
where $R$ is the Ricci scalar and $S_m$ is the matter action for the Standard Model fields, which couple to the scalars only through the Jordan–frame metric\footnote{Models with similar motivations have been proposed, sometimes without the crucial couplings to matter (see e.g. \cite{Bernardo:2022ztc,Alexander:2022own,Toomey:2025yuy}).}
\begin{equation}\label{eq:jf_metric}
    \tilde g_{\mu\nu} = A^2(\chi)\,g_{\mu\nu}, 
    \qquad 
    A(\chi)=e^{\mathbf g\chi}.
\end{equation}
In particular, these definitions ensure that Einstein--frame particle masses depend on $\chi$ through
\begin{equation}\label{eq:mass_universal}
    m_i(\chi) \propto e^{\mathbf g\chi}.
\end{equation}

Taking the function $A(\chi)$ to be exponential ensures $\chi$ couples universally to matter as would a Brans--Dicke scalar, with the coupling $\mathbf g$ related to the usual Brans--Dicke parameter $\omega$ by $2\bfg^2 = (3+2\omega)^{-1}$. In the absence of screening mechanisms, solar--system tests require $|\mathbf g|\!\lesssim\!10^{-3}$ (see {\it e.g.}~\cite{Will:2014kxa}), though we here treat $\mathbf g$ as a free cosmological parameter, allowing for the possibility that the effective coupling on large scales differs from its locally measured value.

The functions $W(\chi)$ and $V(\chi,\mfa)$ specify the internal dynamics of the scalar sector. We choose the kinetic coupling to be exponential, $W(\chi)=e^{-\gdm\chi}$, with $\gdm$ setting the curvature of the axio–dilaton target space. An exponential form can naturally arise from accidental scaling symmetries, for which the target-space metric is proportional to the $\mathrm{SL}(2,\mathbb{R})$–invariant metric on the hyperbolic upper–half plane,
\begin{equation}
     \exd\chi^2 + e^{-2\gdm\chi}\, \exd\mfa^2 = \frac{1}{\zeta^2} \left( \frac{\exd \tau^2 + \exd \mfa^2}{\tau^2} \right),
\end{equation}
with $\tau = \gdm^{-1} e^{\gdm \chi}$. The particular choice $\gdm = \sqrt{2/3} \simeq 0.8$ corresponds to the `no-scale' structure of the Yoga model~\cite{Burgess:2021obw}.
 
For calculational simplicity we choose the scalar potential to be a sum of a dilaton and axion contribution,
\begin{eqnarray}\label{eq:joint_potential}
    V(\chi,\mfa) &\simeq& V_{\rm dil}(\chi) + V_{\rm ax}(\mfa)\nn\\
                  &=& V_p(\chi)\,e^{-\lambda\chi} + \tfrac12 \MP^2 m_\mfa^2 \mfa^2 ,
\end{eqnarray}
although this is really only expected to be approximately true in Yoga models for the small field excursions we in practice encounter in cosmological simulations (see~\cite{QCDAxion} for a fuller discussion of this difference).

For the same reason we take a quadratic approximation to the shift-symmetry-breaking axion potential with $m_\mfa$ treated as an independent low–energy parameter. This quadratic form should be viewed as the leading low-energy expansion about the vacuum of a periodic potential (which, in the case of axions descending from massive form fields under duality, is in particular a consequence of the low-energy derivative expansion on the dual side~\cite{Burgess:2025geh}).

The relaxation mechanism in Yoga models implies $V_{\rm dil} \propto \tau^{-4}$ and so predicts $\lambda = 4\gdm$, and this is part of its appeal: all hierarchies in the model are given by the expectation value of $\tau$, with the electroweak scale given by $M_\EW \sim \MP/\sqrt{\tau}$ and neutrino masses by $m_\nu \sim \MP/\tau$ (which are successful predictions when $\tau \sim 10^{28}$). Without relaxation one finds $V_{\rm dil} \sim \MP^4/\tau^2 \sim \MEW^4$ ({\it i.e.}~not small), but relaxation to $V_{\rm dil} \sim \MP^4/\tau^4$ gives a result that is parametrically a successful Dark Energy estimate: $(\MEW^2/\MP)^4$.

We examine two options for $V_p(\chi)$:

\bitemize{Exponential (EXP) potential:}
The first assumes it to be a constant $V_p(\chi)=V_0$, in which case the dilaton potential reduces to a scale-invariant runaway potential of a standard dilaton/quintessence type~\cite{Wetterich:1987fm, Ratra:1987rm, Peebles:1987ek, Ferreira:1997hj, Copeland:1997et} that has also more recently been investigated~\cite{Rahimy:2025iyj,Russo:2022pgo}. For sufficiently small $\lambda$ the dilaton can drive late–time accelerated expansion as $\chi$ slow-rolls down the potential.

\bitemize{Albrecht--Skordis (Yoga) potentials:}
The other choice entertains the possibility that $V_p(\chi)$ contains a mild polynomial dependence on $\chi$, encoding small loop-generated logarithmic scale-breaking effects. In this case, for concreteness’ sake we choose the form
\begin{equation}\label{eq:albrecht-skordis}
        V_p(\chi)=V_0\!\left[1-u_1\chi+\frac{u_2}{2}\chi^2\right], 
        \;\;\; \lambda = 4\sqrt{\tfrac{2}{3}}\,,
\end{equation}
where, following~\cite{Albrecht:1999rm,Burgess:2021obw}, the coefficients are chosen so that $V_{\rm dil}$ has a local minimum at which $V_{\rm dil} > 0$, separated from the runaway at $\chi \to \infty$ by a small local maximum. This minimum can easily be located at $\tau \sim 10^{28}$ using only values $u_i \sim \cO(50)$ because $V_p$ is a polynomial in $\ln \tau$. In this case the value of $V_{\rm dil}$ at this minimum becomes the late-time Dark Energy density, assuming the dilaton evolution is eventually trapped at this minimum.

\medskip

Later sections compare both choices for $V_p$ to observations.

\subsubsection{Field equations}

The evolution equations for both the axio-dilaton and the metric obtained by varying~\cref{eq:Action} with respect to these fields are 
\begin{align}
\label{eq:einstein_eq}
 & G_{\mu\nu}+\frac{1}{\MP ^2}  V(\chi,\mfa)\, g_{\mu\nu}  -\left(\partial_\mu \chi \, \partial_\nu \chi - \frac{1}{2} g_{\mu\nu} \, \partial^\sigma \chi  \,\partial_\sigma \chi \right) \nonumber
\\&\qquad\qquad-
W^2 \left( \partial_\mu \mfa  \,\partial_\nu \mfa - \frac{1}{2} g_{\mu\nu} \, \partial^\sigma \mfa \, \partial_\sigma \mfa \right) 
 = \frac{T_{\mu\nu}}{\MP^2} \, ,  
\end{align}
\begin{equation}
\label{eq:axion_eom}
\Box\mfa + \frac{2\,W_{,\chi}}{W}\,\partial_\mu\chi\,\partial^\mu\mfa - \frac{V_{,\mfa}}{W^2\,\MP^2} = 0 \,,
\end{equation}
and
\begin{equation}
\label{eq:dilaton_eom}
\Box\chi - W\,W_{,\chi}\,\partial_\mu\mfa\,\partial^\mu\mfa - \frac{V_{,\chi}}{\MP^2}
= -\,\frac{\mathbf g\,T}{\MP^2}\, ,
\end{equation}
where \(T_{\mu\nu}\) is the total Einstein-frame energy–momentum tensor of all matter and radiation species obtained from the matter action and $T := g^{\mu\nu}T_{\mu\nu}$ is its trace. 

Evolution for the cosmic matter and radiation fluids is instead found using the conservation of stress-energy. This states that $T_{\mu\nu}$ for each matter species obeys
\begin{equation}\label{eq:non-conservation}
\nabla_\mu T^{\mu\nu}_{(f)} = \,\mathbf g\,T_{(f)}\,\nabla^\nu\chi  \,.
\end{equation}
The right-hand side vanishes when $f$ is radiation, but for nonrelativistic matter the equation reads $\nabla_\mu T^{\mu\nu}_{(m)} \simeq -\,\mathbf g\,\rho_{m}\,g^{\mu\nu}\partial_\mu\chi$, where $\rho_{m}$ is the matter's Einstein–frame energy density.

\subsection{Cosmological Evolution}\label{sec:Minimal_cosmology}

We next briefly summarize the specialisation of these field equations to cosmology, for both a homogeneous and isotropic background and linearised fluctuations about this background, following~\cite{Smith:2024ayu, Smith:2024ibv, Smith:2025grk}. We work in Newtonian gauge, for which the background and fluctuating parts of the metric are given (in conformal time) by
\begin{equation}
    {\rm d}s^2
    = a^2(\eta)\Bigl[-(1+2\Phi)\,{\rm d}\eta^2
    + (1-2\Psi)\,\delta_{ij}\,{\rm d}x^i{\rm d}x^j\Bigr],
\end{equation}
where $a(\eta)$ is the background scale factor (with conformal Hubble scale $\cH = a'/a$, where primes denote differentiation with respect to $\eta$). $\Phi$ and $\Psi$ are the Bardeen potentials. The axion and dilaton are in principle similarly given by $\chi = \bar\chi(\eta) + \delta \chi$ and $\mfa = \bar\mfa(\eta) + \delta \mfa$, and all field equations are linearized about the background quantities (denoted by the overbars).

The main technical complication arises when the axion mass is large compared to the Hubble rate, $m_\mfa \gg \cH$, which is the regime of interest if the axion is to be the Dark Matter. In this case the axion undergoes rapid coherent oscillations around the minimum of its potential whose details can be averaged when predicting evolution on cosmological time frames. This average yields an effective pressureless fluid whose mean energy density redshifts as $a^{-3}$, up to corrections of order $(\cH/m_\mfa)^2$~\cite{Turner:1983he,Abbott:1982af,Preskill:1982cy}, and this behaviour underpins the standard identification of massive axions with Dark Matter. 

The coarse-graining over rapid oscillations captures all effects relevant for cosmological observables without needing to explicitly resolve the microscopic oscillations. In detail this happens because the averaging procedure introduces a small effective sound speed and gradient pressure that determine the axionic Jeans length and how this influences the growth of structures. It has been extensively studied in scalar–tensor and modified-gravity contexts~\cite{Brax:2004qh,Amendola:2005ad,vandeBruck:2022xbk,Poulot:2024sex}. Recent analyses have revisited and generalised these results in the context of early- and late-time cosmology, including ultralight axions, interacting axions, and axion dark-energy mixtures (e.g.~\cite{Hlozek:2014lca, Arias:2012az, Marsh:2015xka, Marsh:2015xka, Poulin:2018cxd, Efstathiou:2023fbn, Smith:2024ibv, Smith:2025grk, Gomes:2023dat}). 

In practice the rapid oscillations around the minimum are made explicit by writing
\begin{align}
\label{eq:asplit}
    \mfa(\eta,\mathbf{x})
    = &\frac{1}{\sqrt{2}\MP}
      \Bigl[
        \psi(\eta,\mathbf{x}) \, e^{-i\!\int m(\tilde\eta)\,d\tilde\eta}
        \nn\\&\qquad\qquad\qquad\qquad+ \psi^*(\eta,\mathbf{x}) \, e^{i\!\int m(\tilde\eta)\,d\tilde\eta}
      \Bigr] ,
\end{align}
where the time-dependent mass arises because of the axion's coupling to the time-dependent background dilaton:
\begin{equation}\label{eq:axion_effective_mass}
  m(\eta) \;=\; \frac{m_{\mfa}}{W(\bar\chi)} \,. 
\end{equation}
In the decomposition \cref{eq:asplit} the field $\psi(\eta,\mathbf{x})$ denotes the slowly varying envelope of oscillations whose time evolution happens over Hubble scales rather than typical oscillation times.

In the limit $m(t) \gg \cH$ and $m'/m \ll m$, the evolution of $\psi$ can be written much like the evolution of a fluid by following its modulus and phase separately, through the redefinition
\begin{equation}\label{eq:Madelung}
    \psi(\eta,\mathbf{x})
    = \frac{1}{m(\eta)} \, \sqrt{\rho_\ax(\eta,\mathbf{x})}\; e^{i S(\eta,\mathbf{x})} \,,
\end{equation}
with $\sqrt{\rho_\ax}/m$ playing the role of the amplitude of the oscillations and $S$ their collective phase. One computes the field equations obtained by varying $\rho_\ax$ and $S$ and separates these into the homogeneous background and fluctuations that vary in both space and time.

\subsubsection{Background evolution}
\label{ssec:background_evolution}

Defining the axion fluid's physical energy density as the time-time component of the axion's energy–momentum tensor, $\rho_{\sax} := -T_0^{0(\mfa)}$, one finds the simple background time dependence
\begin{equation}
  \label{eq:axion_background_density}
  \bar\rho_{\sax} \;\equiv\; \bar W^2\,\bar\rho_{\ax}
  \;=\; \frac{C\,m(\eta)}{a^3(\eta)} \,,
\end{equation}
where $C$ is a constant fixed by initial conditions. This expresses a conservation law for which the quantity $C/a^3$ behaves like a conserved axion number density. Equivalently,
\be
   \bar\rho_{\sax}'+3\cH\,\bar\rho_{\sax}=\left( \frac{m'}{m}\right)\,\bar\rho_{\sax}, 
\ee
which makes explicit how departures from pure $a^{-3}$ scaling are controlled by the slow scale $m'/m$.

The dilaton background knows about the background axion fluid through its derivative interactions with the axion. After averaging \cref{eq:dilaton_eom} over the rapid axion oscillations the evolution equation becomes
\begin{equation}
  \label{eq:dilaton_background}
  \bar\chi'' + 2\cH\,\bar\chi' + \frac{a^2}{\MP^2}\,V_{,\chi}
  \;=\; \frac{a^2}{\MP^2}\!\left(-\bfg\,\bar\rho_{\ssB}
  + \frac{W_{,\chi}}{W}\,\bar\rho_{\sax}\right) \,,
\end{equation}
where $V_{,\chi}$ and $W_{,\chi}$ denote the derivatives of $V$ and $W$ with respect to $\chi$ and are evaluated here at the dilaton background $\bar \chi$. The first source term ($\propto \bfg\,\bar\rho_\ssB$) captures momentum exchange with baryons implied by \cref{eq:jf_metric}, while the second ($\propto W_{,\chi}/W$) encodes the backreaction of the axion kinetic sector.

The same dilaton–matter couplings that cause these effects for structure formation also cause ordinary particle masses to depend on the dilaton and so also to vary in time as the dilaton background evolves. It is this feature that underpins the later success of these models in resolving cosmological tensions. Appendix~\ref{sec:constraints_universal} contains a summary of constraints on evolving particle masses, showing in particular how they are evaded within the axio–dilaton picture because of the universal way {\it all} particle masses get scaled the same way.

The expansion rate of the background metric is governed by the usual Friedmann equation, which is given from \cref{eq:einstein_eq} by 
\begin{equation}
  \label{eq:friedmann_background}
  \cH^2 \;=\; \frac{1}{3\MP^2}\!\left[
    \frac{\MP^2}{2}\,\bar\chi'^2 \;+\; a^2 V(\bar\chi)\;+\; a^2\bar\rho
    \;+\; a^2\bar\rho_{\sax}
  \right],
\end{equation}
where $\bar\rho$ collects the standard matter and radiation fluids (photons, neutrinos, and baryons), and $\bar\rho_{\sax}$ is given by~\cref{eq:axion_background_density}. 

The evolution of the remaining background fluids follows from the stress–energy (non)conservation as expressed by \cref{eq:non-conservation}. For relativistic species the trace vanishes, $T_r=0$, leading to the usual expressions \footnote{
We assume no direct coupling between neutrinos and the dilaton, whose impact depends on the detailed form of the neutrino mass function \cite{Burgess:2021obw}.
If such a coupling is present, neutrinos begin to exchange energy with the dark-energy sector once
they become non-relativistic, leading primarily to modifications of the late-time
ISW effect and mild changes to structure growth. In representative mass-varying
neutrino models these effects are concentrated at low multipoles
($\ell \lesssim 100$) and are typically at the few-percent level for order-unity
couplings and sub-eV neutrino masses, while larger couplings are constrained by
perturbative stability \cite{Brookfield:2005td, Brookfield:2005bz,Bjaelde:2007ki}. The detailed phenomenology is model dependent and we
therefore neglect these effects here.
}

\begin{equation}
    \bar\rho_\gamma' + 4\cH\,\bar\rho_\gamma = 0, 
    \qquad
    \bar\rho_\nu' + 4\cH\,\bar\rho_\nu = 0 \,.
\end{equation}
By contrast, the baryon density exchanges energy with the dilaton according to
\begin{equation}\label{eq:baryon_background_density}
    \bar\rho_{\ssB}' + 3\cH\,\bar\rho_{\ssB} = \,\bfg\,\bar\chi'\,\bar\rho_{\ssB} \,.
\end{equation}
The axion background is described by \cref{eq:axion_background_density}, with deviations from pure $a^{-3}$ controlled by $m'/m$ through the evolution of $W(\bar\chi)$.

These equations specify the background evolution and show how these now depend on the two new parameters $\bfg$ and $\gdm$ (which appears through $W_{,\chi}$ and $V_{,\chi}$), and once integrated are known to produce acceptable background cosmologies.

\subsubsection{Linear perturbations}
\label{sssec:linear_perturbations}

The evolution equations for first–order perturbations of the axion, dilaton, radiation, and baryon sectors are obtained in a similar way, together with the metric potentials that mediate their gravitational interactions. Axion fluctuations are computed using the coarse–grained fluid description described above, valid in the regime $m \gg \cH$ and $m'/m \ll m$, and for a spatially flat FRW background geometry we transform to Fourier space using the convention
\begin{equation}
    f(\mathbf{x},\eta)
    = \int\!\frac{d^3k}{(2\pi)^3}\,f(\mathbf{k},\eta)\,e^{i\mathbf{k}\cdot\mathbf{x}},
\end{equation}
with $k^2 \equiv \delta^{ij} k_i k_j$.

For each fluid we define the velocity perturbation from the momentum density of its energy–momentum tensor. Writing the perturbed stress tensor in Newtonian gauge as $T^{0}{}_{i} = -(\bar{\rho}+\bar{p})\,v_i$, the scalar velocity potential $v$ is introduced through $v_i = -i k_i v$, and the velocity divergence is then
\begin{equation}\label{eq:velocity_divergence}
    \Theta \equiv k^2 v
    = i\,k_j v^j .
\end{equation} 

\bitemize{Axion perturbations:}
At linear order the axion fluid turns out to be an \emph{almost} pressureless, CDM–like fluid with density contrast\footnote{Notice that $\delta_\mfa$ is defined as the relative density contrast of $\rho_\mfa = \Bar{\rho}_\mfa \delta_\mfa$ of \cref{eq:Madelung}, which should not be confused with the physical axion energy density $\rho_\sax = - T_0^{0(\mfa)}$ defined in \cref{eq:axion_background_density}. These are related by $\delta_\sax = \delta_\ax - \Phi - \Phi_\chi$, with $\Phi_\chi$ defined in \cref{eq:PhiQPhiChidefs}.} $\delta_\mfa$ and velocity divergence $\Theta_\mfa$. Their evolution is found in the same manner as for the background axion evolution, by substituting the ansatz \cref{eq:asplit} into the action \cref{eq:Action} and expanding all terms up to second order in perturbation theory and averaging over the fast–oscillating regime, leading to the perturbed continuity and Euler equations for the axion fluid:
\begin{equation}\label{eq:cont_ax}
    \delta_\mfa' - 2\Phi_\chi' + \Theta_\mfa = 3\Psi' + \Phi' ,
\end{equation}
\begin{equation}\label{eq:euler-equation}
    \Theta_\mfa' + \left(\cH + \frac{m'}{m}\right)\Theta_\mfa
    = k^2\left(\Phi + \Phi_\chi + \Phi_Q\right) ,
\end{equation}
where 
\begin{equation} \label{eq:PhiQPhiChidefs}
    \Phi_Q := \frac{k^2\,\delta_\mfa}{4a^2 m^2(\eta)}
    \qq{and}
    \Phi_\chi := -\,\frac{W_{,\chi}}{W}\,\delta\chi \, .
\end{equation}
See~\cite{Smith:2024ibv, Smith:2025grk} for the details of this derivation.

The term $\Phi_Q$ is the quantum–pressure contribution sourced by the scalar’s gradient energy.  
It is equivalent to a scale–dependent effective sound speed,
\begin{equation} \label{eq:csforaxion}
    c_s^2(k,\eta) = \frac{k^2}{4a^2 m^2(\eta)}\,,
\end{equation}
so that $\delta P_\mfa \simeq c_s^2\,\bar\rho_\sax\,\delta_\mfa$.  
On scales well above the axion Compton wavelength, $k \ll a m$, one has $c_s^2 \to 0$ and the axion behaves as standard CDM, whereas below that scale the gradient pressure opposes collapse and suppresses power.

The fluid equations above can be equivalently obtained by averaging the Klein–Gordon \cref{eq:axion_eom} in the $m\gg\cH$ regime, using a periodic ansatz
\begin{equation}
    \mfa(\eta,\mathbf{x})
    \;\simeq\;
    \frac{\sqrt{\rho_a(\eta,\mathbf{x})}}{m(\eta)\MP}\,
    \cos\!\bigg(\int m(\tilde\eta)\,d\tilde\eta+\vartheta(\eta,\mathbf{x})\bigg),
\end{equation}
where $\rho_a$ and $\vartheta$ vary slowly compared to the fast oscillation with frequency $m(\eta)$. In the associated hydrodynamic (Madelung) description one rewrites the dynamics in terms of a density and a velocity potential, and the Euler equation acquires a Bohm–potential term whose linearisation reproduces $\Phi_Q$~\citep{Brax:2019sdm,GalazoGarcia:2022SelfSimilar,Hui:2016ltb}.

This mapping clarifies the physical role of the quantum–pressure contribution. The term $\Phi_Q$ originates from gradients of the field and therefore behaves as an effective, scale–dependent pressure: it is negligible on large scales but becomes important near the axion Compton wavelength, where gradients are sizeable. In this regime the axion field can no longer be described as a perfectly smooth fluid; wave interference generates quasi-stationary patterns and, at high densities, supports soliton–like central cores within gravitationally bound objects~\citep{Hu:2000ke,Brax:2020Solitons,GalazoGarcia:2024SolitonsHalos}. As the effective mass $m(\eta)$ increases, the associated de~Broglie wavelength shrinks and the characteristic size of these cores decreases, so heavier axions produce more compact dark–matter halos at fixed mass.

The validity of the averaged fluid description requires $m\gg\cH$ for the background and $k\ll a m(\eta)$ for CDM–like clustering. Outside these limits one must retain the full field dynamics and the corresponding $\Phi_Q$ contribution rather than relying on the coarse–grained fluid equations.

\bitemize{Einstein constraints:}
The perturbed Einstein equations \cref{eq:einstein_eq} can be written in fluid form once the axion sector has been averaged, as described in \Cref{ssec:perutrbed_einstein_eqs}. In Fourier space the $(00)$ and $(0i)$ components give the usual constraint equations
\begin{align}
    k^2\Psi + 3\cH(\Psi' + \cH \Phi) &= -\,\frac{a^2}{2\MP^2}\,\delta\rho_{\rm tot}\,, \label{eq:ein00_fluid}\\[2mm]
    k^2(\Psi' + \cH \Phi) &= \frac{a^2}{2\MP^2}\,(\bar\rho_{\rm tot}+\bar P_{\rm tot})\,\Theta_{\rm tot}\,, \label{eq:ein0i_fluid}
\end{align}
where $\delta\rho_{\rm tot}$, $\bar\rho_{\rm tot}$, $\bar P_{\rm tot}$, and $\Theta_{\rm tot}$ denote the total density, pressure, and velocity divergence of all species (radiation, baryons, dilaton dark energy, and the axion fluid). The difference of the $(ij)$ equations determines the relation between the Bardeen potentials,
\begin{equation}
    \Psi - \Phi = \frac{a^2}{\MP^2}\,\pi_{\rm tot}\,,
\end{equation}
with $\pi_{\rm tot}$ the total anisotropic stress, dominated by photons and neutrinos and negligible at late times.

The contributions of the effective axion fluid defined in \cref{eq:axion_background_density} are
\begin{align}
    \delta\rho_\sax &= \bar\rho_\sax\,\delta_\sax\,, \\
    \delta P_\sax &= \bar\rho_\sax\,c_s^2(k,\eta)\,\delta_\mfa\,, \\
    (\bar\rho_\sax+\bar P_\sax)\,\Theta_\sax &\simeq \bar\rho_\sax\,\Theta_\mfa\,,
\end{align}
with the scale–dependent sound speed given by \cref{eq:csforaxion} and effective axion mass $m(\eta)=m_\mfa/W(\bar\chi)$. These expressions are what enter $\delta\rho_{\rm tot}$, $\delta P_{\rm tot}$, and $\Theta_{\rm tot}$ in \cref{eq:ein00_fluid}–\cref{eq:ein0i_fluid}. The explicit form of the axio–dilaton contributions in the $(00)$, $(0i)$, and $(ij)$ Einstein equations after averaging is given in Appendix~\ref{ssec:perutrbed_einstein_eqs}.

\bitemize{Dilaton fluctuation:}
Averaging the axion sector shifts the effective mass of $\delta\chi$ and sources it through $W_{,\chi}$. At linear order the dilaton fluctuation obeys
\begin{align}\label{eq:dilaton_perturbed}
& \delta\chi'' + 2\mathcal{H}\delta\chi' - \bar{\chi}' \left(\Phi'+3\Psi'\right)
 -\frac{W_{,\chi}}{W}\, \delta\!\left\langle \bar{W}^2\mfa'^2 \right\rangle
\nn\\
&\quad
+ \left[k^2 - \frac{a^2\bar\rho_{\sax}}{\MP^2}\left(\frac{W_{,\chi}^2}{W^2} + \frac{W_{,\chi\chi}}{W}\right)
 + \frac{a^2}{\MP^2} V_{,\bar{\chi}\bar{\chi}}\right] \delta\chi
\nn\\
&\qquad\qquad
+ 2\frac{a^2}{\MP^2} V_{,\chi}\,\Phi
= -\,\frac{a^2}{\MP^2}\,\bfg\left( \delta_\ssB + 2\Phi \right)\bar{\rho}_\ssB \, ,
\end{align}
where
\begin{equation}
    \delta\big\langle \bar W^2 \mfa'^2\big\rangle
    = a^2\frac{\bar\rho_{\sax}}{\MP^2}\,\delta_\mfa
      - 2a\,\frac{\bar\rho_{\sax}}{\MP^2}\,\frac{S'}{m(\eta)} \,,
\end{equation}
with $m(\eta)$ given by \cref{eq:axion_effective_mass} and the phase evolution equation gives $S'$ through
\begin{equation}\label{eq:phase_equation}
    \frac{S'}{a\,m(\eta)} = -\,\Phi_\chi - \Phi_\ssQ - \Phi \,.
\end{equation}
This equation plays the role of a Hamilton–Jacobi equation for the axion phase $S$ in the WKB/Madelung description. The Euler equation \cref{eq:euler-equation} follows from \cref{eq:phase_equation} upon acting with $\nabla^2$ and using the definition of the axion velocity divergence \cref{eq:velocity_divergence}. The terms proportional to $W_{,\chi}$ generate the source $\Phi_\chi = -(W_{,\chi}/W)\,\delta\chi$ that appears in the axion Euler equation, so that the dilaton perturbations are jointly sourced by the baryon coupling $\bfg$ and by the axion kinetic mixing $W_{,\chi}/W$.

\bitemize{Baryons and Electrons:}
Nonrelativistic particles like baryons and electrons have $\chi$-dependent masses and so their background and linear dynamics acquire source terms proportional to \(\bfg\). At linear order we have
\begin{equation}
\label{eq:baryondeltaevo}
     \delta_\ssB' + \Theta_\ssB - 3\Psi' = \mathbf{g}\,\delta\chi',
\end{equation}
and
\begin{equation}
 \label{eq:baryonthetaevo}
     \Theta'_\ssB + \Theta_\ssB\mathcal{H} - k^2\Phi
     = -\,\mathbf{g}\left(\bar{\chi}'\,\Theta_\ssB - k^2\delta\chi\right).
\end{equation}
The baryon Euler equation \cref{eq:baryonthetaevo} contains the two physical channels by which the dilaton affects the baryons: firstly through a direct momentum exchange (terms \(\propto \bfg\,\bar{\chi}'\,\Theta_\ssB\)) and secondly via a fifth–force correction to the potential gradient (\(\propto \bfg\,k^2\delta\chi\)). 

In our implementation the above equations are coded into \texttt{CLASS} and evolved numerically. As quantities like the electron mass evolve (due to their dependence on $\chi$), the microphysical rates relevant at recombination are also updated, ensuring the visibility function, sound horizon, and diffusion scale are updated consistently.

\subsubsection{Quasi--static description}\label{sssec:QSA_details}

Although not required for the numerical evolution, it is worth pausing also to describe how physics on sub-horizon scales simplifies within the quasi-static approximation, since this helps build intuition about what it is the numerics is doing. 

For a constant axion mass $m$ this approximation produces the familiar quantum–pressure cutoff in the matter power spectrum and a relation between the de~Broglie wavelength and the characteristic size of equilibrium cores in virialised halos; when $m$ is ultralight this reduces to the usual fuzzy–dark–matter phenomenology. In the minimal axion–dilaton framework two additional effects appear on top of this baseline:

\bitemize{Time–varying axion mass:}
Because $m(\eta)=m_\mfa/W(\chi)$, both the effective sound speed $c_s^2 \propto k^2/m^2(\eta)$ and the associated Jeans scale drift in time as the dilaton rolls. Episodes of rapid evolution in $W(\chi)$ temporarily enhance $c_s^2$ and suppress small-scale growth, while in periods where $W$ is nearly constant the axion again clusters like standard CDM. The resulting matter transfer function can therefore develop smooth, redshift–dependent breaks or changes in slope tied to epochs of significant dilaton evolution, rather than a single fixed cutoff scale.

\bitemize{Dilaton backreaction and mixed couplings:}
The coupling term $\Phi_\chi = -\big(W_{,\chi}/W\big)\,\delta\chi$ links axion and dilaton perturbations, sourcing the axion velocity divergence even when the intrinsic sound speed is negligible. Together with the $(m'/m)\Theta_\mfa$ term in the Euler equation, this acts as a time–dependent modification of the effective drag and gravitational response, shifting both the position and the sharpness of the small–scale cutoff compared to an uncoupled axion.

\smallskip

These effects are captured compactly in the quasi-static regime, which is defined by
\begin{equation}
    |X'| \;\lsim\; \cH\,|X|
    \qquad\hbox{and}\qquad 
    k^2 \gg \cH^2 \,,
\end{equation}
for the scalar and metric sectors, so that $\delta\chi''$ and $\cH\,\delta\chi'$ are subdominant to spatial-gradient and mass terms while the full time evolution of the axion and baryon fluids is retained. We also assume we are far enough into the late universe for the potential energy of the dilaton to be dominating over its kinetic energy contribution $(a^2 V(\chi) \gg \chi'^2)$, consistent with the slow-roll tracker phase in which the dilaton behaves as a dark-energy–like component coupled to matter. In this limit the coupled axion–dilaton–baryon system reduces to a single, closed evolution equation for the axion density contrast. Carrying out this reduction and eliminating $\delta\chi$ as described in \Cref{ssec:qsa_details} yields

\begin{align}\label{eq:quasistatieq}
    \delta''_\sax 
    &+ \delta'_\sax\left(\mathcal{H} - \frac{m'}{m}\right)
    + \frac{k^4\,\delta_\sax}{4a^2 m^2(\eta)}
    \nn\\
    &= 4\pi a^2 \GN\Biggl[
        \left(
            1 + 2\left(\frac{W_{,\chi}}{W}\right)^2 
            \frac{k^2}{a^2 k_\chi^2}
        \right)\bar{\rho}_\sax \delta_\sax
        \nn\\
        &\qquad\qquad\qquad\qquad
        + \left(
            1 - 2\frac{W_{,\chi}}{W}\,\mathbf{g}
            \frac{k^2}{a^2 k_\chi^2}
        \right)\bar{\rho}_\ssB \delta_\ssB
    \Biggr],
\end{align}
where 
\begin{equation}\label{eq:kchi}
    k_\chi^2 \equiv k^2 + a^2 m_{\chi,\mathrm{eff}}^2
\end{equation}
is the characteristic inverse length scale of dilaton interactions,
separating the screened and unscreened regimes of the fifth force,
and from \cref{eq:dilaton_perturbed} one can identify an effective mass-squared for $\delta\chi$,
\begin{equation}\label{eq:effmass}
    m_{\chi,\mathrm{eff}}^2 \;\equiv\; 
    \frac{1}{M_{\!P}^2} V_{,\bar\chi\bar\chi}
    \;-\;
    \frac{\bar\rho_{\sax}}{\MP^2}\frac{W_{,\chi\chi}}{W}.
\end{equation}

The structure of \cref{eq:quasistatieq} illuminates how axion fluctuations deviate from standard CDM behaviour. The time-dependent mass modifies the effective drag term so that when $m'/m>0$ the friction is reduced and infall temporarily enhanced, while $m'/m<0$ increases damping. The $k^4$ term encodes the quantum–pressure.

The terms on the right-hand side are a consequence of dilaton mediation and can be interpreted in terms of an effective gravitational coupling that applies to between-species attraction:
\begin{align}
    G_{\mfa\mfa}(k) &\simeq \GN\!\left[1 + 2\!\left(\frac{W_{,\chi}}{W}\right)^{\!2}\frac{k^2}{a^2 k_\chi^2}\right],
\end{align}
and
\begin{align}
    G_{\mfa B}(k) &\simeq \GN\!\left[1 - 2\,\frac{W_{,\chi}}{W}\,\mathbf g\,\frac{k^2}{a^2 k_\chi^2}\right],
\end{align}
with interaction range set by $k_\chi^{-1}$ in \cref{eq:kchi}. These correspond to \emph{effective} gravitational couplings that govern the linear response of perturbations, while the fundamental Newton constant in the Einstein–Hilbert term remains fixed.

For the axio–dilaton models considered here, $W_{,\chi}/W<0$ and $\mathbf g<0$, so the axion–axion Newton's constant $G_{\mfa\mfa}$ is amplified while the axion–baryon gravitational constant $G_{\mfa B}$ is reduced. This implies the self-gravity of the axion field is \emph{enhanced} due to the presence of the dilaton field, whereas the axion–baryon attraction is \emph{reduced}, suggesting the dilaton is effectively mediating a repulsive fifth force between the axion and baryons. The latter effect weakens the baryonic infall into axion-dominated potential wells, slightly offsetting the clustering enhancement from the axion sector itself.

For $k \ll a k_\chi$ these corrections are Yukawa–screened and general relativity is recovered, 
while for $k \gg a k_\chi$ the effective couplings saturate to scale–independent limits set by the instantaneous 
background values of $W_{,\chi}/W$ and $k_\chi(\eta)$, producing a scale–dependent axion–baryon bias.

Taken together, \cref{eq:quasistatieq,eq:kchi} show that axion growth in this framework is governed not only by the familiar quantum–pressure cutoff of fuzzy dark matter, but also by the evolving axion mass and the dilaton-mediated interactions. The resulting phenomenology is therefore distinct from either mechanism in isolation, leading to richer, time-dependent modifications to the small-scale matter power spectrum and potentially testable departures from standard CDM evolution.\footnote{Understanding how these deviations in small scale structure can affect key cosmological observables such as the lensing amplitude is even more complicated, see \cite{Costa:2025kwt} for a discussion.}

\subsubsection*{The Jeans scale}

Within the averaged heavy–axion and quasi–static regime
($m\!\gg\!\cH$, $k\!\gg\!a\cH$, $k\!\ll\!am$), the competition between the
$k^4$ gradient term and the gravitational source in \cref{eq:quasistatieq}
defines a characteristic wavenumber called the Jeans scale $k_J$ that separates CDM–like from
suppressed axion clustering. Neglecting baryons for clarity and so using
$\bar\rho_\mfa\simeq\bar\rho_{\mathrm m}$ during matter domination, the
balance condition
\begin{equation}\label{eq:jeans_balance}
    \frac{k_J^4}{4a^2 m^2(\eta)}\,\delta_\sax
    \;\simeq\;
    4\pi a^2\,G_{\mfa\mfa}(k_J)\,\bar\rho_\sax\,\delta_\sax,
\end{equation}
implies
\begin{equation}\label{eq:jeans_scale}
    k_J^4(a)
    \;\simeq\;
    16\pi\,G_{\mfa\mfa}(k_J)\,a^4 \bar\rho_\sax\,m^2(\eta)\,.
\end{equation}

In the screened regime ($k_J\ll ak_\chi$) one has $G_{\mfa\mfa}\simeq \GN$,
and \cref{eq:jeans_scale} reduces to the standard fuzzy–DM result,
\begin{equation}\label{eq:kJ_screened}
    k_J(a)
    \;\simeq\;
    \bigl[16\pi \GN\,a^4 \bar\rho_\sax\,m^2(\eta)\bigr]^{1/4},
\end{equation}
now with the time–dependent mass $m(\eta)=m_\mfa/W(\chi)$ inherited from the
dilaton background. Within the force range ($k_J\gg ak_\chi$), the kinetic
coupling enhances the axion self–gravity,
\begin{equation}\label{eq:kJ_enhanced}
    k_J(a)
    \;\simeq\;
    \Bigl\{16\pi \GN\bigl[1+2(W_{,\chi}/W)^2\bigr]\,
    a^4 \bar\rho_\sax\,m^2(\eta)\Bigr\}^{1/4}.
\end{equation}

Including baryons replaces the source in \cref{eq:jeans_balance} by
$4\pi \GN a^2\!\left[G_{\mfa\mfa}\bar\rho_\sax\,\delta_\sax
+ G_{\mfa B}\bar\rho_B\,\delta_B\right]$ (with $G_{\mfa B}$ from the
couplings in \cref{eq:quasistatieq}); the resulting shift in $k_J$ is mild
for $\bar\rho_B/\bar\rho_\sax\ll 1$. In all cases, the Jeans scale evolves
with $W(\chi)$ through both the effective axion mass
$m(\eta)=m_\mfa/W(\chi)$ and the coupling slope $(W_{,\chi}/W)$ that governs
the $\chi$–mediated force. As $W(\chi)$ varies in time, $k_J(a)$ drifts
accordingly, imprinting smooth, redshift–dependent departures from a single
warm–DM–like cutoff in the matter transfer function and in small–scale
lensing.

Previous work has shown that kinetic couplings between an ultralight axion and a second scalar can significantly enhance the fuzzy–DM Jeans scale and tighten small–scale–structure bounds on the model parameters~\cite{Toomey:2025mvx}. In the present scenario the axion is by construction heavy enough to behave as standard CDM, so its quantum–pressure cutoff lies far below the scales probed by Ly$\alpha$ and dwarf–galaxy observables. Consequently, the specific numerical constraints derived in those ultralight setups do not directly apply; instead, our limits are dominated by CMB, BAO, and lensing data together with the dilaton’s couplings to baryons and electrons.

\subsubsection*{Stability of perturbations}

Since both $k_\chi(\eta)$ in \cref{eq:kchi} and $k_J(\eta)$ in \cref{eq:jeans_scale} evolve with the dilaton background, the locations at which the corresponding scale–dependent effects appear in the transfer function $T(k)$ also drift with time. The observable Fourier modes span a fixed range of comoving wavenumbers ($k \sim 10^{-3}$\,$-$\,1$\,h\,{\rm Mpc}^{-1}$) probed by CMB lensing and large–scale–structure surveys~\cite{Planck:2018lbu,DESI:2024mwx}. As $k_J(\eta)$ or $k_\chi(\eta)$ sweep across this interval, different subsets of $k$-modes transition between CDM-like growth, gradient suppression, and dilaton-mediated enhancement. The result is a mild, redshift-dependent departure from a single power-law transfer function, often appearing as gentle changes in slope rather than sharp features~\cite{Hu:2000ke}.

If $k_\chi^2$ were to become negative, the dilaton acquires a tachyonic effective mass and long-wavelength modes experience transient growth instead of Yukawa suppression. The second term in \cref{eq:effmass} represents a negative susceptibility induced by the axion background.  
Large $\bar\rho_{\ax}$ and/or a steep kinetic coupling $W(\chi)$ (i.e., large $W_{,\chi}$ or $W_{,\chi\chi}$) decrease $m_{\chi,\mathrm{eff}}^2$ and hence $k_\chi^2$.  
A tachyonic window opens whenever
\begin{equation}
\frac{a^2}{\MP^2}\bar\rho_{\sax}\tfrac{W_{,\chi\chi}}{W}
>\;
k^2 + \frac{a^2}{\MP^2} V_{,\bar\chi\bar\chi},
\end{equation}
that is, on sufficiently large scales (small $k$) if the axion-induced term dominates over the stabilising curvature of $V(\chi)$.  
In this regime the fluctuation equation for $\delta\chi$ contains a negative
$m_{\chi,\mathrm{eff}}^2$, so modes with $k^2 < a^2|m_{\chi,\mathrm{eff}}^2|$ grow
exponentially until the instability is quenched by the background evolution
or non-linear effects.

This instability then feeds into the axion and baryon sectors through the $W_{,\chi}/W$ and $\mathbf{g}$ couplings appearing on the right-hand side of \cref{eq:quasistatieq}.  
Physically, the interaction ceases to be Yukawa-suppressed and becomes an infrared enhancement of the force rather than a screened correction.

To ensure stability across the linear scales of interest, one must require
\begin{equation}
k_\chi^2(k,z) > 0
\quad \text{for} \quad
k\in[k_{\min},k_{\max}],
\end{equation}
corresponding approximately to $k_{\min}\!\sim\!10^{-3}\,h\,\mathrm{Mpc}^{-1}$ and $k_{\max}\!\sim\!1\!-\!10\,h\,\mathrm{Mpc}^{-1}$ for large-scale-structure and CMB-lensing observables~\cite{Planck:2018lbu,DESI:2024mwx}.  
A sufficient (though not strictly necessary) condition for this is
\begin{equation}
V_{,\bar\chi\bar\chi}
\;\gtrsim\;
\bar\rho_{\sax}
\!\tfrac{W_{,\chi\chi}}{W},
\end{equation}
to hold throughout the redshift range probed.  
In practice, this demands either (i) a potential $V(\chi)$ with sufficient local curvature ($V_{,\bar\chi\bar\chi}>0$, avoiding shallow inflection regions near the background trajectory), or (ii) a coupling function $W(\chi)$ whose logarithmic slope and curvature are moderate enough that the axion backreaction does not overwhelm the stabilising dilaton bare mass $V_{,\bar\chi\bar\chi}$ term when weighted by $\bar\rho_{\ax}$.

For the exponential ansatz used in our axio-dilaton model, and approximating the axion as the dominant cold component,
$\bar\rho_{\sax}\simeq 3\,\Omega_{\ax}(\eta)\,\cH^2(\eta)M_P^2/a^2$, the stability
condition can be written as
\begin{equation}
    \frac{a^2 V_{,\bar\chi\bar\chi}}{M_P^2\,\cH^2(\eta)}
    \;\gtrsim\;
    3\,\Omega_{\ax}(\eta)\,\zeta^2 \, .
\end{equation}
Along the tracker solutions of interest for light dilaton fields the dilaton potential satisfies
$V(\bar\chi)\sim 3\,\cH^2 M_P^2/a^2$, and hence
$a^2 V_{,\bar\chi\bar\chi}/M_P^2 = \mathcal{O}(\cH^2)$, so the left-hand side is
naturally of order unity during matter domination. This implies that
tachyonic instabilities on linear scales are avoided provided the kinetic slope
satisfies $|\zeta|\lesssim\mathcal{O}(1)$ over the redshift range we probe. In
our analysis we treat $\zeta$ as a free parameter and find that the data prefer
$|\zeta|\ll 1$, well inside this conservative stability regime.

\section{Datasets \& Analysis}
\label{sec:data_analysis}

This section switches gears towards the mechanics of comparing the above theoretical model with the extant cosmological datasets.

\subsection{Analysis}
\label{ssec:analysis}

In the interest of reproducibility, in this section we outline the main datasets used in our Bayesian cosmological analysis as well as the methodology for the numerical implementation of our models into the Boltzmann solver \texttt{CLASS}~\cite{Diego_Blas_2011}.

\subsubsection{Datasets}
\label{sssec:datasets}

We confront the full axio–dilaton model using complementary early- and late-time probes chosen to test recombination microphysics, the sound horizon $r_s$, and late-time distances/growth, thereby breaking CMB-only degeneracies. Specifically, we employ the following dataset combinations:

\medskip\noindent\textbf{\emph{Cosmic Microwave Background.}}
We consider two main combinations of CMB datasets:

\bitemize{Planck+ACT (CMB-A):}
We use the \emph{Planck} 2018 legacy likelihoods—high-$\ell$ TT/TE/EE from \texttt{Plik} together with low-$\ell$ TT (\texttt{Commander}) and low-$\ell$ EE (\texttt{SimAll})—as described in Refs.~\cite{Planck:2018nkj,Planck:2019nip}.
We also include the \texttt{actplanck\_baseline v1.2} lensing combination, which jointly uses the \emph{Planck} PR4~\cite{Carron:2022eyg} and ACT DR6 lensing reconstruction data~\cite{ACT:2023kun,ACT:2023dou}.

\bitemize{Planck+ACT+SPT (CMB-B):}
This setup extends CMB-A with the \emph{SPT-3G D1} TT/TE/EE spectra, derived from the 2019–2020 observations of the 1500\,deg$^2$ SPT-3G Main Field~\cite{SPT-3G:2025bzu}, and with the corresponding SPT-3G lensing reconstruction (\textsc{MUSE})~\cite{SPT-3G:2025zuh}.
The D1 release provides high-resolution temperature and polarization measurements that sharpen the small-scale damping-tail leverage; in particular, its lensed EE and TE spectra at $\ell \simeq 1800$--$4000$ deliver the tightest constraints at those multipoles, exceeding the ACT DR6 sensitivity at the smallest angular scales~\cite{SPT-3G:2025bzu}.
In our analysis we use the \texttt{sptlite} likelihood, which marginalizes over the experiment's nuisance parameters and offers a faster implementation without degrading cosmological constraints~\cite{SPT-3G:2025bzu,SPT-3G:2025zuh}. We include this SPT extension because our axio-dilaton model is expected to leave clear imprints on small angular scales---via the mass-rescaling effects on recombination microphysics and baryon loading, and via axion-sector perturbations that modify the damping tail and lensing at $\ell \gtrsim 2000$ (see \Cref{fig:power_spectra}).

\medskip\noindent\textbf{\emph{Baryon Acoustic Oscillations.}} 
We use BAO distances from either DESI DR2 or the SDSS/eBOSS compilation, denoted as follows:

\bitemize{DESI DR2 (DESI):} 
We adopt the DESI DR2 BAO sample, which provides sixteen anisotropic BAO measurements of $D_M/r_d$ and $D_H/r_d$ spanning $0.4 < z < 4.2$~\cite{DESI:2025zpo}, together with a single low-redshift isotropic BAO constraint on $D_V/r_d$ over $0.1 < z < 0.4$ (see Table~IV of Ref.~\cite{DESI:2025zgx}). These results, drawn from ELG, LRG, QSO, and Ly$\alpha$ tracers, currently provide the most precise BAO distance ratios across low-to-intermediate redshifts~\cite{DESI:2025zpo,DESI:2025zgx}.

\bitemize{eBOSS/SDSS (SDSS):} 
For cross-checks, we also use the SDSS/eBOSS BAO compilation~\cite{eBOSS:2020yzd}, which provides both isotropic and anisotropic measurements of distances and expansion rates across multiple tracers and redshifts (see Table~3 of Ref.~\cite{eBOSS:2020yzd}).

\medskip\noindent\textbf{\emph{Supernovae.}}
We use the following Type~Ia supernova datasets:

\bitemize{Pantheon+ (PP):} 
The \textit{Pantheon+} compilation~\cite{Brout:2022vxf} comprises 1701 light curves for 1550 unique SNe~Ia over $0.001 < z < 2.26$, with updated photometric calibration and a full systematic covariance.

\bitemize{Pantheon+SH0ES (PPSH0ES):}
The combined \textit{Pantheon+} and SH0ES calibration likelihood augments the \textit{Pantheon+} compilation with the Cepheid-based absolute-magnitude calibration from the SH0ES distance ladder~\cite{Riess:2021jrx,Brout:2022vxf}. This likelihood simultaneously fits the SN~Ia light-curve distances and the local $H_0$ prior, allowing for a consistent propagation of calibration uncertainties into cosmological parameters.

\subsubsection{Methodology}
\label{sssec:methods}

In order to model the dynamics of the multi-field cosmology studied here, we use our modified version of \texttt{CLASS}~\cite{Diego_Blas_2011}, designed to handle couplings between light dark-energy scalars and different particle species. The code has also been extended to implement a dynamical evolution of the electron mass within the \texttt{HyRec} recombination framework~\cite{Ali_Ha_moud_2011}, such that binding energies, scattering cross sections, and recombination coefficients are updated consistently at each time step. In practice, the electron mass is dynamically rescaled through the conformal coupling defined in \cref{eq:mass_universal}, so that the background evolution of the scalar field $\chi$ feeds directly into the recombination microphysics.

\texttt{CLASS} is run with non-linear corrections to the matter power spectrum computed using the \texttt{halofit} prescription, and a standard neutrino sector with one massive eigenstate of $m_\nu = 0.06~\mathrm{eV}$ and an effective number of relativistic species consistent with $N_{\rm eff} \simeq 3.044$. This setup ensures that any modification to the small-scale matter power spectrum and CMB damping tail is driven by the coupled axion--dilaton dynamics and the induced variation in $m_e(z)$, rather than by changes in the neutrino sector or ad hoc rescalings.

Parameter estimation is performed with the \texttt{Cobaya} framework~\cite{Torrado:2020dgo}, interfaced with our modified version of \texttt{CLASS}. Priors on all parameters are listed in \Cref{tab:models} and chosen wide enough to encompass both the $\Lambda$CDM limit and the regime of interest for non-zero couplings, with $H_0$, $\Omega_B h^2$, $\Omega_c h^2$, $n_s$, $A_s$, and $\tau_{\rm reio}$ treated as free cosmological parameters alongside the scalar-sector parameters. We employ the Metropolis--Hastings MCMC sampler with adaptive proposal learning and the \texttt{drag} option to improve mixing along curved degeneracies.

In some of the models considered here, the posterior distributions of the initial dilaton field value $\chi_i$, the matter coupling $g$, and, to a lesser extent, the axion coupling $\zeta$, exhibit bimodality. For such cases, chains are allowed to converge until the Gelman--Rubin $R - 1$ coefficients for the individual non-bimodal parameters fall below 0.03. For chains without strong bimodalities, we require the global Gelman--Rubin statistic for the full parameter set to satisfy $R - 1 < 0.03$ before declaring convergence.

The resulting chains are then processed using \texttt{GetDist}~\cite{Lewis:2019xzd} using a smoothing scale of 0.4, which is used to compute marginalized one- and two-dimensional posterior distributions, as well as derived parameters such as $\sigma_8$.

\subsubsection{Model Implementation}

We assume that the axion is oscillating throughout the entire late-time, post-BBN cosmology, corresponding to the heavy-field regime $m(\eta)\gg\mathcal{H}$. This is equivalent to imposing $m(\eta)\gtrsim 10^{-15}\,\mathrm{eV}$. In this limit, the axion can be treated as an effective pressureless fluid whose mean energy density redshifts as $a^{-3}$, up to $\mathcal{O}(\mathcal{H}^2/m^2)$ corrections, with a small scale-dependent sound speed that is negligible on the linear scales probed here. Practically, we therefore identify the axion with the CDM component in \texttt{CLASS} and sample its abundance by identifying $\Omega_c$ with the axion abundance $\Omega_{\rm ax}$.

The prefactor of the polynomial dilaton potential $V_0$ is rescaled at each parameter point to ensure that spatial curvature remains flat, such that $\Omega_{\chi,0} = 1-\Omega_{\rm ax,0}-\Omega_{B,0}$ today. This choice fixes the total energy density to be exactly critical in all models, isolating the impact of the scalar couplings and the evolution of $m_e(z)$ on the expansion history and the growth of structure at fixed spatial curvature.

\begin{table}
\centering
\begin{minipage}{0.45\textwidth}
\centering
\begin{ruledtabular}
\begin{tabular}{c c c}
Model & Parameters & Priors \\
\hline
$\Lambda$CDM & $\Omega_B h^2$ &$\left[0.005, 0.1\right]$ \\ & $\Omega_c h^2$ & $\left[0.001, 0.99\right]$\\ 
& $\tau_{reio}$& $\left[0.01, 0.8\right]$\\ &$A_s$&$\left[5\times10^{-10}, 5\times10^{-9}\right]$\\ &$n_s$& $\left[0.8, 1.2\right]$\\ & $\theta_s$ & $\left[0.5, 10\right]$\\
\addlinespace[2pt]\cmidrule(lr){1-3}\addlinespace[2pt]
+$m_e$ &$m_e$& $\left[0.9, 1.1\right]$\\
\addlinespace[2pt]\cmidrule(lr){1-3}\addlinespace[2pt]
Minimal Yoga         & $\gdm$ &$\left[-0.2, 0.2\right]$\\ 
Exponential Yoga         & $\mathbf{g}$ &$\left[-0.3, 0.3\right]$\\ 
\addlinespace[2pt]\cmidrule(lr){1-3}\addlinespace[2pt]
Minimal Yoga + VI & $\chi_i$&$\left[73.5, 75\right]$\\
& &\\
\end{tabular}
\end{ruledtabular}
\caption{
Free parameters, their labels, and the flat prior ranges adopted for each model.
For runs including the \texttt{PPSH0ES} likelihood, the coupling priors are widened to 
$\gdm \in [-0.4,0.4]$ and $\mathbf{g} \in [-0.8,0.8]$.
}
\label{tab:models}
\end{minipage}
\end{table}

\bitemize{Exponential potential:}
In the exponential case the scalar field has no true minimum, and the overall dark-energy scale is fixed at each parameter point by retuning $V_0$. Because this leaves the present-day value of $\chi$ undetermined, we instead adjust the initial condition so that the field always satisfies $\chi_0 = \chi_{\min}$ today, where $\chi_{\min}$ denotes the location of the would-be minimum in the corresponding Yoga models. This prescription guarantees that the universal mass rescaling $A(\chi)$ takes the same value today across all samples, ensuring consistent particle masses, and therefore identical local particle physics, in every model realisation.

\bitemize{Albrecht-Skordis potential: }
When $V_p$ is taken to be a quadratic function of $\chi$ as in \cref{eq:albrecht-skordis}, the position of its local minimum is determined by the requirement that Standard Model particle masses adjust to the electroweak scale as their coupling to the dilaton varies. For computational convenience, we choose to fix the position of this local minimum, even though in principle it should shift as the matter coupling $\mathbf{g}$ varies. This choice is physically consistent because moving the minimum to larger/smaller values of $\chi_{\min}$ as $\mathbf{g}$ decreases/increases would yield exactly the same cosmology once $V_0$ is retuned to raise the dilaton potential energy to the dark-energy value required to close the universe. We then take the overall mass scale associated with the dilaton today, given schematically by $M_{\rm Pl}\,e^{\mathbf{g}\chi_{\min}}$, to reproduce the electroweak scale.

\begin{figure*}
    \centering
    \includegraphics[width=\linewidth]{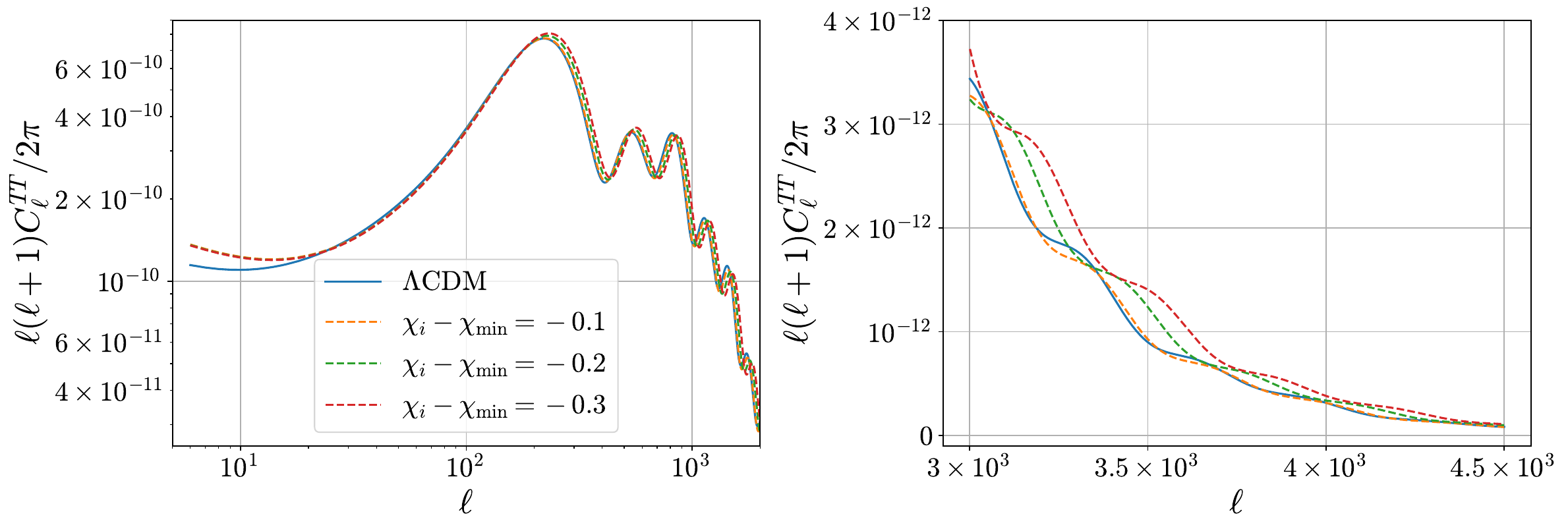}
\caption{CMB angular power spectra for different initial displacements of the dilaton field in the Yoga model variants. 
The quantity $\chi_i - \chi_{\rm min}$ parametrises the initial distance from the local minimum of the Albrecht--Skordis dilaton potential, given by \cref{eq:albrecht-skordis}. 
We fix the dilaton--matter coupling to $\mathbf{g} = -0.3$, the axion--dilaton kinetic coupling to $\zeta = 0.1$, and set the six standard $\Lambda$CDM cosmological parameters to their Planck 2018 best-fit values~\cite{Planck:2018vyg}.
}
    \label{fig:power_spectra}
\end{figure*}

Figure~\ref{fig:power_spectra} illustrates the impact of varying the initial dilaton displacement on the CMB angular power spectrum in the Yoga model variants, spanning both the linear regime and the quasi--non-linear damping tail at high multipoles.  
For couplings of order $\mathcal{O}(0.1)$, the Planck data exert their strongest constraining power through the first few acoustic peaks, which are measured close to the cosmic-variance limit and shift appreciably when the additional scalar-sector parameters are varied while holding the remaining cosmological parameters fixed.  
Although Planck also measures the higher-order acoustic peaks and the damping tail, its sensitivity in this regime is comparatively weaker and increasingly limited by instrumental noise and foreground modelling.  
Compensating adjustments in degenerate parameters such as $\Omega_c$ and $\Omega_b$ can therefore restore agreement with the low- and intermediate-$\ell$ spectrum, while generically leaving residual distortions in the higher acoustic peaks and at large multipoles.  
In this high-$\ell$ regime, where SPT measurements provide substantially tighter constraints than Planck alone, the $\Lambda$CDM best-fit spectrum obtained from low-$\ell$ data exhibits increasing tension with the observed power, whereas the Yoga models predict systematic, coupling-dependent deviations from $\Lambda$CDM that allow the additional scalar dynamics to improve the fit to the small-scale CMB data when SPT observations are included.

% \begin{figure}
%     \begin{subfigure}[b]{0.99\textwidth}
%         \centering
%     \includegraphics[width = \linewidth]{quad_background.png}
%     \end{subfigure}
%     \hfill
%     \begin{subfigure}[b]{0.99\textwidth}
%         \centering
%     \includegraphics[width = \linewidth]{quad_particle_masses.png}
%     \end{subfigure}
%     \begin{subfigure}[b]{0.99\textwidth}
%         \centering
%     \includegraphics[width = \linewidth]{quad_power_spectra.png}
%     \end{subfigure}
%     \hfill
%     \hfill
%     \caption{Background and perturbative level plots of the axio-dilaton cosmology with the inclusion of the dilaton potential well. Top row shows the evolution of the background energy densities when $\gdm = 0.1$. Bottom row shows the evolution of the dilaton field and associated baryon masses for a range of $\gdm$ In all cases $\mathbf{g}=-10^{-3}$. }
%     \label{fig:Well}
% \end{figure}

% \begin{figure*}
%     \begin{subfigure}[b]{0.99\textwidth}
%         \centering
%     \includegraphics[width = \linewidth]{fsigma8.png}
%     \end{subfigure}
    
%     \caption{Left panel shows the structure growth parameter $f\sigma_8$ for the pure exponential potential and the right panel shows the same when the dilaton is trapped in a quadratic minimum. The $\Lambda$CDM best-fit is shown in black. The data points in grey are taken from~\cite{Marulli:2020uyy}.}
%     \label{fig:fsigma8}
% \end{figure*}

\subsection{Results}
\label{sec:results}

In this section we present cosmological constraints on the two coupled-scalar frameworks introduced above: the Minimal Yoga model, featuring a shallow polynomial dilaton potential with an optional axionic companion, and the Exponential model, corresponding to the monotonic tracker limit of the same theory. Both scenarios predict a redshift-dependent electron mass $m_e(z) \propto e^{\mathbf{g}\chi/M_{\rm Pl}}$, which modifies recombination microphysics and the late-time expansion history, but they differ in whether the scalar field stabilises (Yoga) or continues to roll (Exp).

We first summarise the constraints obtained for each model separately, using the dataset combinations listed in \Cref{tab:CMB-A}, and then compare their joint performance across different likelihood bundles, with constraints taken from \Cref{tab:CMB-B} and \Cref{tab:ppshoes}.

\begin{table*}
\makebox[\textwidth][c]{%
\begin{tabular}{lc|ccccc}
\toprule
Model & Dataset & $H_0~[\mathrm{km\,s^{-1}\,Mpc^{-1}}]$ & $\mathbf{g}$ & $\zeta$ & $\chi_i$ & $\Delta \chi^2$ \\
\midrule
Yoga-VI & CMB-A DESI PP & $69.18^{+0.63}_{-0.81}$ (69.71) & $0.00 \pm 0.10$ (0.14) & $0.002 \pm 0.052$ (-0.061) & $74.00 \pm 0.13$ (74.10) & -2.1 \\
 & CMB-A DESI & $69.38^{+0.68}_{-0.83}$ (69.61) & $0.00 \pm 0.10$ (0.15) & $0.000 \pm 0.052$ (-0.060) & $74.00 \pm 0.14$ (74.09)  & -3.5 \\
\specialrule{0.2pt}{0pt}{0pt}
(no-$m_e$) & CMB-A DESI PP & $68.50 \pm 0.35$ (68.51) & $0.006 \pm 0.083$ (-0.009) & $-0.001 \pm 0.054$ (-0.050) & $74.01^{+0.18}_{-0.16}$ (73.88) & -2.1 \\
\specialrule{0.2pt}{0pt}{0pt}
Yoga & CMB-A DESI PP & $68.42 \pm 0.35$ (68.33) & $-0.038 \pm 0.086$ (-0.055) & $0.020^{+0.077}_{-0.094}$ (0.032) & -- & -0.7 \\
 & CMB-A DESI & $68.53 \pm 0.36$ (68.54) & $-0.041 \pm 0.088$ (-0.078) & $0.023^{+0.076}_{-0.097}$ (0.050) & -- & -1.8 \\
\specialrule{0.6pt}{2pt}{2pt}
EXP & CMB-A DESI PP & $69.10^{+0.64}_{-0.76}$ (68.81) & $-0.003 \pm 0.099$ (0.127) & $0.005 \pm 0.049$ (-0.058) & -- & -2.7 \\
 & CMB-A DESI & $69.35^{+0.69}_{-0.78}$ (69.96) & $0.004^{+0.12}_{-0.098}$ (-0.167) & $-0.001 \pm 0.049$ (0.077) & -- & -3.8 \\
 & CMB-A & $68.03^{+0.78}_{-1.6}$ (67.43) & $0.000 \pm 0.089$ (-0.013) & $0.001 \pm 0.051$ (0.036) & -- & -1.1 \\
\specialrule{0.6pt}{2pt}{2pt}
w0-wa + me & CMB-A DESI PP & $68.26 \pm 0.87$ (68.29) & -- & -- & -- & -8.2 \\
 & CMB-A DESI & $63.9^{+1.9}_{-2.7}$ (64.78) & -- & -- & -- & -8.6 \\
\specialrule{0.2pt}{0pt}{0pt}
w0-wa  & CMB-A DESI PP & $67.64 \pm 0.60$ (67.71) & -- & -- & -- & -6.9 \\
\specialrule{0.6pt}{2pt}{2pt}
$\Lambda$CDM+me & CMB-A DESI PP & $69.62 \pm 0.69$ (69.91) & -- & -- & -- & -2.6 \\
 & CMB-A DESI & $69.85 \pm 0.69$ (70.12) & -- & -- & -- & -4.2 \\
\specialrule{0.2pt}{0pt}{0pt}
$\Lambda$CDM & CMB-A DESI PP & $68.30 \pm 0.28$ (68.03) & -- & -- & -- & 0.0 \\
 & CMB-A DESI & $68.39 \pm 0.29$ (68.53) & -- & -- & -- & 0.0 \\
 & CMB-A & $67.30 \pm 0.54$ (67.27) & -- & -- & -- & 0.0 \\
\bottomrule
\end{tabular}
}
\caption{
Posterior means with $1\sigma$ confidence intervals, best-fit values in parentheses, and $\Delta\chi^2$ relative to the corresponding $\Lambda$CDM run for each dataset combination.  
The table summarises how the different axio--dilaton (Yoga and EXP), $w_0$--$w_a$, and varying-$m_e$ models shift $H_0$, the scalar couplings $(\mathbf{g},\zeta)$, and the overall goodness of fit relative to $\Lambda$CDM across the CMB-A, CMB-A DESI, and CMB-A DESI PP datasets.
} 
\label{tab:CMB-A}
\end{table*}

\subsubsection{Minimal Yoga Models}
\label{sec:quad_results}

The constraints for the minimal yoga case are summarised in \Cref{tab:CMB-A}. 
We have considered both fixed initial condition cases, where the dilaton field is placed close to the minimum of its effective potential at early times (Yoga), and runs with a free initial condition $\chi_i$ (Yoga-VI).

For the CMB-A DESI PP combination, the Yoga run prefers 
$H_0 = 68.42 \pm 0.35~\mathrm{km\,s^{-1}\,Mpc^{-1}}$, 
while the Yoga-VI run with a displaced initial condition reaches 
$H_0 = 69.18^{+0.63}_{-0.81}~\mathrm{km\,s^{-1}\,Mpc^{-1}}$. 
Comparing this to the SH0ES result of 
$H_0 = 73.04 \pm 1.04~\mathrm{km\,s^{-1}\,Mpc^{-1}}$~\cite{Riess:2021jrx} 
shows a reduction of the Hubble tension from $4.2\sigma$ to $3.0\sigma$ when the initial condition is allowed to vary.

The (no-$m_e$) case shown in \Cref{tab:CMB-A} gives results for runs where we allow the dilaton initial condition to vary but turn off the recombination readjustment when the electron mass evolves. This is done mostly to compare with expectations for a ``plain'' quintessence analysis that ignores recombination feedback. The result yields 
$H_0 = 68.50 \pm 0.35~\mathrm{km\,s^{-1}\,Mpc^{-1}}$ 
and a $4.1\sigma$ tension, illustrating that it is specifically the coupled $m_e(z)$ dynamics, rather than a generic change in the dark energy sector, that is responsible for the modest upward shift in the Hubble rate.

\begin{figure}
    \centering
    \includegraphics[width=\linewidth]{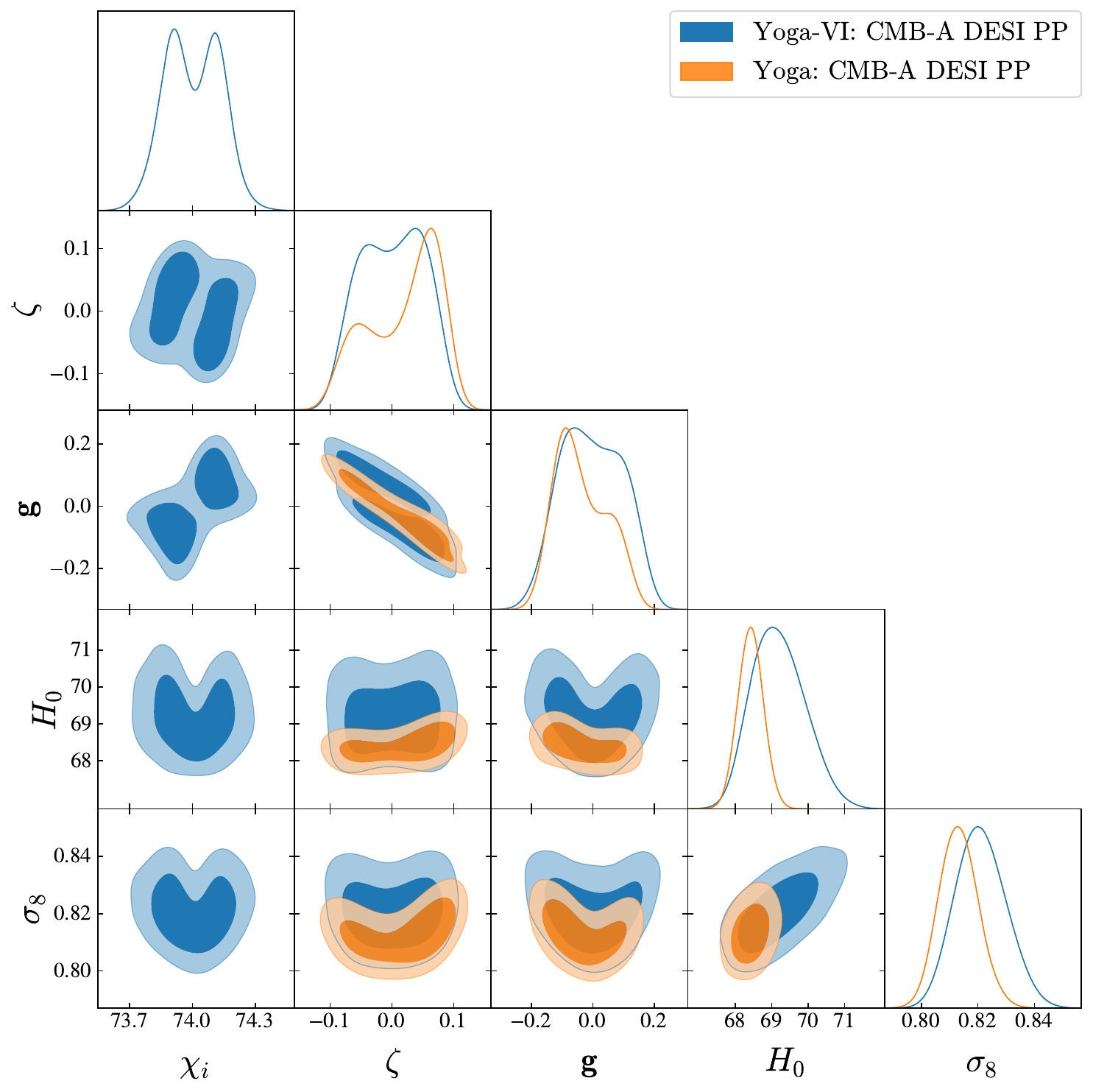}
\caption{
Triangle plot comparing the minimal Yoga model with fixed initial field values and varying initial conditions (VI).  
Both analyses use the same CMB-A DESI PP dataset combination. The two setups exhibit the same correlation directions between $\mathbf{g}$ and $\zeta$, but differ in the correlation with $H_0$, with large $|\mathbf{g}|$ (i.e., values far from zero, either positive or negative) correlating with large $H_0$ only in the VI case.
}
    \label{fig:fixed vs varying}
\end{figure}

As shown in \Cref{fig:fixed vs varying}, the posteriors reveal a clear anti-correlation between the baryon coupling $\mathbf{g}$ and the CDM coupling $\zeta$. A large negative $\mathbf{g}$ can be compensated by a positive $\zeta$, such that the combined source term on the RHS of the dilaton Klein-Gordon \cref{eq:dilaton_background} cancels, leaving the background expansion and effective dark energy close to $\Lambda$CDM.
In the Yoga runs, this compensation mechanism produces slightly bimodal one-dimensional posteriors for $\mathbf{g}$ and $\zeta$, with one peak (negative $\mathbf{g}$, positive $\zeta$) mildly preferred over the other. However, the two peaks are close enough that the two-dimensional $\zeta$-$\mathbf{g}$ contour remains connected and does not display clear bimodality, even though the 1D marginals are visibly non-Gaussian.
In contrast, for the Yoga-VI runs the bimodality becomes more pronounced: the two peaks in the one-dimensional posteriors are both within the 68\% CL region but merge into a single allowed region at 95\% CL. Because the peaks are very close to one another and the central minimum near $\mathbf{g}=\zeta=0$ is relatively high, the two-dimensional $\zeta$-$\mathbf{g}$ posterior develops the characteristic ``butterfly'' shape, with two separated 68\% CL contours but a single connected 95\% CL contour aligned along the anti-correlation direction (positive $\mathbf{g}$, negative $\zeta$).

The same bimodality is present in the dilaton's initial condition, but in this case it is more pronounced: the two peaks in the $\chi_i$ posterior have almost the same height, with a deeper minimum between them around $\chi_i \simeq 74$. This behaviour produces a butterfly-shaped two-dimensional posterior, with $\zeta$ aligned along the positive-correlation direction and $\mathbf{g}$ along the negative-correlation direction. Allowing $\chi_i$ to vary changes the inference in two complementary ways.  
First, it raises the preferred expansion rate from the fixed initial condition value to $H_0 = 69.18~\mathrm{km\,s^{-1}\,Mpc^{-1}}$ in the Yoga-VI branch.  
Second, and equally important, the freedom in $\chi_i$ broadens the $H_0$ posterior, so that the resulting constraint reduces the tension with SH0ES to $3\sigma$.

The Yoga-VI model therefore largely alleviates the Hubble tension within this minimal framework, even though the associated improvement in the overall fit ($\Delta\chi^2 = -2.1$ for CMB-A DESI PP) remains modest compared to $\Lambda$CDM given the three additional parameters. As discussed in \Cref{Including spt}, this conclusion changes once the SPT high-$\ell$ data are included.

The triangle plot in \Cref{fig:fixed vs varying} additionally shows a bimodal, U-shaped correlation between $\mathbf{g}$ and $H_0$. Achieving $H_0 \sim 70$ typically requires $|\mathbf{g}| \sim 0.1$, which would violate Solar-System tests on Earth in the absence of a screening mechanism for the dilaton in dense environments. Exploring such screened completions is beyond the scope of this work and left for future study; however, we note that there are candidate screening mechanisms for these types of light dilaton fields, e.g.\ the BBQ dilaton mechanism discussed in~\cite{Brax:2023qyp}, where the dilaton can be coupled strongly to a heavy axion field through a non-trivial field-space metric, leading to an overall suppression of the effective charge of the full two-field system in the Solar System.

\begin{figure}
    \centering
    \includegraphics[width=\linewidth]{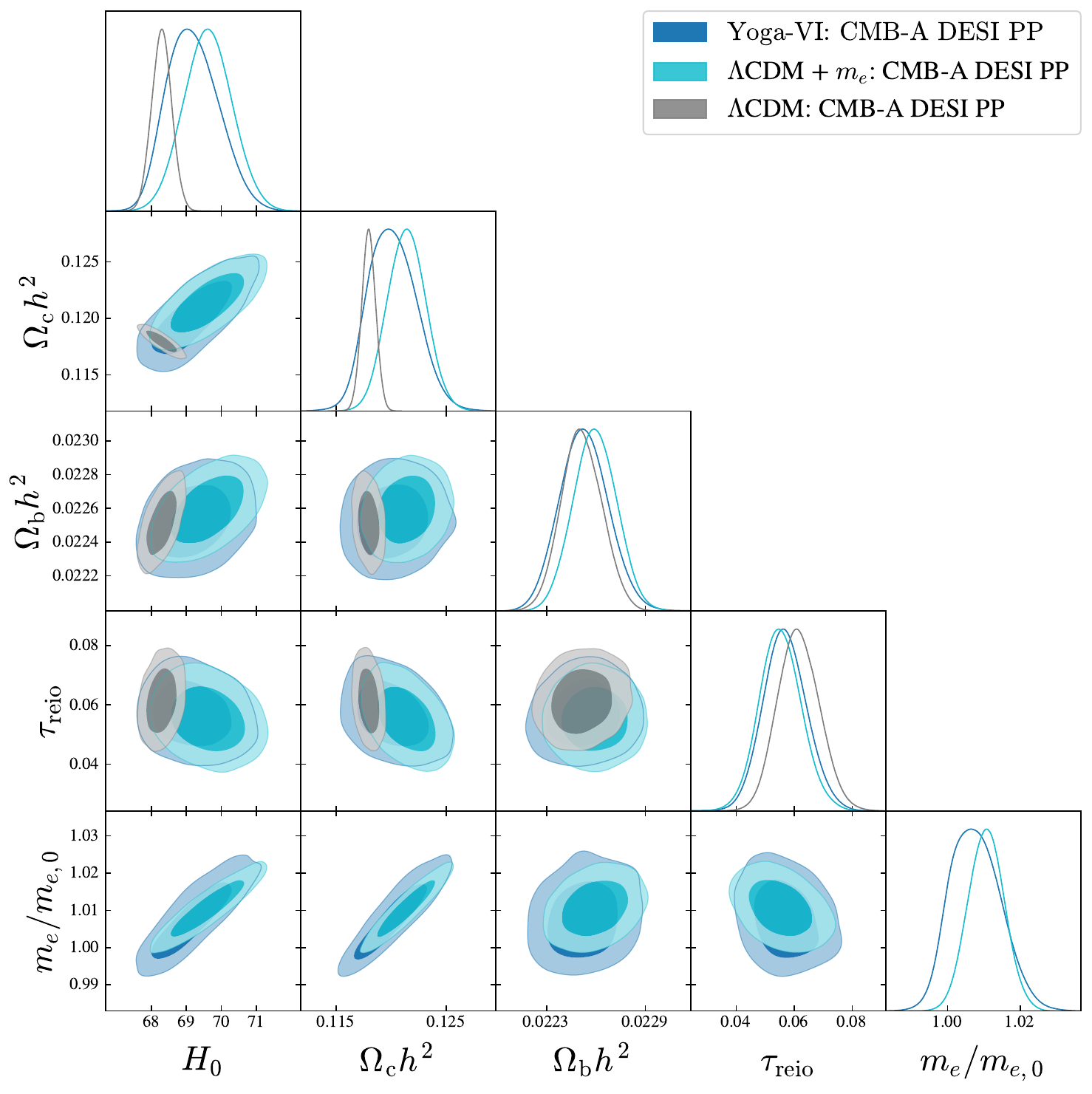}
\caption{
Triangle plot comparing the standard cosmological parameters of $\Lambda$CDM and $\Lambda$CDM+$m_e$ with the minimal Yoga model.  
All cases use the same CMB-A DESI PP dataset combination.
}
    \label{fig:LCDM VS quadratic minimum}
\end{figure}

For comparison, \Cref{fig:LCDM VS quadratic minimum} contrasts the minimal yoga results with $\Lambda$CDM and $\Lambda$CDM+$m_e$.  
In the Yoga-VI runs, the one-dimensional posterior for $m_e$ after recombination is broader and centred closer to the expected $\Lambda$CDM value $m_e = m_{e, 0} \equiv 511\,\mathrm{keV}$ than in the purely phenomenological $\Lambda$CDM+$m_e$ extension. Because larger deviations of $m_e$ from unity are less favoured in Yoga-VI, the model is correspondingly less effective at raising $H_0$.
In the $\Lambda$CDM+$m_e$ case, an increase in $m_e$ can be absorbed by a compensating rise in $\Omega_c h^2$ to preserve the acoustic and damping scales. The same $\Omega_c h^2$-$H_0$ correlation is present in the minimal dilaton model, but the constraints on $\Omega_c h^2$ in Yoga-VI are more relaxed and therefore more compatible with the $\Lambda$CDM values. We return to this comparison, and to the reasons why the phenomenological extensions perform better, in \Cref{sec:discussion}.

\begin{figure*}
    \begin{subfigure}[b]{0.99\textwidth}
        \centering
    \includegraphics[width = \linewidth]{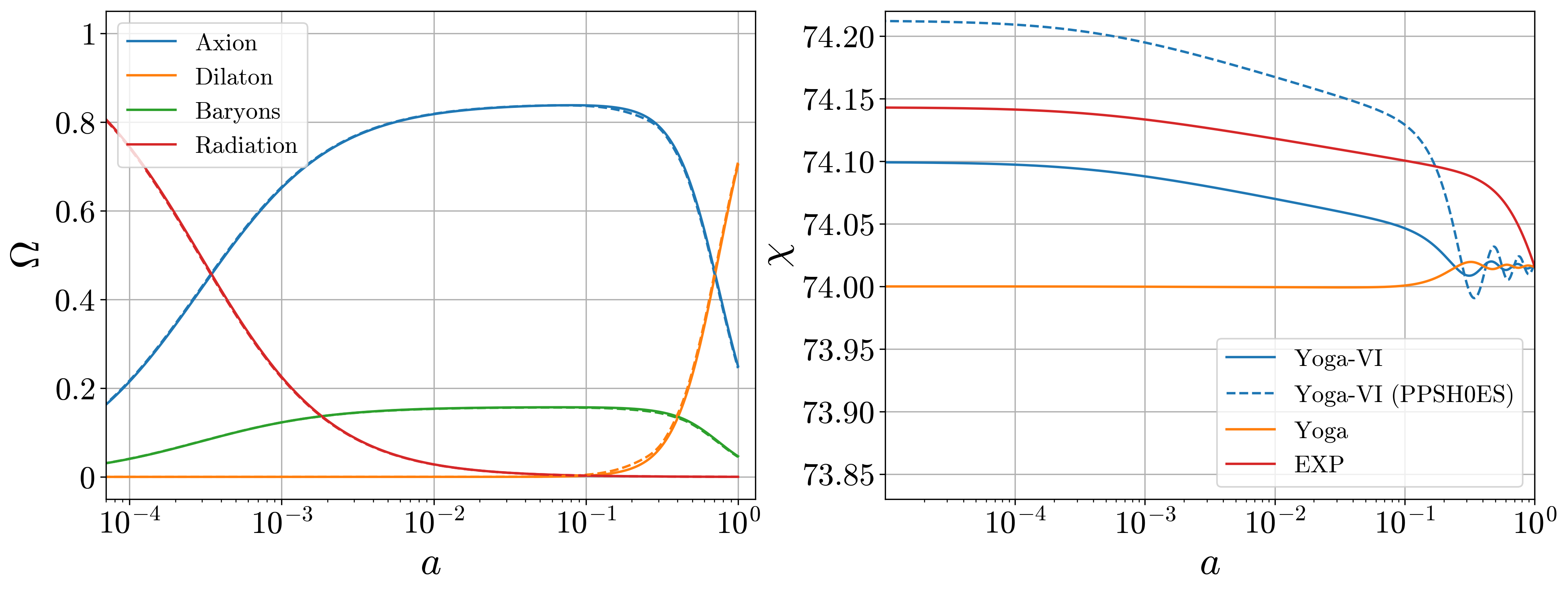}
    \end{subfigure}
    \hfill
    \begin{subfigure}[b]{0.99\textwidth}
        \centering
    \includegraphics[width = \linewidth]{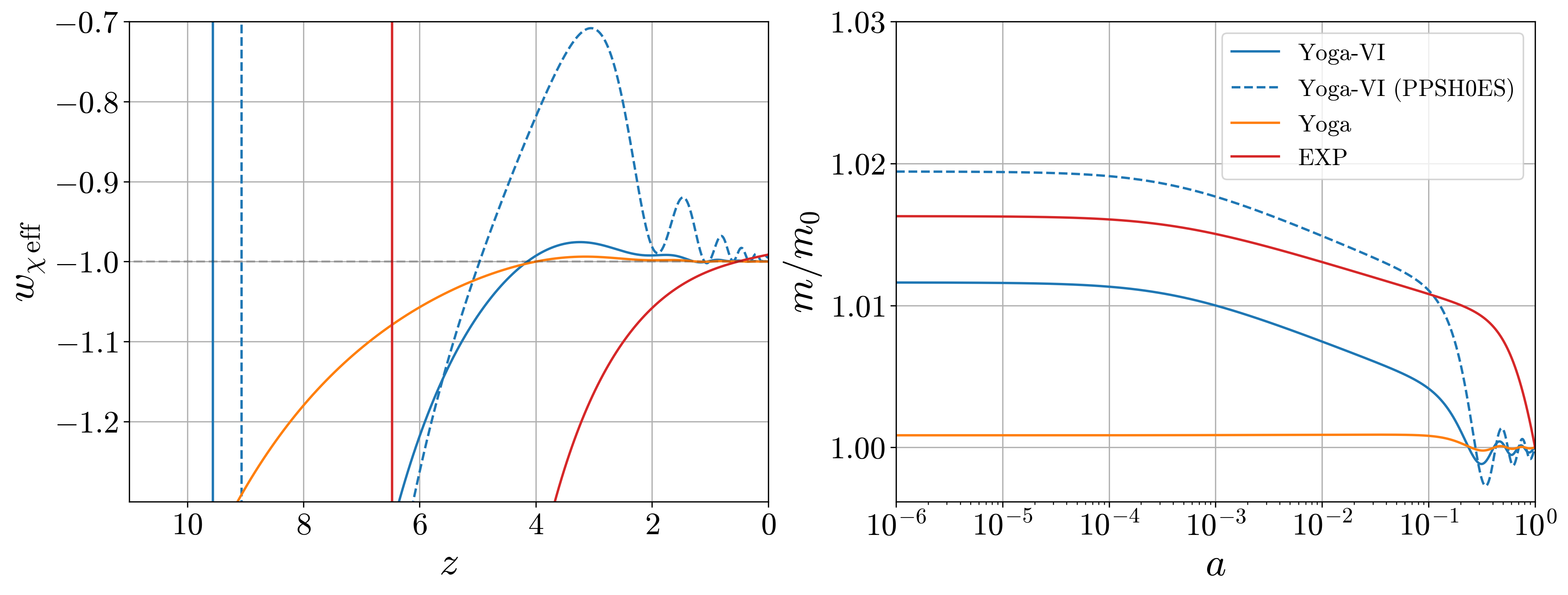}
    \end{subfigure}

\caption{
Best-fit background evolution in different model variants using the CMB-A DESI and PP datasets (calibrated with the SH0ES prior where stated). 
\textbf{Top left:} Fractional energy densities $\Omega_i(a)$ for the Yoga-VI model in solid and with the SH0ES prior in dashed. 
\textbf{Top right:} Dilaton field evolution $\chi(a)$. 
\textbf{Bottom left:} Effective equation of state $w_{\chi}^{\mathrm{eff}}(z)$ showing deviations from $-1$ and damped late-time oscillations in the Yoga cases. 
\textbf{Bottom right:} Relative electron and other Standard Model particle mass variation $m/m_{0}$; the PPSH0ES fit exhibits the largest excursion before relaxing at late times.
}
\label{fig:best-fit_yoga}
\end{figure*}

The best-fitting cosmology for the two Yoga model variants at the background level is also highlighted in \Cref{fig:best-fit_yoga}. 
The top left panel shows that the fractional energy densities closely track a standard matter-radiation history, with the axion playing the role of CDM and the dilaton becoming important only at late times. 
The top right panel illustrates the evolution of the dilaton itself: in the Yoga run the field sits very close to the minimum and only begins to move appreciably at late times, whereas in the Yoga-VI cases the initial condition $\chi_i$ is displaced from the minimum and the field rolls before executing small oscillations around the present-day minimum. In the bottom right panel this translates into a larger electron mass at higher redshifts for the best-fit solutions, where the combination of the initial condition $\chi_i$ and couplings $\mathbf{g}$ and $\zeta$ is always arranged such that $m_e(z)/m_{e,0} > 1$ over the redshift range relevant for the sound horizon and drag epoch, before relaxing back towards unity today.

The bottom left panel in \Cref{fig:best-fit_yoga} shows the corresponding effective dark-energy equation of state that would be inferred for this model if the dark matter were mistakenly assumed to dilute as $1/a^3$ (as is done in the DESI analysis). This is given explicitly by
\begin{equation}
    w_{\chi\,\text{eff}} =
    \frac{w_\chi(\chi)}{
    1
    + \left[\frac{W(\chi)}{W(\chi_0)} - 1\right]\frac{\rho_{\text{ax}0}}{a^3\rho_\chi}
    + \left[\frac{A(\chi)}{A(\chi_0)} - 1\right]\frac{\rho_{b0}}{a^3\rho_\chi}
    } ,
\end{equation}
which exhibits a clear phantom crossing in all best-fit trajectories, despite the true equation-of-state parameter for the dilaton always satisfying $w \geq -1$. This shows that these models achieve phantom crossing in the same manner reported in~\cite{Khoury:2025txd}, through either the conformal matter coupling or the kinetic axion–dark-matter coupling.

The condition for a phantom crossing driven by the baryonic coupling is that the conformal factor $A(\chi)$ be smaller in the past, i.e.\ $A(\chi)/A(\chi_0) = e^{\mathbf{g}(\chi-\chi_0)} < 1$ at some point during matter domination when $\rho_{b0}/a^3 \gtrsim \rho_\chi$.  
Equivalently, the same behaviour can be realised through the axion sector when the non-canonical kinetic prefactor satisfies $W(\chi)/W(\chi_0) = e^{\zeta(\chi_0 - \chi)} < 1$ at epochs where $\rho_{\text{ax}0}/a^3 \gtrsim \rho_\chi$.  
For the baryonic coupling this requires either starting the dilaton at smaller values with a positive $\mathbf{g}$, or at larger values with a negative $\mathbf{g}$. This is opposite to what is required for a successful realisation of a varying electron mass, since it would imply that the electron mass was smaller in the past than today. However, the inevitable oscillations of the dilaton around the local minimum approached at late times ensure that the field can still reside on the correct side of the minimum to generate a phantom crossing even for initial conditions that would not naively correspond to it.

Even when these oscillations occur extremely late because of a weak dilaton–matter coupling, as in the Yoga-VI case with $\mathbf{g} = 0.14$, and the dilaton is initially on the ``wrong'' side of the local minimum, the dilaton–axion kinetic coupling nevertheless guarantees a phantom crossing thanks to the stronger allowed negative coupling $\zeta = -0.061$.  
The best-fitting cosmologies therefore prefer phantom crossings that occur during matter domination at $z \sim 4$, rather than near the present epoch as in typical $w_0$-$w_a$ parametrisations, providing an example where phantom behaviour is tied to matter couplings and axion kinetics rather than to an ad hoc low-redshift deformation of the dark-energy equation of state.

\subsubsection{Exponential Yoga Models}
\label{sec:exp_results}

The constraints for the exponential dilaton model are summarised in \Cref{tab:CMB-A} (referred to as EXP).  
In this case the field evolves on a featureless potential $V(\chi) \propto e^{-\lambda\chi/M_{\rm Pl}}$, with its slope fixed by the yoga-like relation $\lambda = 4\,\zeta$. When allowing $\zeta$ to vary freely, the best-fitting points in the chains when using CMB-A DESI PP datasets correspond to $\zeta < 0$, so the slope of the potential is inverted and is actually preferred to be exponentially growing. However, since the data restrict $|\zeta| \lesssim 0.05$, the potential is extraordinarily shallow and the intrinsic force $V'(\chi)$ is too weak to influence the evolution appreciably. Thus this preference for $\zeta < 0$ should not be interpreted as a preference for an exponentially growing potential, but rather as a preference for the coupling between the axion and dilaton.

Consequently, the dilaton dynamics in the exponential model are driven almost entirely by the matter-coupling source terms rather than by the potential itself. This behaviour is clearly visible in the background trajectories of \Cref{fig:best-fit_yoga}, where the EXP best-fit (green) shows a smooth monotonic drift of $\chi$ down the inverted exponential, while the Yoga-VI solution executes only shallow oscillations around a local minimum, even though the fractional energy densities remain almost indistinguishable.

\begin{figure}
    \centering
    \includegraphics[width=\linewidth]{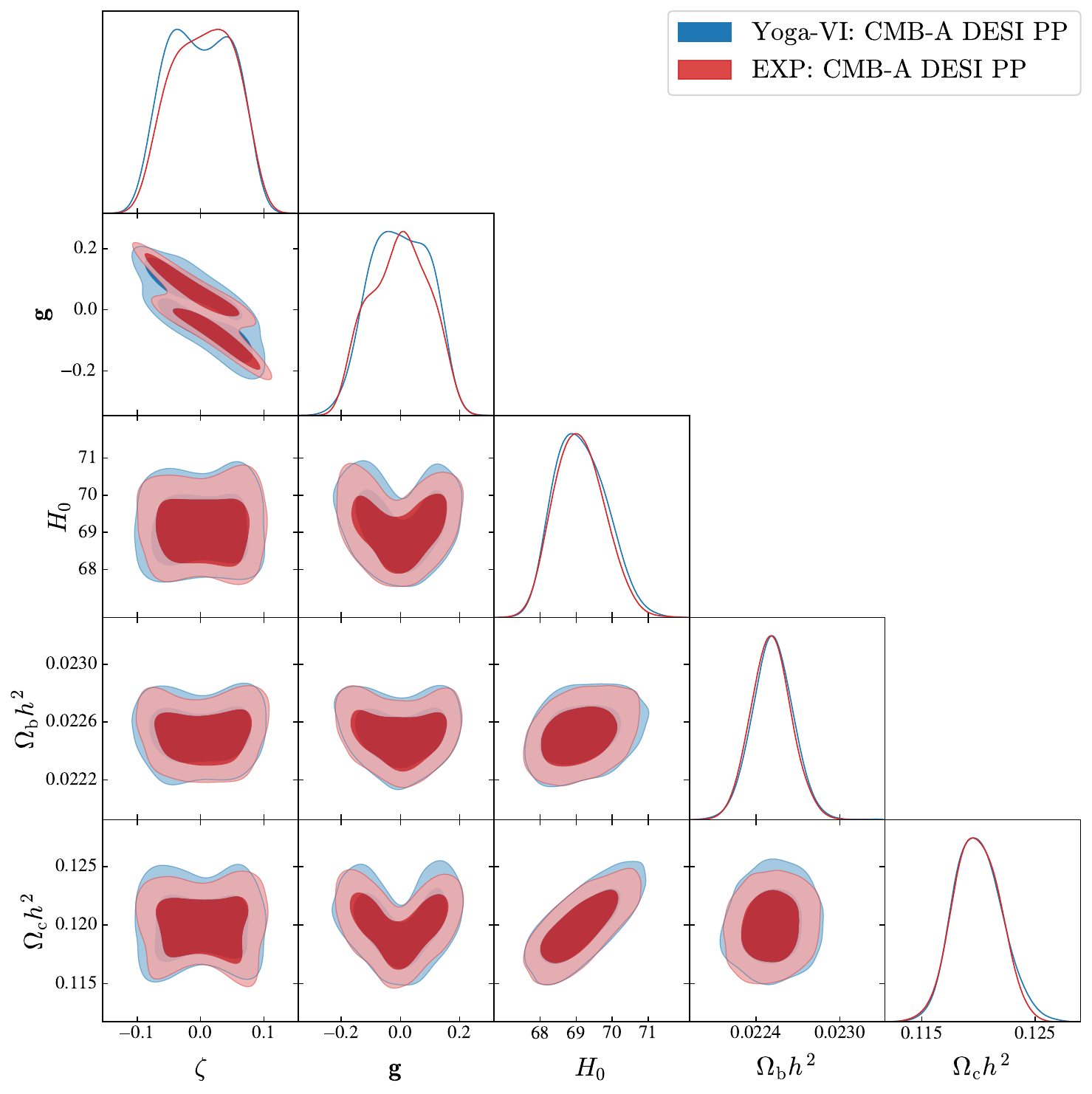}
\caption{
Comparison of the main common parameter constraints for the Exponential and Yoga axio-dilaton models.
}
    \label{fig:EXP vs Yoga}
\end{figure}

Despite this very different underlying structure, the preferred cosmological histories of the 
exponential model closely mirror those of the displaced initial condition Yoga-VI branch, as shown in 
\Cref{fig:EXP vs Yoga}.  
The triangle plot demonstrates that the preferred regions of parameter space are nearly 
identical: both models select a narrow band in $(\Omega_B h^2,\Omega_c h^2,H_0)$ space and 
display the same characteristic bimodality in the two-dimensional $(\mathbf{g},\zeta)$ plane, while the 
corresponding one-dimensional posteriors remain unimodal.
In fact, the exponential model exhibits a slightly more extended pair of islands in the 
$(\mathbf{g},\zeta)$ plane, aligned along the same correlation direction as in the Yoga-VI case, but 
with overall tighter two-dimensional constraints. For the CMB-A DESI PP dataset, the EXP best-fit 
reaches $(\mathbf{g},\zeta)\simeq(0.127,-0.058)$, whereas the corresponding Yoga-VI best-fit sits at 
$(0.14,-0.061)$, with both points lying along the same degeneracy line.

This agreement extends directly to the marginalised constraints in \Cref{tab:CMB-A}.  
For the CMB-A DESI PP combination, the exponential run yields 
$H_0 = 69.10^{+0.64}_{-0.76}$ and $(\mathbf{g},\zeta) = (-0.003 \pm 0.099,\; 0.005 \pm 0.049)$, which is 
statistically indistinguishable from the Yoga-VI solution 
$H_0 = 69.18^{+0.63}_{-0.81}$ with $(\mathbf{g},\zeta) = (0.00 \pm 0.10,\; 0.002\pm0.052)$.
In both cases, the best-fit values lie along the same compensation ridges in the 
$(\mathbf{g},\zeta)$ plane, but they give rise to qualitatively different histories for $m_e(z)$ and  
$w_{\chi,\mathrm{eff}}(z)$, as shown in \Cref{fig:best-fit_yoga}.

The EXP solution shows a somewhat larger preferred early-time enhancement of $m_e(z)/m_{e,0}$, which also survives much later in the cosmology, not reaching 0.1\% agreement with today's value until well into dark energy domination. The phantom phase of the dilaton also survives to much lower redshifts in the cosmology compared to Yoga-VI, crossing the phantom-divide line much later at $z \sim 0.5$. Numerically, this is reflected in a slightly better fit for EXP, with 
$\Delta\chi^2 = -2.7$ (CMB-A DESI PP) and $-3.8$ (CMB-A DESI), compared to $-2.1$ and $-3.5$
for Yoga-VI, although in both cases the improvement over $\Lambda$CDM remains statistically modest
given the 2/3 additional parameters, respectively.

\subsubsection{Inclusion of SPT}\label{Including spt}

We now modify the previous CMB-A dataset combinations by adding the SPT high-$\ell$ temperature and polarisation spectra (our CMB-B bundle), with the corresponding constraints collected in \Cref{tab:CMB-B}.  

For both the Minimal Yoga-VI and Exponential models, the inferred Hubble constant remains essentially unchanged: using CMB-B DESI PP we obtain
$H_0 = 69.19 \pm 0.70~\mathrm{km\,s^{-1}\,Mpc^{-1}}$ for Yoga-VI and
$H_0 = 69.16 \pm 0.68~\mathrm{km\,s^{-1}\,Mpc^{-1}}$ for the Exponential case,
very close to the corresponding CMB-A values.  

The matter couplings likewise remain small and consistent with zero at the $1\sigma$ level, with
$\mathbf{g} \sim \mathcal{O}(10^{-2})$ and $\zeta \sim \mathcal{O}(10^{-2})$ in all CMB-B combinations, and the qualitative structure of the posteriors is unchanged. Although we do not show the corresponding triangle plots for the CMB-B case, we have explicitly checked that the $(\mathbf{g},\zeta)$ contours retain the same approximate symmetry about the origin, and that the best-fit points still lie on the non-zero approximate $\mathbf{g}$-$\zeta$ cancellation ridges that cancel the respective baryon and axion CDM sourcing of the dilaton and generate a narrow excursion in $m_e(z)$ around recombination.

\begin{table*}
\centering
\makebox[\textwidth][c]{%
\begin{tabular}{lc|ccccc}
\toprule
Model & Dataset & $H_0~[\mathrm{km\,s^{-1}\,Mpc^{-1}}]$ & $\mathbf{g}$ & $\zeta$ & $\chi_i$ &  $\Delta \chi^2$ \\
\midrule
Yoga-VI & CMB-B DESI PP & $69.19 \pm 0.70$ (69.13) & $0.003 \pm 0.095$ (-0.052) & $0.002 \pm 0.050$ (-0.003) & $74.02 \pm 0.15$ (73.844) & -7.2 \\
 & CMB-B DESI & $69.42 \pm 0.71$ (69.50) & $-0.005 \pm 0.098$ (-0.037) & $-0.003 \pm 0.048$ (-0.032) & $73.98^{+0.21}_{-0.18}$ (73.814) & -6.8 \\
\specialrule{0.6pt}{2pt}{2pt}
EXP & CMB-B DESI PP & $69.16 \pm 0.68$ (69.99) & $0.002 \pm 0.093$ (0.104) & $-0.001 \pm 0.044$ (-0.022) & --  & -7.3 \\
 & CMB-B DESI & $69.39 \pm 0.71$ (70.16) & $-0.013^{+0.083}_{-0.12}$ (-0.164) & $0.001 \pm 0.046$ (0.071) & -- & -6.2 \\
 & CMB-B & $67.47^{+0.68}_{-1.2}$ (67.78) & $-0.004 \pm 0.074$ (0.041) & $0.001 \pm 0.048$ (-0.002) & -- & -- \\
\specialrule{0.6pt}{2pt}{2pt}
w0-wa + me & CMB-B DESI PP & $68.40 \pm 0.84$ (68.51) & -- & -- & -- & -12.3 \\
 & CMB-B DESI & $64.2^{+1.9}_{-2.6}$ (63.78) & -- & -- & --  & -10.5 \\
\specialrule{0.2pt}{0pt}{0pt}
w0-wa  & CMB-B DESI PP & $67.65 \pm 0.59$ (67.52) & -- & -- & -- & -9.6 \\
\specialrule{0.6pt}{2pt}{2pt}
$\Lambda$CDM & CMB-B DESI PP & $68.04 \pm 0.26$ (68.25) & -- & -- & -- & 0.0 \\
 & CMB-B DESI & $68.13 \pm 0.26$ (68.23) & -- & -- & -- & 0.0 \\
\bottomrule
\end{tabular}
}
\caption{
Posterior means with quoted $1\sigma$ marginal uncertainties and best-fit values in parentheses for all models fit to the CMB-B dataset combinations including SPT-3G. Columns list the inferred Hubble constant, the dilaton coupling $\mathbf{g}$, the axion CDM kinetic coupling $\zeta$, the initial dilaton value $\chi_i$ when present, and the change in best-fit $\Delta\chi^2$ relative to the corresponding $\Lambda$CDM run.
}
\label{tab:CMB-B}
\end{table*}

The main impact of including SPT is therefore not to shift the preferred parameter values, but to sharpen the fit preference for these models over $\Lambda$CDM.  
For CMB-B DESI PP the Yoga-VI and Exponential runs now achieve
$\Delta\chi^2 \simeq -7.2$ and $\Delta\chi^2 \simeq -7.3$, respectively, compared with the 
$\Delta\chi^2 \simeq -2$ to $-3$ obtained from CMB-A DESI PP.  
A similar pattern holds when Pantheon+ is omitted: with CMB-B DESI alone we find
$\Delta\chi^2 \simeq -6.8$ (Yoga-VI) and $-6.2$ (Exp), versus $-3.5$ and $-3.8$ for the CMB-A DESI case.

The SPT high-$\ell$ spectra therefore substantially increase the statistical weight of the preference for a small recombination-era step in $m_e(z)$ and the accompanying changes in baryon and CDM dynamics, even though the inferred $H_0$ and coupling values themselves remain nearly identical to the CMB-A solutions.

The physical reason for this strengthened preference is that SPT’s high-$\ell$ measurements constrain the damping tail far more tightly than Planck alone.  
A small upward excursion in $m_e(z_\star)$ simultaneously increases the hydrogen binding energy and reduces the Thomson cross section, shifting both the visibility function and the diffusion (Silk) scale.  
Such correlated changes allow the axio-dilaton trajectories to better reproduce the detailed high-$\ell$ TT/TE/EE structure at $\ell \gtrsim 2000$ in the angular power spectrum of the CMB, whereas $\Lambda$CDM is forced to accommodate the same features using only the standard six parameters. The deviation of the angular power spectrum for the Yoga-VI models using illustrative parameters is shown in \Cref{fig:power_spectra}.

This is essentially the same mechanism by which $w_0$-$w_a$ and $w_0$-$w_a$+$m_e$ extensions gain their even larger improvements in $\chi^2$ with respect to $\Lambda$CDM. For CMB-B DESI PP we find 
$\Delta\chi^2 \simeq -9.6$ for $w_0$-$w_a$ and $\Delta\chi^2 \simeq -12.3$ for $w_0$-$w_a$+$m_e$, again relative to $\Lambda$CDM. However, in the present scenarios the pattern of changes is not freely switched on, with the resulting cosmology across different epochs tied to a single rolling scalar field and its couplings to baryons and dark matter.  
The fact that the dynamical dilaton models improve the fit almost as effectively as these phenomenological extensions, despite having a much more constrained structure, indicates that SPT is responding specifically to the recombination-era microphysics induced by $m_e(z)$ rather than to arbitrary late-time freedom in $w(z)$.

Taken together, the CMB-B results demonstrate that the same dilaton-driven $m_e(z)$ excursion which alleviates the Hubble tension is also favoured by the small-scale SPT data, and that this preference is robust across both the Minimal Yoga and Exponential potentials.

\subsubsection{SH0ES Prior}
\label{sec:shoes}

We finally assess the impact of imposing a local $H_0$ prior by using the SH0ES calibration of the
Pantheon$+$ supernova sample (PPSH0ES), with results summarised in \Cref{tab:ppshoes}.  
Because both the Minimal Yoga-VI and Exponential models already reduce the Planck-SH0ES
discrepancy below $3\sigma$ in the CMB-A DESI PP analysis, the distance-ladder prior is not in sharp
conflict with the CMB in these scenarios. It is therefore meaningful to include the SH0ES
calibration through the full PPSH0ES likelihood, which consistently propagates the
calibration information into the cosmological parameter inference.

With the SH0ES calibration imposed, both axio-dilaton models are driven into the higher-$H_0$
regions of parameter space already present in the CMB BAO PP posteriors.
For CMB-A DESI PPSH0ES we obtain
$H_0 = 70.92 \pm 0.61~\mathrm{km\,s^{-1}\,Mpc^{-1}}$ for Yoga-VI and
$H_0 = 70.80 \pm 0.58~\mathrm{km\,s^{-1}\,Mpc^{-1}}$ for the Exponential model.

\begin{figure}
    \centering
    \includegraphics[width=\linewidth]{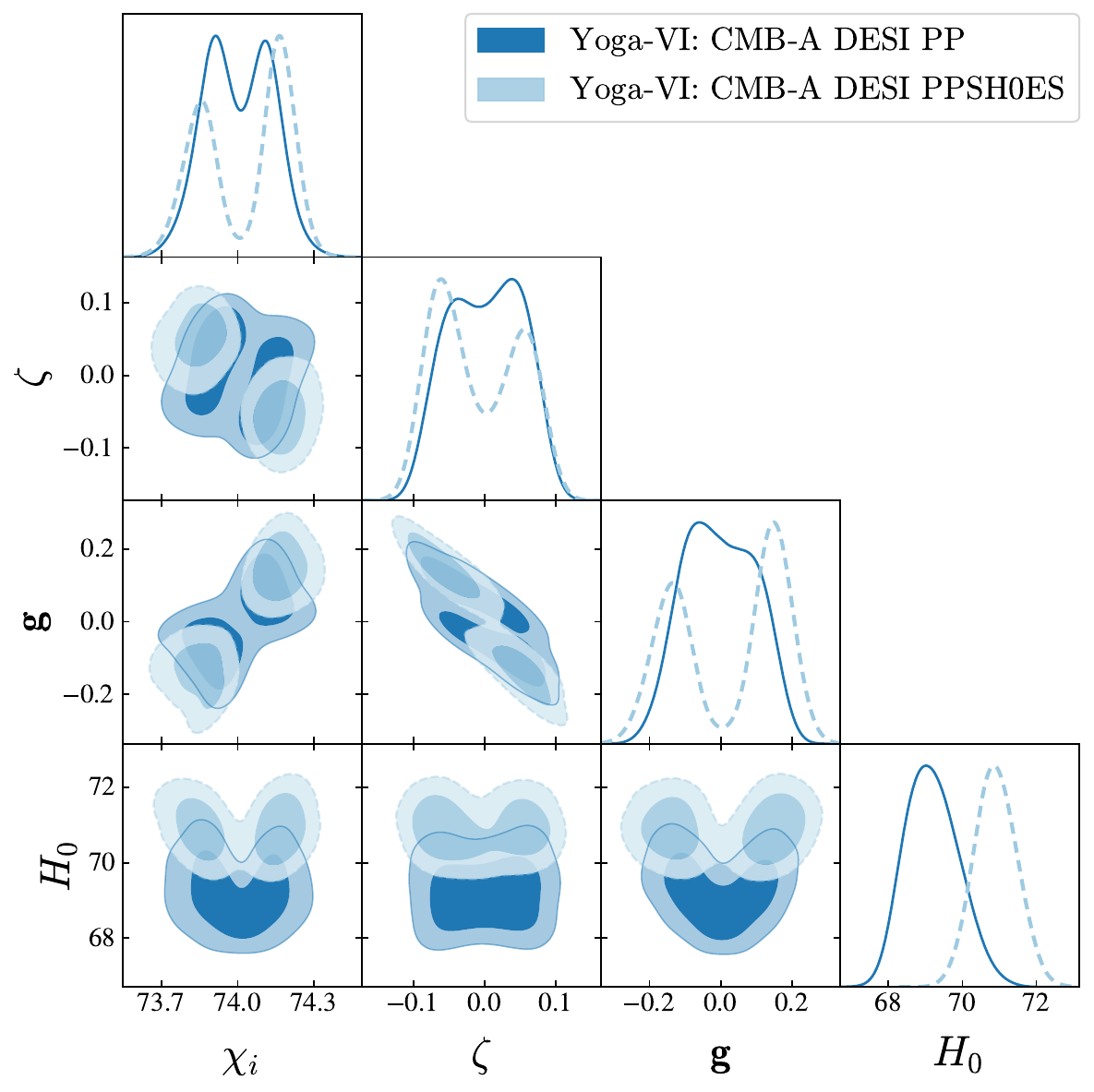}
\caption{
Triangle plot showing constraints on the minimal Yoga-VI model with and without the SH0ES calibration using the CMB-A DESI PP datasets. Imposing the SH0ES calibration amplifies the existing correlations among $\mathbf{g}$, $\zeta$, and $H_0$, with much larger couplings and initial dilaton displacements being preferred.
}
    \label{fig:Yoga Shoes}
\end{figure}

\begin{figure}
    \centering
    \includegraphics[width=\linewidth]{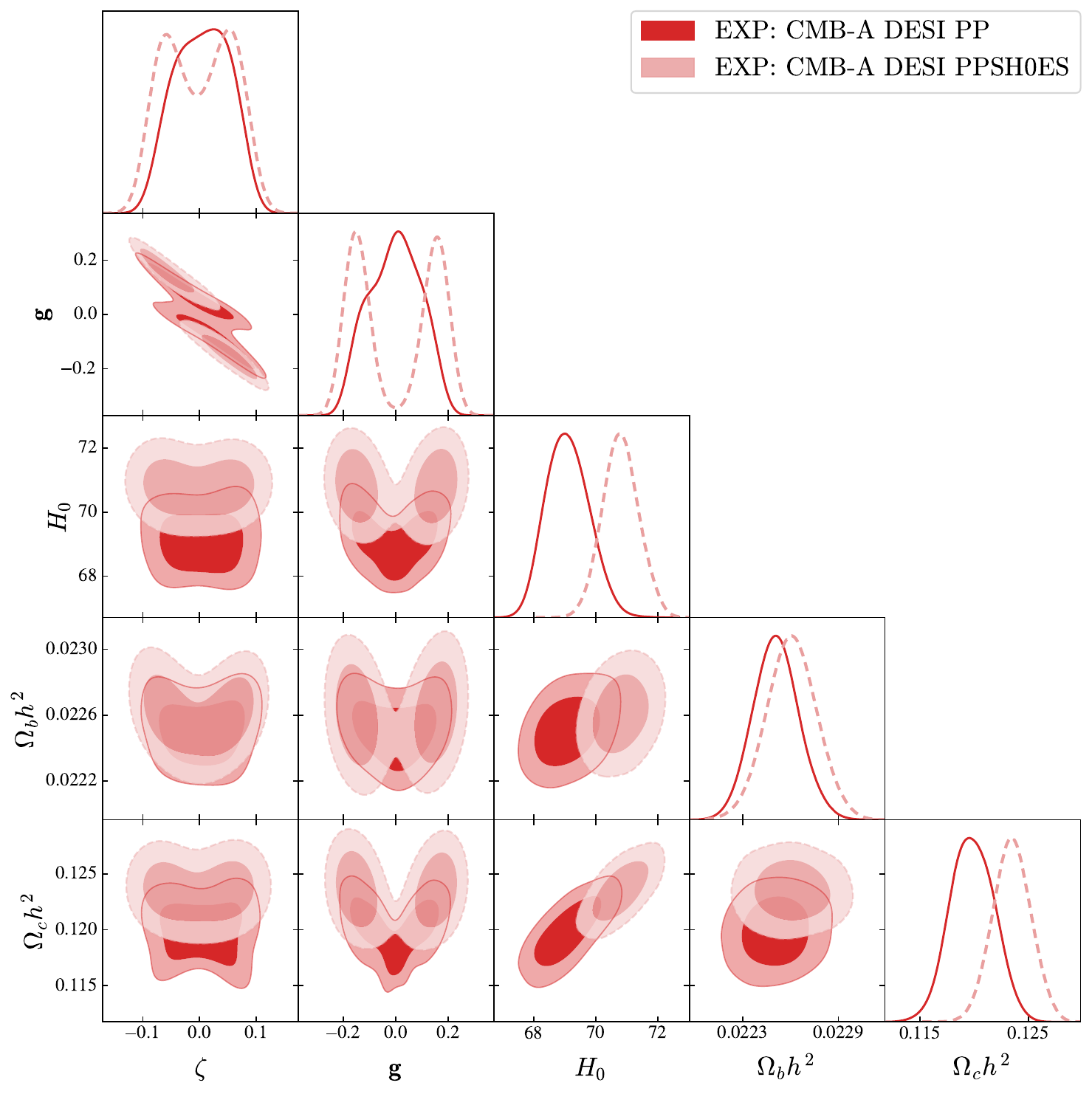}
\caption{
Triangle plot showing constraints on the minimal Exponential model with and without the SH0ES calibration using the CMB-A DESI PP datasets.
}
    \label{fig:Exp Shoes}
\end{figure}

By construction, these values are consistent with the SH0ES calibration and lie within the
pre-existing high-$H_0$ solutions visible in \Cref{fig:Yoga Shoes} and \Cref{fig:Exp Shoes}.
In particular, the SH0ES calibration selects the high-$H_0$ branches of the U-shaped
correlations already present in the CMB BAO PP posteriors, both in the
$\mathbf{g}$--$H_0$ and $\chi_i$--$H_0$ planes. As a result, the two-dimensional posteriors
develop two separate $1\sigma$ regions, symmetric about $\mathbf{g}=0$ and
$\chi_i \simeq 74$, which remain connected within the $2\sigma$ contour.

The matter couplings remain statistically compatible with zero at the $1\sigma$ level,
but their best-fit values are nevertheless shifted away from the origin within the allowed
uncertainties: for Yoga-VI we find $\mathbf{g} = -0.01^{0.19}_{-0.17}$ and
$\zeta = -0.006\pm 0.059$, while the Exponential run yields
$\mathbf{g} = 0.00 \pm 0.16$ and $\zeta = -0.001 \pm 0.061$, with best-fit values at
$(\mathbf{g},\zeta) \simeq (0.099, -0.031)$ and $(-0.199, 0.082)$, respectively.
From the same figures, \Cref{fig:Yoga Shoes} and \Cref{fig:Exp Shoes}, it is also clear
that the butterfly-shaped correlations between $\zeta$ and $\chi_i$, and between
$\mathbf{g}$ and $\chi_i$, are now selected in two separate regions at more than the
$2\sigma$ level, symmetric along the directions of the correlations already discussed in
the absence of the $H_0$ calibration.

The $\Delta\chi^2$ values reflect the strength of the models combined with the SH0ES
calibration relative to $\Lambda$CDM with the same dataset calibration added. The dilaton
models achieve $\Delta\chi^2 = -19.2$ (Yoga-VI) and $-19.5$ (Exponential), a substantial
jump from the $\Delta\chi^2 \simeq -2$ to $-3$ obtained without SH0ES. This improvement is
driven almost entirely by the much better agreement with the local $H_0$ determination,
since in $\Lambda$CDM the SH0ES calibration is in tension with the CMB at the $5$--$6\sigma$
level.

The phenomenological $w_0$--$w_a+m_e$ and $\Lambda$CDM+$m_e$ extensions show comparable
improvements ($\Delta\chi^2 \simeq -19.4$ and $-19.2$), as expected when an external $H_0$
calibration is imposed. However, in the axio--dilaton case the shift to high $H_0$ is
realised through the controlled dynamics of a single rolling scalar whose couplings
simultaneously determine the $m_e(z)$ excursion, the associated changes in baryon loading,
and the correlated response of the dark-matter sector. This stands in contrast with
$w_0$--$w_a$ and $w_0$--$w_a+m_e$, where $w(z)$ and $m_e(z)$ can vary independently.

\begin{table*}
\centering
\makebox[\textwidth][c]{%
\begin{tabular}{l|ccccc}
\toprule
Model  & $H_0~[\mathrm{km\,s^{-1}\,Mpc^{-1}}]$ & $\mathbf{g}$ & $\zeta$ & $\chi_i$ & $\Delta \chi^2$ \\
\midrule
Yoga VI & $70.92 \pm 0.61$ (70.86) & $-0.01^{+0.19}_{-0.17}$ (0.099) & $0.006 \pm 0.059$ (-0.031) & $74.00^{+0.23}_{-0.20}$ (74.213) & -19.2 \\
\specialrule{0.6pt}{2pt}{2pt}
EXP & $70.80 \pm 0.58$ (71.13) & $0.00 \pm 0.16$ (-0.199) & $-0.001 \pm 0.061$ (0.082) & -- & -19.5 \\
\specialrule{0.6pt}{2pt}{2pt}
w0-wa + me & $70.51 \pm 0.73$ (70.37) & -- & -- & -- & -19.4 \\
\specialrule{0.6pt}{2pt}{2pt}
$\Lambda$ CDM+me & $70.97 \pm 0.57$ (71.34) & -- & -- & -- & -19.2 \\
\specialrule{0.2pt}{0pt}{0pt}
$\Lambda$ CDM & $68.72 \pm 0.28$ (68.77) & -- & -- & -- & 0.0 \\
\bottomrule
\end{tabular}
}
\caption{
Posterior means ($\pm 1\sigma$) with best-fit values in parentheses, and $\Delta\chi^2$ relative to the $\Lambda$CDM run all using the CMB-A DESI PPSH0ES datasets. The bottom two rows are separated by a line to illustrate that these cases are included purely for comparison and, in general, should not be viewed as reliable model constraints due to the combination of datasets in more than $3\sigma$ tension with each other (Planck and SH0ES).
}
\label{tab:ppshoes}
\end{table*}

A noteworthy feature of the SH0ES-calibrated runs is that the large-$H_0$ solution typically
requires $|\mathbf{g}| \sim \mathcal{O}(0.1)$, which is similar to the size
$|\mathbf{g}| \simeq \sqrt{1/6} \simeq 0.4$ of the dilaton--matter couplings chosen for other
reasons within the Yoga model (which is itself similar to those that commonly appear in
extra-dimensional dilaton constructions).

The main puzzle with such couplings is that, if they are of the same strength in the
present-day Universe, then they should have been detected by Solar-System tests of GR. We
do not understand how this is ultimately resolved but find it intriguing that cosmology
prefers scalar couplings that present-day measurements seem to forbid. This issue was
already recognised as a potential problem when the Yoga model was initially
proposed~\cite{Burgess:2021qti,Brax:2022vlf}.

This suggests the need for a screening mechanism:
a situation in which nonlinear or matter-dependent effects allow the effective coupling
carried by a macroscopic object such as the Sun to differ from the sum of couplings to each
of its constituent atoms, perhaps along the lines that have been proposed for these models
in~\cite{Brax:2023qyp}. Work is underway to find a modification of the theory for which
microscopic reasoning, cosmological observations, and Solar-System phenomenology can agree
on the same values of $\mathbf{g}$.

Once the SH0ES calibration is included, the axio--dilaton models move cleanly onto their
high--$H_0$ region without degrading the quality of the CMB and BAO fits.
The SH0ES-calibrated solutions should therefore be viewed as consistency checks: the same
regions of parameter space that alleviate the tension in CMB BAO PP dataset combinations also remain compatible
with the local distance ladder, and do so via a tightly correlated physical mechanism.

\subsection{Model comparisons}\label{sec:discussion}

In this section we compare the axio--dilaton framework to the two phenomenological benchmarks:
$\Lambda$CDM+$m_e$ and CPL ($w_0$–$w_a$).  
We first explain why $\Lambda$CDM+$m_e$ can push $H_0$ to such high values while remaining
compatible with CMB and BAO, and then analyse why CPL fits late--time distance and growth
data so well.  
We identify which pieces of this behaviour axio--dilaton models can reproduce, and where
they necessarily fall short, clarifying why the minimal constructions do a marginally
worse job in both the $H_0$ and late--time sectors despite sharing many of the same
qualitative mechanisms.

\subsubsection{$\Lambda$CDM + $m_e$ vs.\ axio–dilaton}
\label{sec:discussion_vem}

In the axio--dilaton implementation the rolling scalar modifies both the
background and the recombination microphysics. At the level of homogeneous
densities we have from \cref{eq:mass_universal} and
\cref{eq:axion_background_density}
\begin{equation}
  \rho_B(a)\propto A(\chi)\,a^{-3},\;\;\;
  \rho_{\rm ax}(a)\propto \frac{a^{-3}}{W(\chi)},\;\;\;
  m_e(\chi)\propto A(\chi),
\end{equation}
so a brief excursion of $\chi$ around recombination, $a_\star$, simultaneously
shifts the electron mass (and hence the visibility and diffusion scales), the
baryon loading $R(z_\star)=3\bar\rho_B/4\bar\rho_\gamma$, and the CDM density
entering matter--radiation equality. This correlated evolution is precisely
what distinguishes the axio--dilaton fits from phenomenological
$\Lambda$CDM+$m_e$, where $m_e$ can change while $\rho_B\propto a^{-3}$ and
$\rho_C\propto a^{-3}$ remain fixed.

Early--recombination constructions such as those based on $\Lambda$CDM+$m_e$
exploit this extra freedom. In those models one raises $H_0$ by shrinking the
sound horizon while keeping the primary CMB spectra essentially unchanged,
moving along an approximate early--recombination degeneracy direction (see,
e.g.\ Ref.~\cite{Sekiguchi:2020teg})\footnote{Here $\omega_i = \Omega_i h^2$ is the
Hubble-rate-independent, non-comoving physical density, and
$\omega_m = \omega_B + \omega_c$ is the total matter density.}
\begin{equation}
  \Delta\!\ln\omega_B \;=\; \Delta\!\ln\omega_m \;=\; -\,\Delta\!\ln a_*,
  \qquad
  \theta_*~\text{fixed},
  \label{eq:ERdeg}
\end{equation}
so that $R(x)$ and $a\mathcal{H}$ remain essentially unchanged, while the sound
horizon, the drag scale, and the matter--radiation equality scale all co--move,
$r_s\propto r_d \propto r_{eq}\propto a_*$. In this way the early acoustic ruler
is recalibrated without disrupting the detailed shape of the primary CMB
anisotropies, and a higher $H_0$ can then be absorbed by late--time distance
data.

By contrast, in the axio--dilaton case the same dilaton excursion that modifies
$m_e(\chi_\star)$ also drives non--universal rescalings of the baryon and CDM
sectors. The background Klein--Gordon \cref{eq:dilaton_background} can be written schematically with a matter source
\begin{equation}
  S(a)\;\equiv\;-\,\mathbf{g}\,\bar\rho_B(a)\;-\;\zeta\,\bar\rho_{\rm ax}(a),
  \label{eq:Ssource}
\end{equation}
and the CMB+BAO data prefer a short, controlled excursion of $\chi$ for which
this source remains small near recombination. Since
$\bar\rho_{\rm ax}\gg\bar\rho_B$ at $a_\star$, keeping
$S(a_\star)\simeq 0$ typically requires $|\zeta|<|\mathbf{g}|$
(weighted by the density ratio), so that the fractional changes in the two
matter components are not equal and satisfy
\begin{equation}
  \Delta\!\ln\bar\rho_{B\star} \;=\; \Delta\!\ln A(\chi_\star)
  \;\not\approx\;
  \Delta\!\ln\bar\rho_{{\rm ax}\star} \;=\;
  \Delta\!\ln\!\big[1/W(\chi_\star)\big].
\end{equation}

\begin{figure}
    \centering
    \includegraphics[width=\linewidth]{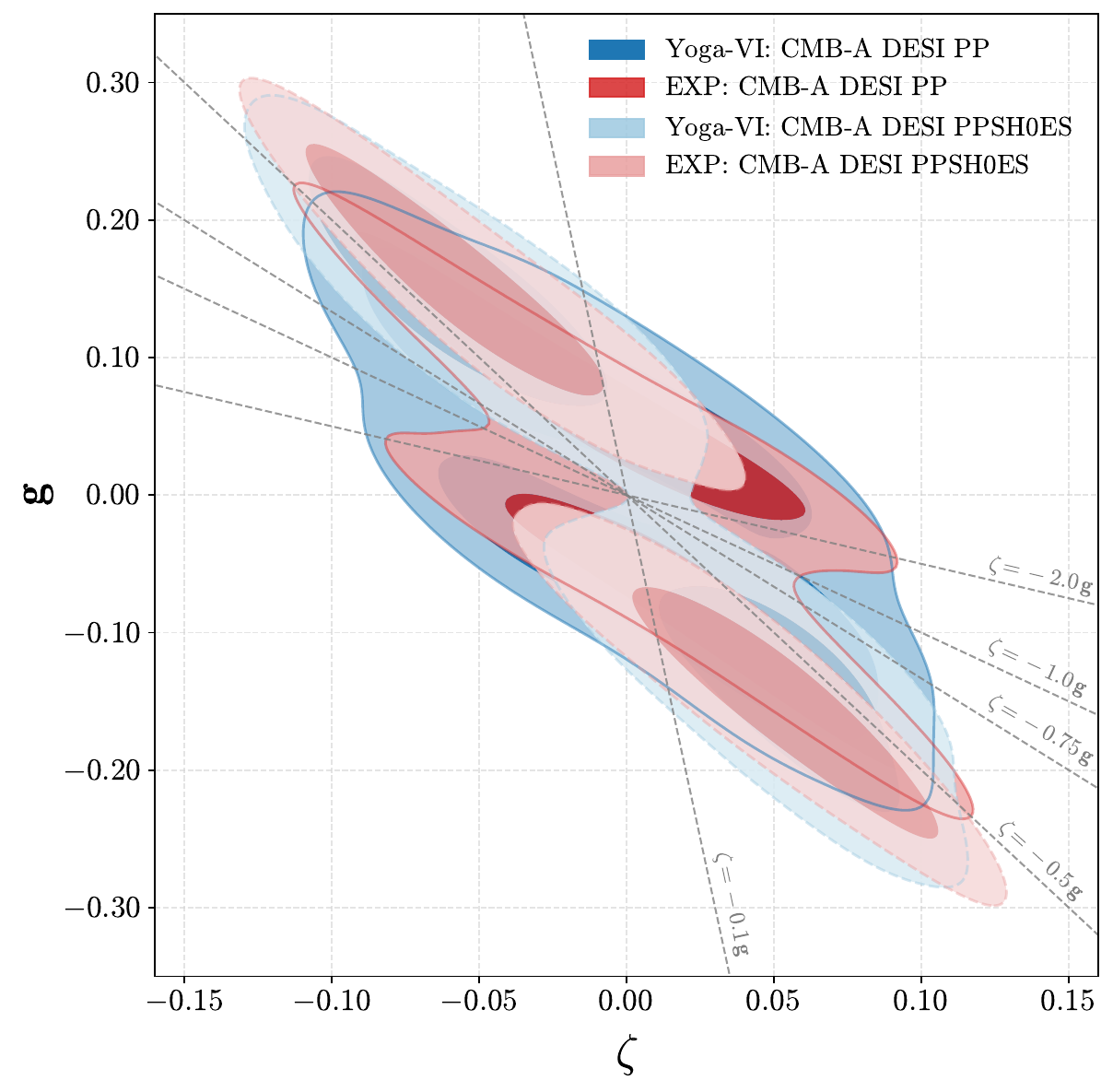}
\caption{Posteriors in the $\zeta$--$\mathbf{g}$ plane for the relevant models with and without the SH0ES calibration.}
    \label{fig:zeta-g plane}
\end{figure}

The equal--rescaling pattern \cref{eq:ERdeg} therefore cannot be realised at the
same time as the near--cancellation of the two terms in
\cref{eq:Ssource} when $\bar\rho_{\rm ax}\gg\bar\rho_B$. This is illustrated in
\Cref{fig:zeta-g plane}, where we can see that the posterior contours lie
explicitly displaced from the $\zeta = -\mathbf{g}$ line that would ensure a
universal rescaling of the matter densities.

In practice, trying to shift the recombination epoch more strongly to raise
$H_0$ further inevitably induces correlated changes in $R(z_\star)$ and
$z_{\rm eq}$, which are tightly constrained by the detailed shape of the
TT/TE/EE spectra, the high--$\ell$ damping tail, CMB lensing, and BAO.

These constraints show up directly in the parameter posteriors. For the CMB--A
DESI PP combination, the dynamical Yoga--VI model with $m_e$ variation prefers
$H_0\simeq 69.2~\mathrm{km\,s^{-1}\,Mpc^{-1}}$ with
$\Delta\chi^2\simeq -2.1$ relative to $\Lambda$CDM, and best--fit values lying
in the off--zero regions of $(\mathbf{g},\zeta)$ space. Turning off the
electron--mass variation in an otherwise identical ``no--$m_e$'' run lowers the
inferred Hubble constant to $H_0\simeq 68.5~\mathrm{km\,s^{-1}\,Mpc^{-1}}$ and
tightens the baryon coupling to $g\simeq 0.006\pm 0.083$,
while the CDM coupling $\zeta$ changes very little. The reason is that, once $m_e$ is
fixed, the CMB peak positions and damping tail can no longer be retuned
through the visibility and diffusion scales, so any dilaton--induced change in
the baryon sector must be absorbed directly through $R(z_\star)$ and
$r_s(z_d)$. Planck+BAO then penalise deviations in $\bar\rho_B$ more strongly,
constraining $\mathbf{g}$ toward zero, whereas the $\zeta$ constraint
remains set mainly by equality and growth or lensing through the background
scaling \cref{eq:axion_background_density}.

The same comparison explains why phenomenological $\Lambda$CDM+$m_e$ can lift
$H_0$ slightly more than the axio–dilaton fits.  In $\Lambda$CDM+$m_e$ (with
CMB–A DESI PP) the recombination lever alone can shrink $r_s$ and raise $H_0$
while keeping $\rho_B\propto a^{-3}$ and $\rho_c\propto a^{-3}$ fixed, so the
data tolerate a larger step in $m_e(\chi_\star)$; numerically we find
$H_0\simeq 69.6~\mathrm{km\,s^{-1}\,Mpc^{-1}}$ and
$\Delta\chi^2\simeq -2.6$.  In the axio-dilaton case the same excursion in
$\chi$ simultaneously moves $m_e$, $\bar\rho_B$ and $\bar\rho_c$, so BAO and
the CMB acoustic pattern restrict the joint response of $(r_s,H_0)$ more
tightly and the posterior settles at
$H_0\simeq 69.2~\mathrm{km\,s^{-1}\,Mpc^{-1}}$, with marginalised couplings
still overlapping $\mathbf{g}=\zeta=0$ despite off–zero best–fits.

Imposing the exponential potential $V(\chi)\propto e^{-\lambda\chi/M_{\rm Pl}}$
with $\lambda = 4\zeta$ further reduces the freedom in the dilaton sector, but
does not change this basic picture. The data again prefer a narrow excursion
of $\chi$ that produces a small step in $m_e(\chi_\star)$ while keeping the
induced drifts in $\bar\rho_B$ and $\bar\rho_c$ modest, leading to
$H_0\simeq 69.1~\mathrm{km\,s^{-1}\,Mpc^{-1}}$ and
$\Delta\chi^2\simeq -2$ to $-4$ depending on the dataset, still below the
$H_0$ uplift achieved by $\Lambda$CDM+$m_e$ for the reasons above.

In other words, VEM--type solutions work by effectively decoupling recombination
microphysics from matter rescalings and moving along \cref{eq:ERdeg}, whereas
the axio--dilaton model recouples these sectors: the step in $m_e$, the change
in baryon loading, and the shift in equality are fixed in a correlated way by
the same source \cref{eq:Ssource}.

A useful way to view these results is that our model predicts definite ratios
between the derived shifts in the sound horizon, baryon loading, and equality,
for example in $(\Delta\!\ln r_s,\,\Delta\!\ln R,\,\Delta\!\ln z_{\rm eq})$,
whereas VEM scenarios realise the equal--rescaling pattern \cref{eq:ERdeg}.
The modest uplift in $H_0$ and the tightening of $\mathbf{g}$ when $m_e$ is
held fixed are therefore testable signatures of the fact that recombination,
baryons, and CDM all feel the same rolling scalar rather than being dialled
independently.

\subsubsection{\texorpdfstring{Comparison to $w_0w_a$ cosmology}{Comparison to w0–wa cosmology}}
\label{sec:discussion_CPL}

It is also useful to compare the axio--dilaton fits to more agnostic dynamical
dark energy models, such as the CPL parametrisation with equation of state
$w(a)=w_0+w_a(1-a)$, both with and without a time-varying electron mass. In our
chains these are labelled $w_0$--$w_a$ and $w_0$--$w_a+m_e$ in
\Cref{tab:CMB-A,tab:CMB-B,tab:ppshoes}. Across all CMB+BAO combinations the CPL
extensions achieve a larger raw improvement in $\chi^2$ than either the
Minimal Yoga--VI or Exponential dilaton models, despite yielding comparable or
even slightly lower values of $H_0$.

For instance, with CMB--A DESI PP we find $\Delta\chi^2\simeq-6.9$ for
$w_0$--$w_a$ and $\Delta\chi^2\simeq-8.2$ for $w_0$--$w_a+m_e$, compared with
$\Delta\chi^2\simeq-2.1$ (Yoga--VI) and $-2.7$ (Exponential). Including SPT in
the CMB--B bundle improves the fit of all dynamical dark energy models but does
not alleviate the goodness--of--fit hierarchy between them: for CMB--B DESI PP
the CPL+$m_e$ model reaches $\Delta\chi^2\simeq-12.3$, whereas the corresponding
Yoga--VI and Exponential fits saturate around $\Delta\chi^2\sim-7$.

The reason is that the CPL sector can adjust the background geometry and growth
history much more freely than the axio--dilaton. Allowing $w_0$ and $w_a$ to
float effectively decouples the late-time expansion rate $H(z\lesssim 1)$ from
the early-time physics that sets the sound horizon and the damping tail. Small
residual mismatches in the CMB acoustic envelope, the DESI BAO distances, and
the SPT high-$\ell$ spectra can then be absorbed by gentle changes in $w(z)$
and $\Omega_{\rm m}$, without being tied to a specific pattern of changes in
$\rho_B(a)$, $\rho_c(a)$, and $m_e(z)$. In other words, CPL uses the freedom in
$w(z)$ to clean up detailed geometric and lensing residuals, and the data reward
this with a more negative $\Delta\chi^2$.

In the axio--dilaton constructions the situation is much more constrained. For
the Minimal Yoga variants the effective equation of state
$w_{\chi,\rm eff}(z)$ remains extremely close to $-1$ at late times, evolving
only weakly for $z\lesssim 1$, and the main handle on the data is a brief
excursion of $\chi$ around recombination that simultaneously perturbs
$m_e(\chi_\star)$, the baryon loading $R(z_\star)$, and the CDM density
$\rho_c(a)$, as discussed in \Cref{sec:discussion_vem}.
  
In the Exponential model a transient phantom phase can also appear at
intermediate redshifts (as seen in \Cref{fig:best-fit_yoga}), but its timing and
amplitude are \emph{not} freely adjustable in the way that $(w_0,w_a)$ are:
once the couplings $(\mathbf{g},\zeta)$ are chosen to fit the recombination-era
step in $m_e(z)$, the subsequent evolution of $\chi(a)$, and hence of
$w_{\chi,\rm eff}(z)$, is essentially fixed by the Klein--Gordon dynamics.

In both cases, fitting the CMB acoustic scale and damping pattern largely fixes
the allowed $\chi$ histories, and the late-time expansion is then inherited
from this constrained early-time excursion rather than being an independent
degree of freedom. This built-in rigidity is precisely what makes the
axio--dilaton predictive, but it also limits how far the global $\chi^2$ can be
lowered relative to $\Lambda$CDM and to the more flexible CPL sector.

The difference becomes even clearer when $m_e$ variation is included. In the
CPL+$m_e$ runs the electron-mass history can shift the recombination epoch and
sound horizon largely independently of the matter densities, while the CPL
sector adjusts $w(z)$ to keep CMB peak positions, BAO distances, and the SNe
ladder in good agreement. The recombination and late-time degrees of freedom
are only loosely correlated, so the model can exploit both an early-time
recalibration of $r_s$ and a late-time reshaping of $H(z)$.

In the axio--dilaton model, the same $\chi$ excursion that modifies
$m_e(\chi_\star)$ necessarily rescales $\rho_B(a)$ and $\rho_c(a)$, and the
Klein--Gordon source structure constrains their relative changes. As a result,
Planck, DESI, and SPT jointly restrict the allowed $(r_s,H_0)$ shift more
tightly, and the net $\Delta\chi^2$ improvement remains more modest even though
$H_0$ reaches larger values.

The SH0ES-calibrated combinations in \Cref{tab:ppshoes} follow the same pattern.
Once the local $H_0$ calibration is imposed, any model flexible enough to
accommodate $H_0\simeq 70$--$71~\mathrm{km\,s^{-1}\,Mpc^{-1}}$ while preserving
the CMB and BAO fits achieves a large negative $\Delta\chi^2$. The
CPL+$m_e$, $\Lambda$CDM+$m_e$, and axio--dilaton runs all reach
$\Delta\chi^2\simeq -19$ with CMB--A DESI PPSH0ES, and the differences in raw fit
quality become small compared to the effect of the calibration.  

The real distinction is then conceptual: CPL and $\Lambda$CDM+$m_e$ accomplish
this by allowing largely independent deformations of $w(z)$ and $m_e(z)$,
whereas the axio--dilaton achieves a comparable description of the data with a
single scalar field whose couplings tie together the recombination-era
microphysics and the matter sector. This tighter structure explains why the
axio--dilaton fits do not quite match the CPL $\Delta\chi^2$ values on CMB+BAO
alone, while still offering a competitive and more constrained realisation of
dynamical dark energy once the full dataset bundle is considered.

\section{Lessons learned}\label{sec:conclusions}

Measurements are hard, and we do not yet know whether recent cosmological
tensions will evaporate once systematic uncertainties can be more accurately
assessed. But their persistence in the face of repeated reassessment does give
one pause.

The main result of this paper is that these tensions can be resolved by a
relatively simple modification of the $\Lambda$CDM picture. It suffices to
supplement the Standard Model and General Relativity with a minimal
axio--dilaton pair of scalars, designed to model the dark sector in its
entirety---\textit{i.e.} both Dark Energy (the dilaton) and Dark Matter (the
axion).

The key feature that helps resolve the tensions is the universal dependence of
particle masses on the dilaton together with the dilaton's light mass, which
together predict time dependence of particle masses around the epoch of
radiation--matter equality, which is visible because it is still taking place
at recombination.

A major puzzle raised by the success of the cosmological fits of these models
is that the best-fit solutions predict dilaton--matter couplings that appear
too large to have escaped detection in Solar-System tests of gravity. In this
sense, the axio--dilaton framework shifts the primary tension away from a
discrepancy between early- and late-universe cosmology and toward a potential
inconsistency between cosmological observations and local tests of gravity.

This field content and properties of these models were proposed in~\cite{Burgess:2021obw},
motivated by a completely different clue: the cosmological constant
problem~\cite{Weinberg:1988cp,Burgess:2013ara}. This asks how the very small
size of the Dark Energy density can be obtained in a technically natural way,
given that cosmology must emerge as the low-energy limit of a fundamental
theory involving much higher mass scales.

In this picture the dilaton and axion properties are largely suggested by
accidental approximate scaling and shift symmetries that arise commonly in UV
completions of gravity. The small mass of the dilaton is tied to the small size
of the Dark Energy density in these models along the lines suggested
in~\cite{Albrecht:2001xt}. The biggest problem with these models is that they
are predicted to involve dilaton--matter couplings that seem too large to have
escaped detection in Solar-System tests of gravity.

Both fundamental reasoning and the resolution of the cosmological tensions
seem to point in the same direction: new light scalar fields with interestingly
large couplings. Consistency with tests of gravity suggests that the
present-day dilaton either has very different properties (perhaps a larger
mass or weaker couplings) than throughout recent cosmological history, or that
its response to bulk matter in the Solar System is more complicated in a way
that allows screening to occur. In either case one must abandon a beautifully
minimal picture of the dark sector to develop more acceptable gravitational
phenomenology in the current Universe. Work to understand these issues is
underway.

\begin{acknowledgments}
\noindent We thank Niayesh Afshordi, Philippe Brax, Marco Costa, Anne Davis, Neal Delal, Junwu Huang, Jeong Han Kim, Daniel Kessler, Sergio Sevillano Mu\~noz, Sergey Sibiryakov, Zach Weiner and Chan Yeon Won for insightful suggestions and useful discussions. AS is supported by the W.D. Collins Scholarship. CvdB is supported in part by the Lancaster–Sheffield Consortium for Fundamental Physics under STFC grant: ST/X000621/1. CB's research was partially supported by funds from the Natural Sciences and Engineering Research Council (NSERC) of Canada. EDV is supported by a Royal Society Dorothy Hodgkin Research Fellowship. 
We acknowledge the IT Services at The University of Sheffield for the provision of services for High Performance Computing. 
This article is based upon work from the COST Action CA21136 - ``Addressing observational tensions in cosmology with systematics and fundamental physics (CosmoVerse)'', supported by COST - ``European Cooperation in Science and Technology''. MM is supported by Kavli IPMU which was established by
the World Premier International Research Center Initiative (WPI), MEXT, Japan. MM is also grateful for the hospitality of Perimeter
Institute where part of this work was carried out. Her visit to Perimeter Institute was supported by a
grant from the Simons Foundation (1034867, Dittrich). Research at the Perimeter Institute is supported in part by the Government of Canada through NSERC and by the Province of Ontario through MRI.   
\end{acknowledgments}

\bibliographystyle{apsrev4-2}
\bibliography{bibliography}

\appendix

\section{Useful formulae}

\subsection{Perturbed Einstein Equations}\label{ssec:perutrbed_einstein_eqs}

In this appendix we collect the linearized Einstein equations in Fourier space,
written in Newtonian gauge,
$
ds^2 = a^2(\eta)\bigl[-(1+2\Phi)d\eta^2 + (1-2\Psi)\delta_{ij}dx^i dx^j\bigr].
$
We keep both metric potentials $(\Phi,\Psi)$ explicit (so the expressions remain
valid even if small effective anisotropic stresses are present), and treat the
axion field $\mfa$ in the rapid--oscillation regime using oscillation averages
$\langle\cdots\rangle$ defined below. The remaining matter species are packaged
into $(\bar\rho,\delta\rho,\Theta)$, where $\Theta$ is the usual scalar velocity
potential or divergence source appearing in the $0i$ Einstein equation.

The perturbed Einstein equations take the form
\begin{align}
\label{eq:Perturbed_Friedmann2}
&
k^2\Psi+3\mathcal{H}\Psi'+ \frac{1}{2}\Bigl( \Bar{\chi}'\delta\chi'
+\frac{1}{2}\delta\langle\Bar{W}^2\mfa'^2\rangle\Bigr)\nn \\& 
+\frac{a^2}{2\MP ^2}\Bigl( 2\Phi V + V,_{\bar{\chi}}\delta\chi
+ \delta\langle V(\mfa)\rangle\Bigr) \nn\\&
+\frac{1}{2}\frac{W,_\chi}{W}\frac{\bar\rho_{\sax}}{a^2}\delta\chi
=-\frac{a^2}{2\MP ^2} \left(\delta\rho+2\Phi\bar{\rho}\right) \, ,
\end{align}

\begin{equation}
\label{eq:perturbed_0-i2}
k^2(\Psi'+\mathcal{H}\Phi)
-\frac{k^2}{2} \Bar{\chi}'\delta\chi
+\frac{1}{2}\langle W^2 k^i\partial_i \mfa\,\partial^0\mfa\rangle
=\frac{a^2\Bar{\rho}}{2\MP ^2}\Theta \, ,
\end{equation}

\begin{align}
\label{eq:perturbed_i-j2}
&\Psi''+\mathcal{H}\left(\Phi'+2\Psi'\right)
+\left(2\mathcal{H}'+\mathcal{H}^2\right)\Phi
+ \Bigl( \Bar{\chi}'^2 + \frac{\bar\rho_{\sax}}{a^2} \Bigr) \Phi \nn
\\
&\qquad
-\frac{1}{2} \Bigl(
\frac{1}{2}\delta\langle\Bar{W}^2\mfa'^2\rangle
+ \Bar{\chi}'\delta\chi'
+ \frac{W,_\chi}{W}\frac{\bar\rho_{\sax}}{a^2} \delta\chi
\Bigr)  \nn\\
&\qquad\qquad
+\frac{a^2}{2\MP ^2}
\Bigl( V,_{\Bar{\chi}} \delta\chi
+ \delta\langle V(\mfa)\rangle\Bigr)
= 0 \, .
\end{align}
 \Cref{eq:Perturbed_Friedmann2} is the $00$ (energy) constraint,
\cref{eq:perturbed_0-i2} is the $0i$ (momentum) constraint, and
\cref{eq:perturbed_i-j2} is the trace of the spatial $ij$ equation.
The axion sector enters only through the three averaged bilinears
$\delta\langle \bar W^2\mfa'^2\rangle$,
$\delta\langle V(\mfa)\rangle$, and
$\langle W^2 k^i\partial_i\mfa\,\partial^0\mfa\rangle$,
which correspond, respectively, to the perturbations of its kinetic energy
density, potential energy density, and momentum density (energy flux).
These are the combinations that can be recast into an effective
fluid description.

\subsubsection{Averaging identities}
\label{sec:Averaging_identities}

When the axion oscillation frequency is large compared to the Hubble rate, the
stress--energy tensor of $\mfa$ can be averaged over many oscillations at fixed
$\eta$. The resulting fluid variables are the axion density contrast
$\delta_\mfa$ and velocity divergence $\Theta_\mfa$, defined so that the
averaged physical energy density is $\bar\rho_{\sax}$ and the linear
perturbations take the standard ``CDM-like'' form on large scales. The mapping
is implemented by
\begin{align}\label{eq:definitions}
&\delta\langle\Bar{W}^2\mfa'^2\rangle
= a^2\Bar{\rho}_{\sax}\delta_\mfa
- 2a\Bar{\rho}_\sax\frac{S'}{m(\eta)} \,,
\\
&\delta\langle V(\mfa)\rangle
= \frac{\Bar{\rho}_{\sax}}{2}\delta_\mfa
\qq{and}
\langle W^2 k^i\partial_i \mfa\,\partial^0\mfa\rangle
= -\frac{\Bar{\rho}_{\sax}}{\MP^2}\Theta_\mfa \, .
\end{align}
The first identity shows that the averaged kinetic perturbation contains not only
the density-perturbation piece, but also a term proportional to $S'/m(\eta)$,
which encodes the axion momentum (and hence departures from perfectly
pressureless behaviour at finite $k$). The last identity makes the
interpretation of $\Theta_\mfa$ manifest: it is the axion contribution to the
total momentum density that sources the $0i$ Einstein constraint
\cref{eq:perturbed_0-i2}.

Finally, the axion phase obeys the relation
\begin{equation}
    \frac{S'}{a\,m(\eta)} = - \Phi_\chi - \Phi_{\ssQ} - \Phi \, ,
\end{equation}
which makes explicit how gravitational potentials and scalar-sector corrections
source the axion momentum: $\Phi$ is the usual Newtonian potential,
$\Phi_\chi$ parametrizes the dilaton backreaction, and $\Phi_{\ssQ}$ is the
``quantum pressure'' (gradient-energy) contribution responsible for the
scale-dependent sound speed of fuzzy or coherent axion matter.
Substituting \cref{eq:definitions} into
\cref{eq:Perturbed_Friedmann2}--\cref{eq:perturbed_i-j2}, and using the phase
relation to eliminate $S'$, yields a closed system written purely in terms of
$(\Phi,\Psi)$, $\delta\chi$, and the effective axion fluid variables
$(\delta_\mfa,\Theta_\mfa)$. This is the starting point for the quasi-static
reduction in \Cref{ssec:qsa_details}, where one algebraically combines the
constraints to obtain the effective Poisson equation and the modified growth
equation quoted in the main text.

The explicit $\chi$-dependence through $W,_\chi/W$ means that $\delta\chi$
sources the metric even at fixed $\delta_\mfa$, providing the microscopic
origin of an effective, time-dependent modification of the gravitational
coupling in the quasi-static regime.

\subsection{Quasi-static Regime Details}
\label{ssec:qsa_details}

In the sub-horizon quasi-static approximation (QSA), $k\gg\mathcal{H}$, we
neglect time derivatives of $\Phi$, $\Psi$, and $\delta\chi$ wherever they are
suppressed relative to $k^2$ terms. In this limit the metric constraint and the
$\chi$-fluctuation equation become algebraic.

From the $00$ Einstein equation in Newtonian gauge
\cref{eq:Perturbed_Friedmann2} we obtain
\begin{equation}
\label{eq:qsa_poisson}
k^2\Phi \;=\; -\frac{a^2}{2\MP^2}\left(
\bar{\rho}_\sax\delta_\sax
+ V_{,\chi}\,\delta\chi
+ \delta\rho_B
\right)\, .
\end{equation}

Keeping only the leading gradient and mass terms in the dilaton
Klein--Gordon \cref{eq:dilaton_perturbed} gives
\begin{equation}
\label{eq:qsa_deltachi_sol}
\delta\chi \;=\; \frac{a^2}{\MP^2\,k_\chi^2}
\left(
\frac{W_{,\chi}}{W}\bar\rho_\sax\,\delta_\sax
-\mathbf{g}\,\delta\rho_B
\right)\, .
\end{equation}

Substituting \cref{eq:qsa_deltachi_sol} into \cref{eq:qsa_poisson} expresses
$\Psi$ entirely in terms of $\delta_\ax$ and $\delta\rho_B$.

In the fluid description for the axion field we use the standard continuity
relation for the physical density perturbation
$\delta_\sax = \delta\rho_\sax/\bar{\rho}_\sax$, which is related to the
perturbed density parameter in the Madelung decomposition by
$\delta_\sax = \delta_\ax - \Phi - \Phi_\chi$ from \cref{eq:cont_ax},
\begin{equation}
\label{eq:qsa_cont}
\delta_\sax' = -\Theta_\ax \, ,
\end{equation}
together with the Euler \cref{eq:euler-equation}, written as a sum of
potentials,
\begin{equation}
\label{eq:qsa_euler}
\Theta_\ax' + \left(\cH + \frac{m'}{m}\right)\Theta_\ax
= k^2\left(\Phi + \Phi_\chi + \Phi_Q\right) \, .
\end{equation}
Here the scalar-mediated contribution is sourced by spatial fluctuations of
$m(\chi)$,
\begin{equation}
\label{eq:Phi_chi_def}
\Phi_\chi \;\equiv\; \frac{m_{,\chi}}{m}\,\delta\chi
\;=\;\frac{m_{,\chi}}{m}\,\frac{1}{\MP^2\,k_\chi^2}
\left(
\frac{W_{,\chi}}{W}\bar\rho_\sax\,\delta_\sax
-\mathbf{g}\,\delta\rho_B
\right) \, ,
\end{equation}
and $\Phi_Q$ denotes the (linearised) quantum-pressure potential of the MFA
(e.g.\ the usual fuzzy-DM term; we keep the explicit form used in the main
text).

After differentiating \cref{eq:qsa_cont} once and inserting
\cref{eq:qsa_euler}, one obtains
\begin{equation}
\label{eq:qsa_delta_mfa}
\delta_\sax'' + \left(\cH + \frac{m'}{m}\right)\delta_\sax'
= -k^2\left(\Phi + \Phi_\chi + \Phi_Q\right) \, ,
\end{equation}
matching the final quasi-static equation quoted in the main text.
Substituting the expressions for the relevant potentials
\cref{eq:qsa_poisson}, \cref{eq:Phi_chi_def}, and \cref{eq:PhiQPhiChidefs}
into the right-hand side of \cref{eq:qsa_delta_mfa} then gives
\cref{eq:quasistatieq}.

\section{Constraints on time-varying couplings}
\label{sec:constraints_universal}

One of the main features of the model that is responsible for the successful
description of cosmological data is the dilaton dependence of ordinary particle
masses, which leads to a mass evolution over cosmological times that the data
appear to favour. But we know a great deal about particle masses, and so this
appendix examines the various constraints on their variation to verify that the
required evolution is not already ruled out.\footnote{For a complementary review on couplings between generic scalar fields and other matter species see \cite{Baryakhtar:2025uxs}}

We argue that it is not, and that the key to understanding why is the
proportionality of all masses to the same universal conformal factor $A(\chi)$
defined in \cref{eq:jf_metric}. It is the evolution of this factor with time
that feeds into atomic physics, baryon loading, and late-time structure, but
crucially drops out of all mass ratios.

\begin{table*}
\centering
\setlength{\tabcolsep}{3pt}
\renewcommand{\arraystretch}{1.20}
\begin{tabular}{@{}llp{3.1cm}p{3.6cm}p{3.8cm}@{}}
\hline
\textbf{Epoch / probe} 
& \textbf{$z$ range} 
& \textbf{Observable} 
& \textbf{Mass dependence} 
& \textbf{Constraint} \\
\hline
BBN 
& $z \sim 10^{9}$--$10^{7}$ 
& $Y_p$, D/H, $^3$He, $^7$Li 
& $\Delta m_{np}$, $Q_{n\beta}$, $B_D$ 
& $|\Delta\!\ln A|_{\rm BBN}\!\lesssim\!10^{-2}$ \\[3pt]

CMB (recombination) 
& $z \simeq 1100$ 
& Peaks, damping tail 
& $m_e$ via $E_H$, $\sigma_T$ 
& $|\Delta\!\ln A|_{z\simeq1100}\!\lesssim\!2\times10^{-2}$ \\[3pt]

CMB (drag) 
& $z \simeq z_d \sim 10^{3}$ 
& Sound horizon $r_s(z_d)$ 
& $m_B$, $m_e$ via $R = 3\rho_B/4\rho_\gamma$ 
& $|\Delta\!\ln A|_{\rm drag}\!\lesssim\!(6$–$14)\!\times\!10^{-3}$ \\[3pt]

CMB (post-rec.) 
& $10^{4}\!\gtrsim\!z\!\gtrsim\!10^{2}$ 
& $\tau$, kSZ, $\mu/y$ distortions 
& $m_e$ through $\sigma_T$, ionisation 
& $|\Delta\!\ln A|_{\rm post}\!\lesssim\!{\cal O}(10^{-1})$ \\[3pt]

BAO / LSS 
& $z \lesssim 3$ 
& $D_H/r_s$, $D_M/r_s$, $f\sigma_8$ 
& $\Omega_B$, $\Omega_m$ via $A(\chi)$ 
& $|\Delta\!\ln A|_{z<3}\!\lesssim\!\text{few}\!\times\!10^{-2}$ \\[3pt]

21\,cm 
& $z \sim 10$--$200$ 
& Global signal, $P_{21}(k)$ 
& $\mu$, residual $x_e(m_e)$ 
& $|\Delta\mu/\mu|\!\sim\!10^{-3}$ (forecast, non-univ.) \\[3pt]

\hline
Astrophysics 
& $z \lesssim 1$ 
& Stellar tracks, SNe~Ia 
& Environment-dependent $A(\chi)$ (screened) 
& None if universal; strong if broken \\[3pt]

Spectroscopy 
& $0 \lesssim z \lesssim 4.5$ 
& Atomic / molecular lines 
& $\mu$ (inversion, vib/rot) 
& $|\Delta\mu/\mu|\!\lesssim\!3\times10^{-8}$ (non-univ.) \\[3pt]

Geochemical 
& $z \simeq 0$ 
& Resonances, decay $Q$ 
& $m_q/\Lambda_{\rm QCD}$ vs.\ $m_e$ 
& $10^{-7}$–$10^{-6}$ (non-univ.) \\[3pt]

Laboratory 
& $z \simeq 0$ 
& Clocks, EP tests 
& $\dot\mu/\mu$, $\dot\alpha/\alpha$ 
& $<\!10^{-17}\,{\rm yr}^{-1}$; $\eta\!\lesssim\!10^{-13}$ \\
\hline
\end{tabular}
\caption{
Main epochs and observables sensitive to universal variations of particle
masses, together with the approximate redshift ranges over which the universal
rescaling factor $A(\chi)$ is constrained.
Here $Y_p$ is the primordial $^4$He mass fraction, D/H the primordial
deuterium-to-hydrogen ratio, and $^3$He and $^7$Li label the corresponding
primordial abundances of these elements.
$\Delta m_{np}\equiv m_n-m_p$ is the neutron--proton mass splitting,
$Q_{n\beta}$ the neutron $\beta$-decay $Q$ value, and $B_D$ the deuteron
binding energy.
``kSZ'' denotes the kinetic Sunyaev--Zel'dovich effect, and $\mu/y$ are the CMB
spectral-distortion monopole parameters.
$D_H$ and $D_M$ are the radial and transverse BAO distance measures~\cite{BOSS:2017fdr},
$f\sigma_8$ is the standard linear-growth observable~\cite{Reid:2014iaa}, and
$P_{21}(k)$ the 21\,cm power spectrum.
In the laboratory row, $\dot\mu/\mu$ and $\dot\alpha/\alpha$ denote present-day
drifts of $\mu$ and of the fine-structure constant $\alpha$, while $\eta$ is the
E\"otv\"os parameter quantifying the fractional differential acceleration of two
test bodies~\cite{MICROSCOPE:2022doy}.
Quoted bounds should be interpreted as indicative
$\mathcal{O}(1\sigma$--$2\sigma)$ constraints on the mean drift in $\ln A(\chi)$
over the indicated redshift windows. Spectroscopic, geochemical, and many local
bounds apply only if the coupling becomes non-universal or environmentally
screened.
}
\label{tab:mass_constraints}
\end{table*}

Because $A(\chi)$ appears only through a rescaling of the matter metric, it does
not appear at leading order in the electromagnetic and strong gauge couplings
$\alpha$ and $\alpha_s$, while all dimensionful Standard-Model masses rescale
coherently with the same factor of $A(\chi)$,
\begin{equation}
m_i(z)=A(\chi)\,m_i^{(0)} \, .
\end{equation}
The same is also true for the QCD scale if its UV reference value scales in the
same way as other masses,
\begin{equation}
\Lambda_{\rm QCD}(z)=A(\chi)\,\Lambda_{\rm QCD}^{(0)} \, .
\end{equation}

When this is the case, all dimensionless ratios are $A(\chi)$-independent, such
as the proton-to-electron mass ratio and the ratio of light quark masses to the
QCD scale,
\begin{equation}
\mu(z) := \frac{m_p}{m_e}=\text{const} \, ,
\qquad
\frac{m_q}{\Lambda_{\rm QCD}}=\text{const} \, ,
\end{equation}
at least up to loop-suppressed threshold and running effects that are
observationally negligible for the range of $\Delta\!\ln A$ relevant here (see,
e.g.,~\cite{Uzan:2010pm,Martins:2017yxk,Sherrill:2023zah,Vorotyntseva:2024xzv}).

As a consequence, well-measured observables that depend only on these ratios are
ineffective at constraining by how much $A(\chi)$ can change. In particular,
any spectroscopic, molecular, or geochemical bounds on variations of $\mu$ or
$m_q/\Lambda_{\rm QCD}$~\cite{Sherrill:2023zah,Vorotyntseva:2024xzv} are
automatically satisfied in the strictly universal limit, provided these tests
rely on the variation of dimensionless combinations evaluated at the same
epoch.

Cosmological sensitivity to the $z$-dependence of masses can arise from
observables that depend on the ratio of a particle mass to the Planck scale, or
when particle scales are compared at different epochs. Examples include the
competition between particle reaction rates and the universal expansion rate,
such as during Big Bang Nucleosynthesis or recombination. Other examples
discussed below involve the drag epoch (baryon loading in $r_s$), and
post-recombination probes such as the optical depth $\tau$, kSZ effects, and CMB
spectral distortions, as we now summarise (with conclusions listed in \Cref{tab:mass_constraints}).

\subsubsection{Big Bang Nucleosynthesis}
\label{sssec:bbn-universal}

In the strictly universal, composition-independent limit, where all
Standard-Model masses scale with a common factor, BBN is insensitive provided
$A(\chi)$ does not vary during the BBN window
$T \simeq 3~\mathrm{MeV}\!\to\!30~\mathrm{keV}$.
If $A(\chi)$ is time independent over this period, the overall mass scale can
be absorbed by a change of units (frame), and the nuclear reaction network is
unchanged~\cite{Clara:2020efx,Coc:2006sx}.
For a universal conformal coupling, BBN therefore constrains only the time
variation of $A(\chi)$ across this window.

We parameterize the relevant effect by the net drift of $A(\chi)$ between weak
freeze-out and the opening of the deuterium bottleneck,
\begin{equation}
\Delta\ln A(\chi)\big|_{\rm BBN}
\;\equiv\;
\ln\!\frac{A(\eta_{\rm nuc})}{A(\eta_{\rm fr})} \, ,
\end{equation}
where $\eta_{\rm fr}$ approximately labels the epoch when the weak
interconversion rate $\Gamma_{n\leftrightarrow p}$ falls below the Hubble rate
$H$, fixing the neutron-to-proton ratio $n/p$, and $\eta_{\rm nuc}$ is the epoch
when photons no longer efficiently destroy deuterium, enabling light-element
formation.
A non-zero $\Delta\ln A|_{\rm BBN}$ shifts the weak-rate--expansion balance and
the onset of deuterium survival, thereby altering $Y_p$ and D/H.
With the baryon density $\rho_B$ fixed by the CMB, this \emph{single} parameter
captures the universal BBN response and is constrained at the percent
level\footnote{Helium and deuterium data limit such universal drift across the
BBN window to the percent level; see the time-dependent window-integral
constraint in Ref.~\cite{Sibiryakov:2020eir} and the universal Higgs--vev
rescaling bound $|\delta v/v|\!\sim\!0.01$ obtained in
Ref.~\cite{Burns:2024ods}.}:
\begin{equation}
|\Delta\ln A|_{\rm BBN} \ \lesssim\ 10^{-2} \, .
\end{equation}

When the coupling is universal, all masses drift together and BBN cannot
distinguish between sectors (electrons, quarks, nuclei).
Composition-dependent effects (e.g.\ isolated changes in $m_e/T$, $Q_{np}/T$,
or $B_D/T$) arise only if the coupling is slightly
non-universal~\cite{Flambaum:2007mj,Bedaque:2010hr}.
Additionally, because the coupling is conformal, it leaves the radiation energy
density, and hence $N_{\rm eff}$, unchanged, so BBN sensitivity arises
primarily through the weak-rate--Hubble balance and deuterium survival.

\subsubsection{CMB era}
\label{sssec:cmb}

The universal conformal scenario studied here affects recombination
microphysics through atomic binding energies, modifies the baryon loading and
hence the sound horizon before the drag epoch, and impacts
post-recombination observables through the same mass dependence of the Thomson
cross section. We summarise the relevant physics below.

\subsubsection*{Recombination ($z\simeq1100$): peaks and damping tail}
\label{sssec:cmb-recomb}

With $\alpha$ fixed, the atomic energy scales depend on the reduced mass
\begin{equation}
\mu_{\rm red} \equiv \frac{m_e m_p}{m_e+m_p} \;\;\propto\;\; A(\chi) \, ,
\end{equation}
so the hydrogen (and helium) binding energies scale as
\begin{equation}
E_{\rm H}\propto \mu_{\rm red}\,\alpha^2 \;\;\propto\;\; A(\chi) \, .
\end{equation}
Thomson scattering depends on the electron mass alone through the Thomson cross
section,
\begin{equation}\label{eq:Thompson_section}
\sigma_T=\frac{8\pi}{3}\frac{\alpha^2}{m_e^2} \;\;\propto\;\; A^{-2}(\chi) \, .
\end{equation}
A smooth drift in $A(\chi)$ across the (narrow) visibility window therefore
shifts the ionisation history, moves the acoustic scale $\theta_*$, and
rescales the photon-diffusion (Silk) damping scale.
A larger $A(\chi)$ raises $E_{\rm H}$, so recombination happens earlier and
photons have less time to diffuse, reducing small-scale damping.
The same change lowers $\sigma_T$, increasing the photon mean free path and
enhancing damping.
Since these effects go in opposite directions, comparing the acoustic peaks
(position and shape) with the high-$\ell$ damping tail tightly constrains
$A(\chi)$.
Additional atomic rates inherit the same $A(\chi)$ scalings; for example,
two-photon decay rates scale $\propto A(\chi)$.

Current CMB anisotropy analyses show that recombination physics is sensitive to
percent-level shifts in the electron mass. In the simplest case where
$\Delta\!\ln m_e$ is taken to be constant across the narrow visibility window,
Planck~2018 data alone allow several-percent excursions owing to strong
degeneracies with $H_0$ and the physical densities, but adding BAO data brings
the effective electron mass close to its local value, with
$m_e/m_{e,0}=1.0078\pm0.0067 \;\;(1\sigma)$~\cite{Hart:2019dxi}. This corresponds
to a 95\% CL bound $|\Delta\!\ln m_e|\lesssim2\times10^{-2}$ around
$z\simeq1100$. More general analyses that allow correlated changes in
recombination microphysics or include additional scalar degrees of freedom find
a similar percent-level sensitivity to $m_e$ at recombination, even when the
model is tuned to partially relieve the Hubble tension
\cite{Planck:2015fie,Hart:2019dxi,Baryakhtar:2024rky,Schoneberg:2024ynd}.

In the universal-mass scenario, $m_e\propto A(\chi)$, and the CMB
recombination-era constraints can therefore be mapped directly onto $A(\chi)$.
Taking the Planck+BAO constraints above as a representative benchmark, we infer
\begin{equation}
  \big|\Delta\!\ln A(\chi)\big|_{z\simeq 1100}
  \;\lesssim\;
  2\times10^{-2} \, ,
\end{equation}
with the detailed axio--dilaton posterior sharpening this bound once the full
parameter correlations of our model are taken into account.

\subsubsection*{Drag epoch ($z\simeq z_d$): the BAO anchor}
\label{sssec:cmb-drag}

Before the drag epoch the inertia of the photon--baryon fluid is set by the
baryon loading
\begin{equation}
R(z)\equiv \frac{3\rho_B}{4\rho_\gamma}
\ \propto\ A(\chi)\,\frac{n_B}{T_\gamma^{4}} \, .
\end{equation}
so the sound speed is $c_s = [3(1+R)]^{-1/2}$ and the sound horizon is
\begin{equation}\label{eq:r_s}
r_s(z_d) \;=\; \int_{z_d}^{\infty}\frac{c_s(z)}{H(z)}\,dz \, .
\end{equation}
A universal rescaling $A(z)$ therefore changes $r_s$ through two linked effects:
(i) a direct change of $R$ (hence $c_s$), and (ii) a shift of the drag redshift
$z_d$ via the recombination kinetics (earlier drag for larger $A$ because
$E_{\rm H}\!\propto\!A$). Expanding \cref{eq:r_s} to first order in small drifts
($\delta r_s$) and using the Leibniz rule, one finds\footnote{$\langle X\rangle_{\rm pre\text{-}drag}$
denotes a path-weighted average along the sound-horizon integral,
\begin{equation}
\big\langle X \big\rangle_{\rm pre\text{-}drag}\equiv
\frac{\displaystyle \int_{z_d}^{\infty} \frac{c_s}{H}\,X(z)\,dz}
     {\displaystyle \int_{z_d}^{\infty} \frac{c_s}{H}\,dz} \, ,
\end{equation}
and $\partial\ln r_s/\partial\ln(1+z_d)$ captures the dependence on the drag
redshift entering as the upper limit of the integral.}
\begin{align}\label{eq:sound_horizon_scaling}
\Delta\!\ln r_s \;\simeq\;&
-\frac{1}{2}\Big\langle\frac{R}{1+R}\Big\rangle_{\rm pre\text{-}drag}
\,\Delta\!\ln A(\chi) \nn\\
&\qquad+\;
\Big(\frac{\partial\ln r_s}{\partial\ln(1+z_d)}\Big)
\,\Delta\!\ln(1+z_d) \, .
\end{align}
As the dilaton rolls to larger values in our setup, $A(\chi)$ decreases with
redshift, and so the evolution of $A(\chi)$ adds to the natural growth of the
sound horizon.

Near the drag epoch the free–electron fraction $x_e$ is small and, over the
narrow visibility/drag window, is well tracked by the Saha
relation~\cite{Peebles:1970ag,1968ZhETF..55..278Z,Hu:1995en},
\begin{equation}
\frac{x_e^2}{1-x_e}
=\frac{1}{n_B}\left(\frac{m_e T_e}{2\pi}\right)^{3/2}e^{-E_H/T_e} \, ,
\end{equation}
where $x_e$ is the free–electron fraction and $T_e$ is the common
electron–baryon temperature. All the quantities in this expression have
well-defined scaling rules: $m_e\propto A(\chi)$, $E_{\rm H}\propto A(\chi)$,
$n_B\propto(1+z)^3$, and since rapid Compton scattering keeps the electrons
(and hence the baryon gas) in thermal equilibrium with the CMB throughout the
pre–drag window, $T_e\propto T_\gamma \propto (1+z)$.

Holding $x_e$ fixed and taking a total differential,
\begin{align}
0
&=\frac{3}{2}\,(d\ln m_e+d\ln T)-d\ln n_B-d(E_H/T) \nn\\
&=\Big(\frac{3}{2}-\frac{E_H}{T}\Big)d\ln A
+\Big(-\frac{3}{2}+\frac{E_H}{T}\Big)d\ln(1+z) \, ,
\end{align}
means that during the drag epoch we have
\begin{equation}
\Delta\!\ln(1+z_d)=\Delta\!\ln A \, ,
\end{equation}
implying that the characteristic redshift of the drag epoch shifts
proportionally to the scale of particle masses. Corrections from small
departures from Saha equilibrium, He\,I/He\,II details, and the mild
$A(\chi)$ dependence of $c_s/H$ are subleading across the narrow drag window
and induce only small departures from
$d\ln(1+z_d)/d\ln A = 1$.

Inserting this into \cref{eq:sound_horizon_scaling} gives
\begin{equation}\label{eq:sound_horizon}
\Delta\!\ln r_s
\simeq -\,\kappa_{rs}\,\Delta\!\ln A \, ,
\end{equation}
with $\langle R/(1+R)\rangle_{\rm pre\text{-}drag}\approx 0.20$ and
$\partial\ln r_s/\partial\ln(1+z_d)\approx -(0.6\text{--}0.7)$ verified
numerically for $\Lambda$CDM-like cosmologies, this yields
$\kappa_{rs}\sim 0.7\pm0.3$.

BAO observations measure distances in units of $r_s(z_d)$, while the CMB fixes
the acoustic angle $\theta_*$ to $0.03\%$ precision~\cite{Planck:2018vyg}.
Together, these data calibrate the physical sound horizon at the drag epoch to
sub-percent accuracy (e.g.\ $r_s = 147.2\pm0.3~\mathrm{Mpc}$ in a
Planck-anchored analysis~\cite{Arendse:2019hev}).

Using \cref{eq:sound_horizon_scaling}, a smooth, monotonic drift of the universal
mass scale across the pre-drag window therefore translates directly into a
fractional shift of the sound horizon. For a given response coefficient
$\kappa_{rs}\equiv -\partial\ln r_s/\partial\ln A$ we can write
\begin{equation}
  \big|\Delta\!\ln A\big|_{\rm drag}
  \;\lesssim\;
  \frac{2\sigma(r_s)}{r_s}\,\frac{1}{\kappa_{rs}}\,,
  \qquad
  \kappa_{rs}\simeq 0.7\pm0.3,
  \label{eq:A-drag-master}
\end{equation}
where we define $\sigma(r_s)$ as the $1\sigma$ uncertainty on the measured sound
horizon. Adopting current sub-percent calibrations of $r_s$,
$\sigma(r_s)/r_s \simeq (2\text{--}4)\times10^{-3}$, this yields the conservative
estimate
\begin{equation}
  \big|\Delta\!\ln A\big|_{\rm drag}
  \;\lesssim\;
  (6\text{--}14)\times10^{-3}\,,
  \label{eq:A-drag-numeric}
\end{equation}
which is a tight bound on the net mass-scale drift across
$10^4\!\gtrsim z\!\gtrsim z_d$. This estimate assumes a roughly standard
expansion history and a smooth evolution of $A(\chi)$ across the narrow
visibility/drag window. Sharp, instantaneous jumps in the scalar field during
this epoch (for example from an ultra-light species such as neutrinos becoming
non-relativistic and kicking the dilaton, sourcing $A(\chi)$~\cite{Brookfield:2005td,CarrilloGonzalez:2020oac})
would lie outside the regime of validity of this approximation.

\subsubsection*{Post--recombination ($10^4 \gtrsim z \gtrsim 10^2$): $\tau$, kSZ, spectral distortions}
\label{sssec:cmb-post}

After last scattering, a varying universal factor $A(\chi)$ enters CMB
observables mainly through the Thomson cross section and the free--electron
density. The integrated Thomson optical depth is
\begin{equation}
\tau \;=\; \int a\,n_e\,\sigma_T\, d\eta \, ,
\end{equation}
so, using $\sigma_T \propto A^{-2}(\chi)$ as in
\cref{eq:Thompson_section}, yields
\begin{equation}
\Delta\!\ln\tau \;\simeq\; \Delta\!\ln n_e \;-\; 2\,\Delta\!\ln A(\chi)
\end{equation}
for smooth evolution across the (fairly broad) reionisation window.
Large--scale CMB polarisation measures $\tau\simeq0.054\pm0.007$
\cite{Planck:2018vyg}, which conservatively restricts any late, order--unity
drift of $A(\chi)$ between recombination and reionisation:
\begin{equation}
\big|\Delta\!\ln A(\chi)\big|_{\rm post\text{-}rec}
\;\lesssim\;\mathcal{O}(10^{-1}) \, ,
\end{equation}
unless accompanied by a finely tuned change in the ionisation history.
The same $A^{-2}$ scaling controls the overall amplitude of the kinetic
Sunyaev--Zel'dovich signal, so current kSZ measurements at the
$\mathcal{O}(10\%)$ level do not yet improve this bound in the universal case.

At earlier post--recombination times,
$10^4\!\gtrsim\!z\!\gtrsim\!10^2$, a varying $A(\chi)$ also affects the efficiency
of Compton scattering and hence the expected $\mu$- and $y$-type spectral
distortions. COBE/FIRAS limits,
\begin{equation}
|\mu|\lesssim 9\times10^{-5}, \qquad |y|\lesssim 1.5\times10^{-5}
\quad (95\%~\mathrm{CL})
\end{equation}
\cite{Fixsen:1996nj}, constrain any scenario in which $A(\chi)$ changes by order
unity and would thereby boost the tiny $\Lambda$CDM distortions
($\mu\sim10^{-8}$, $y\sim10^{-6}$)~\cite{Chluba:2019kpb}. For the percent-level
drifts in $A(\chi)$ already allowed (and constrained) by recombination and
drag--epoch physics, these post--recombination observables mainly act as a
consistency check and do not drive our bounds. Future spectral--distortion
experiments with orders--of--magnitude better sensitivity would directly probe
$\Delta\!\ln A(\chi)$ at the $10^{-3}$ level in this redshift range,
complementing CMB anisotropy and BAO constraints
\cite{Chluba:2019nxa,Hill:2015tqa}.

\subsubsection{BAO and LSS}
\label{sssec:BAO}

Baryon acoustic oscillation (BAO) measurements provide a geometrical standard
ruler that is calibrated by the comoving sound horizon at the baryon--drag
epoch, $r_s(z_d)$. In the presence of a varying universal mass scale $A(\chi)$,
BAO constrain its evolution both before and after $z_d$ through two physically
distinct effects.

As shown in the previous subsection, variations of $A(\chi)$ prior to the drag
epoch modify the baryon--photon sound speed and hence the value of $r_s(z_d)$.
The combined Planck+BAO calibration of the sound horizon already limits the net
drift of $A(\chi)$ across the pre--drag window
$10^4 \gtrsim z \gtrsim z_d$ to the $\mathcal{O}\!\left(10^{-3}\right)$ level.

Mass variations switched on only after $z_d$ leave $r_s$ fixed but still affect
low--redshift observables by changing the background expansion and the growth
of structure. In our setup the baryon density follows
\begin{equation}
    \bar\rho_B(a) \propto A(\chi)\,a^{-3} \, ,
\end{equation}
while the role of CDM is played by the axion sector, whose averaged energy
density is governed by the time--dependent mass
$m(\eta)=m_a/W(\chi)$ and therefore does \emph{not} simply redshift as $a^{-3}$
(see \Cref{sec:Minimal_cosmology}). As a result, late--time distances and growth depend on both the
universal mass rescaling $A(\chi)$ and the axion mass evolution encoded in
$W(\chi)$, rather than on $A(\chi)$ alone.

BAO measurements constrain the comoving angular--diameter distance
$D_M(z)/r_s$ and the Hubble distance $D_H(z)/r_s$ over
$0.1 \lesssim z \lesssim 3$~\cite{BOSS:2017fdr}, while large--scale structure and
redshift--space distortions constrain the growth rate $f\sigma_8(z)$ in a
similar redshift range. In the axio--dilaton model these observables probe the
combined effect of $A(\chi)$ and the axion sector on $H(z)$ and
$\Omega_m(z)$. For the purposes of this section it is nonetheless useful to
summarise the low--redshift information in terms of an effective,
window--averaged drift of the universal mass scale,
\begin{equation}
    \Delta\!\ln A_{\rm low\text{-}z}
    \equiv
    \ln\!\left[\frac{A\big(\chi(z\sim 2\text{--}3)\big)}{A\big(\chi_0\big)}\right] \, ,
\end{equation}
which measures the net change in $A(\chi)$ across the redshift range where BAO
and LSS are most constraining. In our full joint CMB+BAO+SNe analysis
(\Cref{sec:results}), after marginalising over the axion parameters and the
dark--energy sector, we find that the data are consistent with
\begin{equation}
    \big|\Delta\!\ln A_{\rm low\text{-}z}\big|
    \lesssim \text{few}\times 10^{-2} \, ,
\end{equation}
corresponding to an effectively frozen $A(\chi)$ evolution over
$0 \lesssim z \lesssim 3$.

Taken together, BAO therefore provide two complementary windows on a universal
mass scale: an early--Universe bound on the pre--drag mass drift via the
sound--horizon calibration, and a late--Universe bound on any subsequent
evolution through the observed distance--redshift relation and structure
growth, encoded in $\Delta\!\ln A_{\rm low\text{-}z}$.

\subsubsection{21\,cm probes}
\label{sssec:21cm}

The hyperfine transition frequency scales as
$\nu_{21}\propto \alpha^4 (m_e/m_p) g_p$, making global and fluctuating 21\,cm
signals sensitive to variations in the proton--electron mass ratio $\mu$
during the dark ages and cosmic dawn ($10\lesssim z\lesssim200$).
Time-varying $m_e$ also modifies residual ionisation and Compton coupling
during this epoch.

The primary observable is the differential 21\,cm brightness temperature
relative to the CMB,
\begin{align}
    T_{21}(z,\mathbf{x})
    &\;\equiv\;
    T_S(z,\mathbf{x}) - T_\gamma(z)\nn
    \\&\;\simeq\;
    \frac{T_S(z,\mathbf{x}) - T_\gamma(z)}{1+z}\,
    \bigl(1 - e^{-\tau_{21}}\bigr),
\end{align}
where $T_S$ is the spin temperature, $T_\gamma$ is the CMB temperature, and
$\tau_{21}$ is the optical depth of the hyperfine line.
In the optically thin limit $\tau_{21}\ll1$ this is usually written as
\begin{align}
    T_{21}(z,\mathbf{x})
    \;\simeq\;&
    27\,x_{\mathrm{HI}}(z,\mathbf{x})\bigl[1+\delta_B(z,\mathbf{x})\bigr]
    \left(\frac{\Omega_B h^2}{0.023}\right)
    \nn\\&\times
    \left(\frac{0.15}{\Omega_m h^2}\,\frac{1+z}{10}\right)^{1/2}
    \left(1 - \frac{T_\gamma(z)}{T_S(z,\mathbf{x})}\right)\!{\rm mK},
\end{align}
where $x_{\mathrm{HI}}$ is the neutral hydrogen fraction and $\delta_B$ is the
baryon overdensity.

The spin temperature interpolates between the CMB temperature $T_\gamma$, the
gas kinetic temperature $T_K$, and the effective colour temperature of the
Ly$\alpha$ background $T_c$, according to
\begin{equation}
T_S^{-1}
 = \frac{T_\gamma^{-1} + x_c T_K^{-1} + x_\alpha T_c^{-1}}
        {1 + x_c + x_\alpha},
\end{equation}
where $x_c$ and $x_\alpha$ are the collisional and Ly$\alpha$ coupling
coefficients. Here $T_\gamma(z)$ is the homogeneous CMB temperature, while
$T_S$, $T_K$, $T_c$, $x_c$, and $x_\alpha$ may vary spatially and depend on
redshift.

Variations in particle masses or couplings enter these expressions through
several channels: the hyperfine splitting (via $\nu_{21}$ and hence the mapping
between observed frequency and redshift), the Compton coupling rate
$\Gamma_C\propto n_e\sigma_T\propto \alpha^2/m_e^2$ that sets the thermal
decoupling of baryons from the CMB, and the atomic cross sections and rate
coefficients that determine $x_c$ and $x_\alpha$.
In the strictly universal limit considered in this work, where $m_p$ and $m_e$
track the same factor $A(\chi)$ and $\alpha$ is fixed, $\mu=m_p/m_e$ and
$\nu_{21}$ remain constant, so 21\,cm probes are mainly sensitive to the
indirect impact of $A(\chi)$ on the thermal and ionisation history rather than
to a direct drift of the hyperfine frequency itself.

The complexity involved in evaluating the dependence of this signal warrants a
dedicated study, which is currently in progress~\cite{CHAN}. We do note,
however, that forecasts for constraints on non-universal rescalings of
fundamental constants are projected to be sensitive to
$\mathcal{O}(10^{-3})$ deviations~\cite{Lopez-Honorez:2020lno} in, e.g.,
$m_e$ from the Square Kilometre Array.

\subsubsection{Astrophysics}
\label{sssec:late_astrophysics}

In the minimal axio--dilaton scenario studied here, the dilaton field remains
light and unscreened on all astrophysical scales, so its value is essentially
homogeneous across stars, galaxies, clusters, and the Solar System. As a
result, all particle masses acquire the same cosmological value
$A(\chi_{\rm cos})$ and do not develop the environment--dependence
characteristic of screened scalar--tensor theories. Consequently, nuclear
thresholds, stellar structure, white--dwarf physics, and compact--object
properties remain unchanged relative to standard $\Lambda$CDM, and the strong
astrophysical bounds that apply to chameleon, symmetron, and thin--shell
models do not arise here
\cite{Damour:1994zq,Damour:2010rp,Uzan:2010pm}.

It is nevertheless useful to contrast this behaviour with what would occur if
the scalar were subject to a density--dependent screening mechanism. In
screened theories, the scalar minimises an effective potential
$V_{\rm eff}(\chi) = V(\chi) + A(\chi)\rho$, so that $\chi$ settles at
different equilibrium values inside stars, planets, galaxies, or laboratory
conditions than in the cosmological background. Even with universal
conformal couplings, for which all dimensionful Standard--Model parameters
rescale as $m_i \propto A(\chi)$ and all dimensionless ratios remain fixed, a
screened scalar would nonetheless generate environment--dependent absolute
mass scales. Since nuclear reaction thresholds, fusion rates, opacity,
electron degeneracy pressure, and the Chandrasekhar mass depend on
absolute masses rather than mass ratios, such variations generically modify
stellar evolution and compact--object structure
\cite{Damour:1994zq,Damour:2010rp,Coc:2006rt,Davis:2011qf}.
The absence of observed anomalies in helioseismology~\cite{Saltas:2019ius},
white--dwarf cooling~\cite{Jain:2015edg}, or neutron--star mass
distributions~\cite{Brax:2017wcj} therefore places tight constraints on any
screened scalar that induces environment--dependent mass rescalings.

Because the axio--dilaton considered in this work remains light and \emph{unscreened},
it does not generate such spatial variations. The only relevant constraints on
the universal conformal factor therefore arise from cosmological epochs in which
absolute mass scales directly enter the microphysics.

\subsubsection{Spectroscopy and supernovae}

Spectroscopic probes (stellar atmospheres, quasar absorption systems, molecular
lines) constrain shifts in transition frequencies relative to laboratory
values. These frequencies depend on the reduced mass of the electron–nucleus
system and therefore scale with the ratio $\mu \equiv m_N/m_e$, where $m_N$ is
the nuclear mass. For nuclei one may write $m_N \simeq A m_p - B_A$, with
binding energies $B_A$ set predominantly by $\Lambda_{\rm QCD}$~\cite{Brax:2010uq,Brax:2010gi}.

If the scalar only rescales fermion masses through the Higgs sector while
leaving $\Lambda_{\rm QCD}$ fixed, nucleon masses do not track $m_e$
coherently: the bulk gluonic binding energy and nuclear binding energies remain
unchanged while $m_e$ shifts. A percent-level variation in $m_e$ then induces
a percent-level variation in $\mu$, which is very strongly excluded.
State-of-the-art molecular and quasar absorption spectroscopy constrain
$\mu$ at the $10^{-6}$–$10^{-7}$ level over $0\!\lesssim\!z\!\lesssim\!4.5$:
methanol and ammonia absorption systems at $z\simeq0.7$–$0.9$ yield
$|\Delta\mu/\mu| \lesssim 10^{-7}$~\cite{Malec_2010,Levshakov_2011}, while H$_2$/CO systems at
$z\sim 2$–$3$ give typical limits $|\Delta\mu/\mu|\lesssim\mathcal{O}(10^{-6})$
\cite{Bagdonaite:2013sia,Ubachs:2017zmg,Dapra:2016dqh}. For our purposes we may
summarise these results schematically as
\begin{equation}
    \big|\Delta\mu/\mu\big|(0\lesssim z\lesssim 4.5)
    \;\lesssim\; 10^{-6}\!,
\end{equation}
to be interpreted as an indicative $\mathcal{O}(1\sigma$–$2\sigma)$ bound on
non-universal mass rescalings.

In contrast, in a genuine dilaton scenario where $\Lambda_{\rm QCD}$ rescales
together with fermion masses, both $m_N$ and $m_e$ shift coherently.
Dimensionless ratios such as $\mu$ remain fixed, and all atomic and molecular
levels rescale by a common factor, which is observationally degenerate with
redshift. In this universal limit, spectroscopy does not impose additional
constraints.

Type~Ia supernovae are likewise insensitive in the universal limit.
Universal mass rescalings modify the Chandrasekhar mass and hence the
intrinsic peak luminosity, but the absolute magnitude $M_B$ is not predicted
from first principles and is fully marginalised over as a nuisance parameter
in all SN likelihood analyses. Any shift in $M_B$ induced by $A(\chi)$ is
therefore perfectly degenerate with the calibration of the distance ladder.
Without an external absolute--distance anchor (e.g.\ from CMB or BAO),
SNe~Ia do not provide independent constraints on $A(\chi)$~\cite{Zhao:2018gwk,Alestas:2022gcg}.

\subsubsection{Local tests}

Laboratory clock comparisons constrain present-day drifts in the
proton--electron mass ratio at the level
$|\dot{\mu}/\mu|\lesssim 10^{-17}\,\mathrm{yr}^{-1}$~\cite{Jansen_2014,PhysRevLett.113.210801}.
Such bounds are only relevant when $\mu$ varies; in the universal conformal
scenario considered here, where $m_p$ and $m_e$ track the same factor
$A(\chi)$, $\mu$ remains constant and clock limits impose no constraint.

Geochemical probes provide complementary bounds on historical changes in
nuclear physics. The Oklo natural fission reactor (1.8\,Gyr ago)~\cite{Fujii:1998kn} and
long-lived radioactive decays in meteorites (4.5\,Gyr ago) are sensitive to
shifts in nuclear resonance energies and reaction thresholds, which depend on
the interplay between $\Lambda_{\rm QCD}$, light-quark masses, and binding
energies.
If only fermion masses vary while $\Lambda_{\rm QCD}$ remains fixed, these
nuclear energy levels shift relative to $m_e$ and $m_p$, leading to
percent-level changes that are excluded at the
$10^{-7}$--$10^{-6}$ level
\cite{Petrov:2005pu}.
These bounds are once again evaded in the universal limit.

Equivalence--principle (EP) tests place some of the strongest constraints on
new long-range interactions.
A violation of the weak EP arises whenever the scalar couples
non-universally to different constituents of matter, producing
composition-dependent differential accelerations.
In the universal minimal axio--dilaton scenario studied here, no
composition-dependent force is generated, and laboratory EP tests
(e.g.\ torsion-balance or lunar-laser-ranging limits with
$\eta\lesssim10^{-13}$) impose no additional constraints on these models.
Screened scalar theories, by contrast, generally reintroduce
environment-dependent couplings and can lead to strong EP violations; because
the dilaton here remains light and unscreened on solar-system scales, no such
effects arise~\cite{Brax:2023qyp}.

\section{All Parameter Constraints}\label{Appendix: Removing me}

\begin{figure*}
    \centering
    \includegraphics[width=1.0\linewidth]{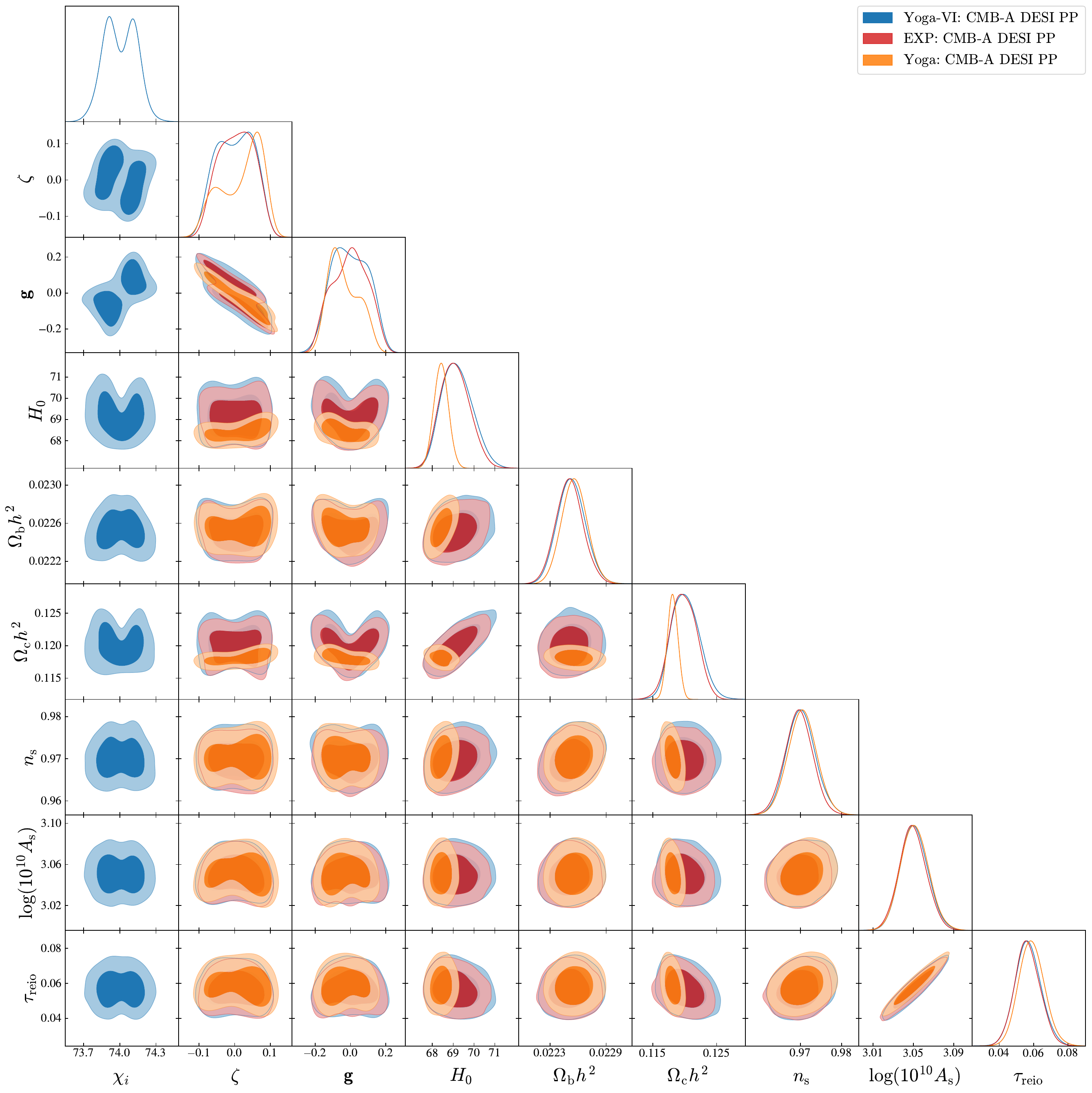}
\caption{Parameter constraints on all nine cosmological parameters of the models
(where applicable) using the CMB-A DESI PP dataset combinations.}
    \label{fig:triange full constraints}
\end{figure*}

\begin{figure*}
    \centering
    \includegraphics[width=1.0\linewidth]{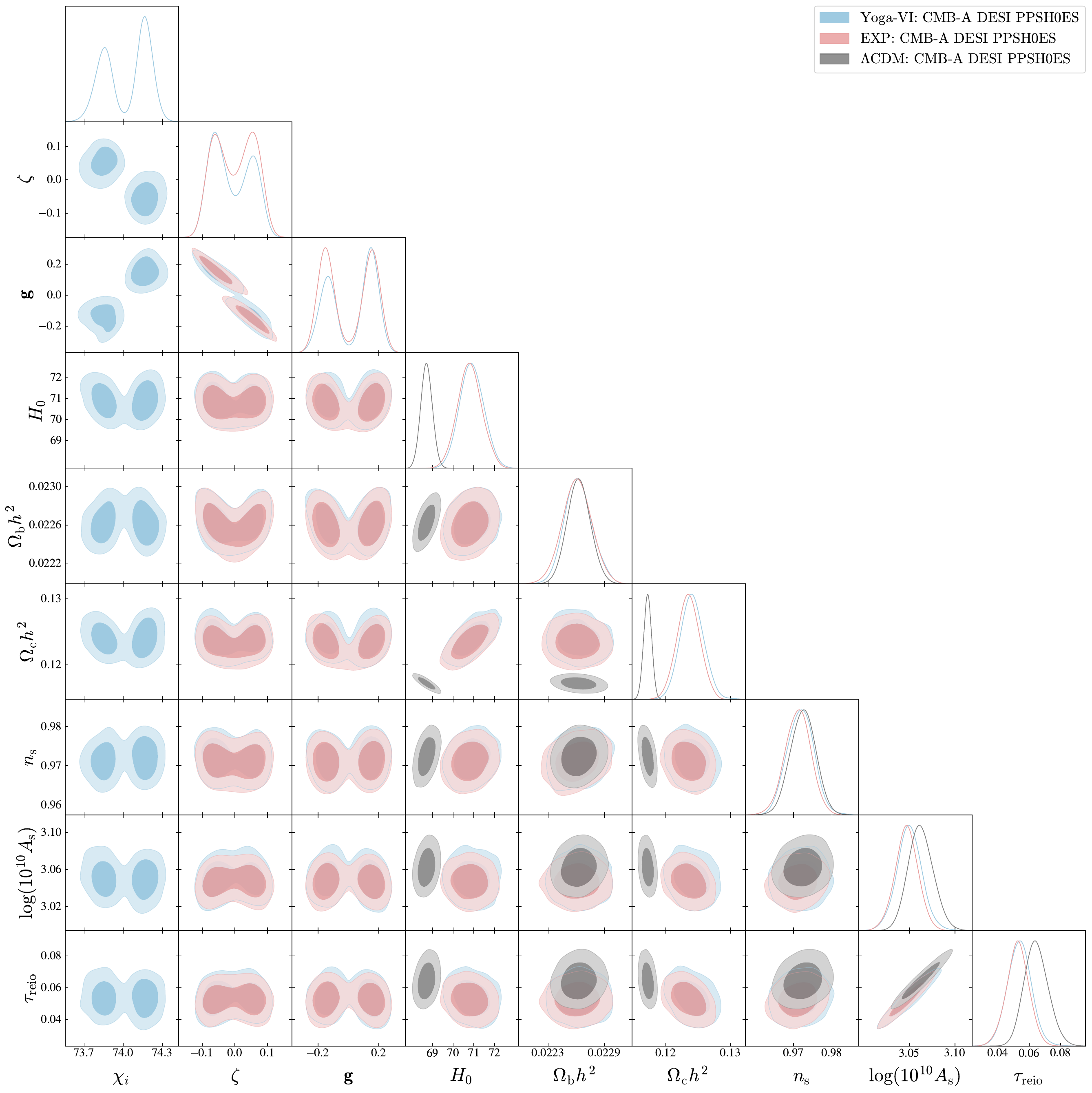}
\caption{Parameter constraints on all nine cosmological parameters of the models
(where applicable) using the CMB-A DESI PPSH0ES dataset combinations.}
    \label{fig:ppshoes_full_constraints}
\end{figure*}

\Cref{tab:yoga-exp full cosntraints} and Figures~\ref{fig:triange full constraints} 
and~\ref{fig:ppshoes_full_constraints} present the full posterior constraints,
including the usual six $\Lambda$CDM parameters, for the Yoga and EXP models
(for CMB-A and CMB-B, each combined with DESI BAO and Pantheon$+$).
The full triangle plots are particularly useful because they make parameter
correlations explicit: the shifts that support larger $H_0$ are not confined
to a single direction in parameter space, but occur together with compensating
changes in $\Omega_c h^2$ that preserve the main CMB structure (notably the
matter--radiation balance and the amount of CMB lensing smoothing) as the
acoustic ruler is altered.

\begin{table*}
\centering
\begin{tabular}{l|cc|cc}
\toprule
Parameter & Yoga-VI (CMB-A DESI PP) & EXP (CMB-A DESI PP)& Yoga-VI (CMB-B DESI PP) & EXP (CMB-B DESI PP) \\
\midrule
$H_0$ & $69.19^{+0.65}_{-0.83}$ (69.712) & $69.17 \pm 0.68$ (69.823) & $69.19 \pm 0.70$ (69.134) & $69.10^{+0.63}_{-0.77}$ (68.814) \\
$\Omega_b h^2$ & $0.0225 \pm 0.0001$ (0.0226) & $0.0224 \pm 0.0001$ (0.0225) & $0.0224 \pm 0.0001$ (0.0224) & $0.0225 \pm 0.0001$ (0.0225) \\
$\Omega_c h^2$ & $0.1201^{+0.0019}_{-0.0024}$ (0.121) & $0.120 \pm 0.002$ (0.121) & $0.121 \pm 0.002$ (0.121) & $0.120 \pm 0.002$ (0.120) \\
$n_s$ & $0.970 \pm 0.004$ (0.973) & $0.969 \pm 0.003$ (0.968) & $0.969 \pm 0.003$ (0.970) & $0.970 \pm 0.003$ (0.971) \\
$A_s$ & $3.051 \pm 0.014$ (3.051) & $3.046 \pm 0.009$ (3.044) & $3.046 \pm 0.009$ (3.054) & $3.049 \pm 0.013$ (3.050) \\
$\tau$ & $0.057 \pm 0.007$ (0.057) & $0.053 \pm 0.004$ (0.050) & $0.053 \pm 0.004$ (0.056) & $0.057 \pm 0.007$ (0.055) \\
$\mathbf{g}$ & $0.00 \pm 0.10$ (0.137) & $0.004^{+0.11}_{-0.095}$ (-0.082) & $0.003 \pm 0.095$ (-0.052) & $-0.003 \pm 0.099$ (0.127) \\
$\zeta$ & $0.001 \pm 0.052$ (-0.061) & $-0.002 \pm 0.045$ (0.010) & $0.002 \pm 0.050$ (-0.003) & $0.005^{+0.055}_{-0.050}$ (-0.058) \\
$\chi_i$ & $74.00 \pm 0.13$ (74.099) & -- & $74.02 \pm 0.15$ (73.844) & -- \\
\bottomrule
\end{tabular}
\caption{Posterior means ($\pm 1\sigma$) with best-fit values in parentheses for the
Yoga-VI and EXP models using the CMB-A/CMB-B DESI PP datasets.} 
\label{tab:yoga-exp full cosntraints}
\end{table*}

\begin{figure}
    \centering
    \includegraphics[width=1\linewidth]{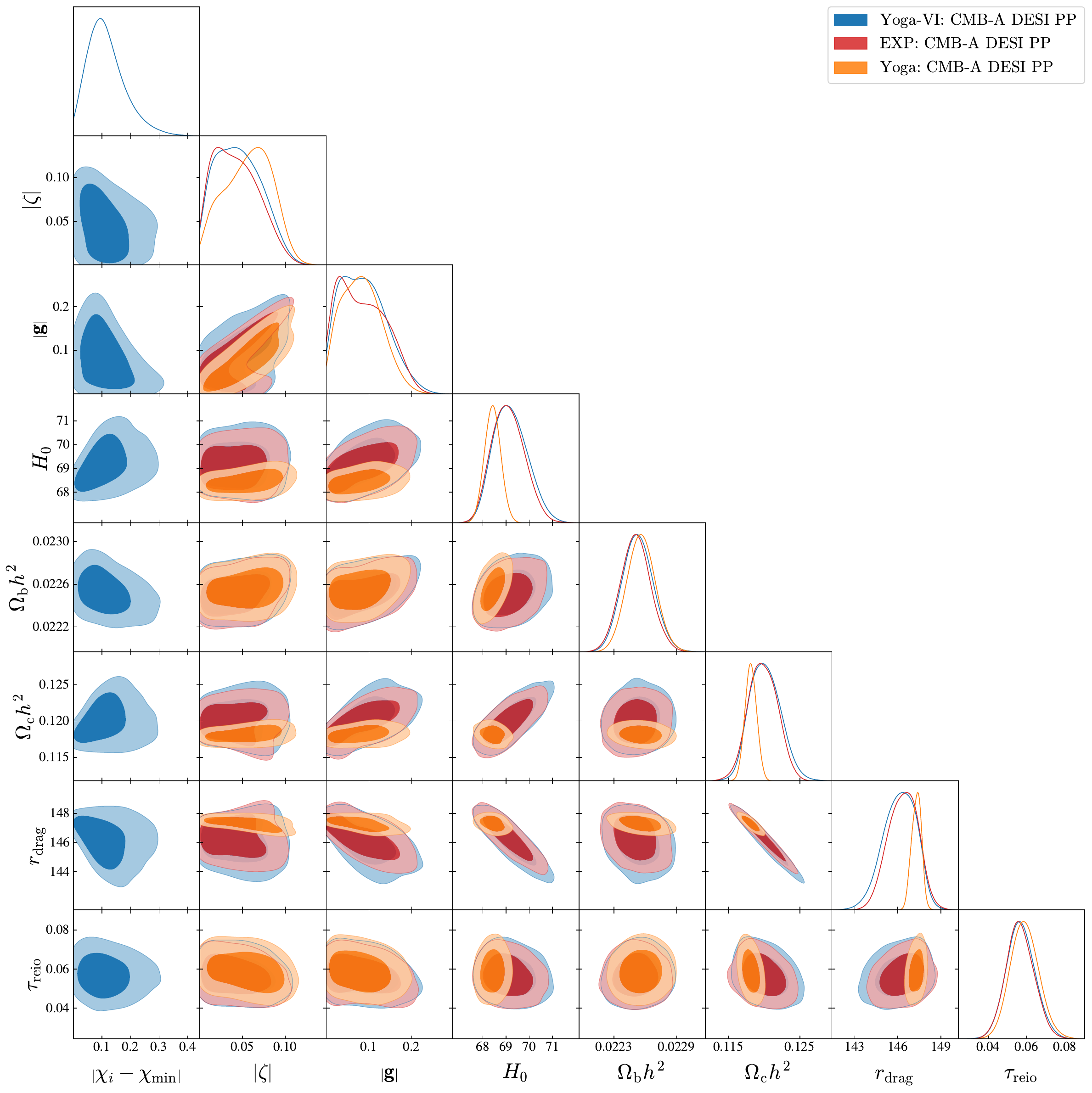}
\caption{Cross constraints on the most recombination-relevant parameters, along with the
derived sound horizon $r_s(z_{\rm drag})$, for the CMB-A DESI PP dataset combinations.
To filter out the symmetric nature of the couplings and field displacement around zero
in this plot, we show the constraints on the magnitude of these three parameters.
Here $\chi_{\rm min}$ is the value of $\chi$ at the bottom of the quadratic local
minimum of its potential.}
    \label{fig:recombination triangle}
\end{figure}

\Cref{fig:recombination triangle} isolates the part of parameter space most
directly responsible for this effect by showing the cross-constraints between
the couplings and the initial displacement, together with the derived drag-epoch
sound horizon $r_{\rm drag}\equiv r_s(z_{\rm drag})$.
Because the likelihood is approximately symmetric under sign flips of the
couplings (and of the displacement about the local minimum), the signed
posteriors can appear bimodal when the data prefer a nonzero effect.
We therefore plot the magnitudes $|g|$, $|\zeta|$, and
$|\chi_i-\chi_{\rm min}|$, which collapses the mirror solutions into a single
``typical amplitude'' constraint and makes the underlying physics more
transparent.

The dominant trend in \Cref{fig:recombination triangle} is a strong
anti-correlation between $H_0$ and $r_{\rm drag}$: moving to higher $H_0$ aligns
with a smaller drag sound horizon, as expected when BAO constrain late-time
distances in units of an early-time standard ruler.
The recombination-relevant magnitudes correlate with $r_{\rm drag}$ and
therefore project onto the same high-$H_0$ direction, indicating that the
preferred region is reached by shifting recombination microphysics (and hence
the ruler) rather than by a purely late-time modification of the expansion
history.
At the same time, $\Omega_b h^2$ remains tightly pinned, while $\Omega_c h^2$
supplies the principal compensating response, shifting along the usual CMB
degeneracy directions so that the acoustic peak pattern remains a good fit as
$r_{\rm drag}$ moves.
The optical depth $\tau_{\rm reio}$ is largely orthogonal to this recombination
sector, confirming that the effect is not being driven by reionisation
modelling.

A useful diagnostic from \Cref{fig:recombination triangle} is that the data
largely constrain a single ``amplitude'' direction in the space spanned by
$|g|$, $|\zeta|$, and $|\chi_i-\chi_{\rm min}|$: increasing one can be partly
compensated by decreasing another, consistent with the CMB being primarily
sensitive to the overall size of the recombination-era distortion rather than
its sign.
Moreover, only models that allow a genuine initial-condition freedom (visible
through a nontrivial $|\chi_i-\chi_{\rm min}|$ direction) populate the broader
high-$H_0$/low-$r_{\rm drag}$ branch, whereas the more restricted variant
occupies a narrower locus in $r_{\rm drag}$ and, correspondingly, a narrower
range of $H_0$.
This provides a simple interpretation of why the full posteriors can look deceptively consistent with zero couplings when marginalised, because the preference is for a nonzero magnitude, but the likelihood does not strongly distinguish the sign.

\subsection{Results with SDSS}\label{Appendix: SDSS}

As a robustness check on our late-time distance information, we repeat the
baseline analysis replacing the DESI DR2 BAO likelihood with the SDSS BAO
compilation, while keeping the CMB likelihoods and priors fixed.
BAO (baryon acoustic oscillations) provide a standard ruler through the sound
horizon at the baryon--drag epoch and therefore constrain geometrical distance
combinations.
The resulting posterior constraints for the Yoga-VI model are summarised in
Table~\ref{tab:yoga-sdss-compare}, shown both with and without Pantheon$+$.

The SDSS-based constraints are qualitatively consistent with our DESI-based
results: the inferred Hubble constant remains in the high-$H_0$ regime,
$H_0\simeq 68$--$69~\mathrm{km\,s^{-1}\,Mpc^{-1}}$, with $\Omega_c h^2$ and the
remaining $\Lambda$CDM parameters tightly constrained and close to their usual
CMB-preferred values.
Moreover, the model-specific parameters remain consistent with the
$\Lambda$CDM-like limit: both the matter coupling $\mathbf{g}$ and the parameter
$\zeta$ are consistent with zero within $1\sigma$ and retain large error bars
($\pm 0.05$).
Including Pantheon$+$ produces only mild shifts in central values and primarily
stabilises the late-time distance inference.

For reference, the final row of
Table~\ref{tab:yoga-sdss-compare} reports the difference in the minimum effective
chi-squared relative to $\Lambda$CDM, indicating a modest improvement of
$\Delta\chi^2\simeq -1.4$ for CMB-A+SDSS+PP.
Overall, our conclusions do not depend sensitively on whether DESI DR2 or SDSS
BAO data are used.

\begin{table*}
\centering
\begin{tabular}{l|cc}
\toprule
Parameter & Yoga-VI (CMB-A SDSS PP) & Yoga-VI (CMB-A SDSS)\\
\midrule
$H_0$ & $68.24^{+0.68}_{-0.96}$ (67.708) & $68.62^{+0.77}_{-1.1}$ (69.146)\\
$\Omega_b h^2$ & $0.0225 \pm 0.0001$ (0.0225) & $0.0225 \pm 0.0002$ (0.0225)\\
$\Omega_c h^2$ & $0.1201^{+0.0016}_{-0.0019}$ (0.117) & $0.1201^{+0.0017}_{-0.0021}$ (0.122)\\
$n_s$ & $0.968 \pm 0.004$ (0.969) & $0.969 \pm 0.004$ (0.967)\\
$A_s$ & $3.048 \pm 0.013$ (3.054) & $3.049 \pm 0.014$ (3.042)\\
$\tau$ & $0.056 \pm 0.007$ (0.059) & $0.056 \pm 0.007$ (0.053)\\
$\mathbf{g}$ & $-0.004 \pm 0.095$ (0.079) & $0.000 \pm 0.100$ (0.104)\\
$\zeta$ & $0.004 \pm 0.051$ (-0.060) & $0.002 \pm 0.052$ (-0.044)\\
$\chi_i$ & $74.02 \pm 0.11$ (73.965) & $74.00 \pm 0.12$ (74.108)\\
\midrule
$\Delta\chi^2$ & -1.4 & -- \\
\bottomrule
\end{tabular}
\caption{Posterior means ($\pm 1\sigma$) with best-fit values in parentheses for
the Yoga-VI model using the CMB-A SDSS PP dataset. The final row shows the
best-fit $\Delta\chi^2$ relative to the corresponding $\Lambda$CDM run.} 
\label{tab:yoga-sdss-compare}
\end{table*}

\end{document}